\documentclass{article}
\usepackage{iclr2027_conference,times}
\usepackage[T1]{fontenc}
\usepackage[utf8]{inputenc}
\usepackage{amsmath,amssymb,amsthm,mathtools}
\usepackage{graphicx,booktabs,array,multirow}
\usepackage{xcolor}
\usepackage{url}
\usepackage{algorithm}
\usepackage{algorithmic}
\usepackage{microtype}
\usepackage{placeins,float,flafter}
\usepackage[hidelinks]{hyperref}

\newcommand{\E}{\mathbb{E}}
\newcommand{\KL}{\operatorname{KL}}
\newcommand{\Ber}{\operatorname{Bern}}

\newcommand{\sg}{\operatorname{sg}}
\newcommand{\op}{\mathrm{op}}
\newcommand{\qt}{\textsc{qtDNN}}
\newcommand{\pre}{\operatorname{Pre}_{\theta}}
\newcommand{\post}{\operatorname{Post}_{\psi}}
\newcommand{\norm}[1]{\left\lVert #1\right\rVert}
\newcommand{\abs}[1]{\left\lvert #1\right\rvert}
\newtheorem{proposition}{Proposition}
\newtheorem{theorem}{Theorem}
\newtheorem{corollary}{Corollary}
\newtheorem{example}{Example}

\theoremstyle{remark}

\newcommand{\MainPQCMedian}{4,588}

\newcommand{\CompactMean}{5,088}

\newcommand{\CompactParameters}{32,261}
\newcommand{\CompactParameterRatio}{14.7}
\newcommand{\STETimeRatio}{18.6}
\newcommand{\STECallRatio}{318}
\newcommand{\EntropyCeiling}{7.716}

\newcommand{\EntropyNFiveThousand}{0.16}
\newcommand{\EntropyNFiveHundred}{1.28}
\newcommand{\EntropyNTenThousand}{1.92}
\newcommand{\EntropyNTenHundred}{4.62}

\title{Training Hybrid Quantum Policies with Classical Surrogates for Actor–Critic Reinforcement Learning}

\author{%
  \textbf{Jeremy Kulcsar}$^{1^\dagger,2}$ \quad
  \textbf{Giulio Giaconi}$^{1^\ddagger}$ \quad
  \textbf{Jie Luo}$^{3}$ \quad
  \textbf{Xueyin Chen}$^{4}$ \quad
  \textbf{Georgios Korpas}$^{1^\S,5}$ \\[2pt]
  $^1$Quantum \& Advanced Compute Group, HSBC, $^\dagger$Hong Kong, $^\ddagger$United Kingdom, $^\S$Singapore \\
  $^2$The Hong Kong University of Science and Technology, Hong Kong \\
  $^3$Anyon Computing Inc., USA \quad $^4$Airbnb Inc., USA \quad $^5$Czech Technical University, Czechia \\[2pt]
    \small\texttt{jeremy.kulcsar@hsbc.com.hk, giulio.giaconi@hsbc.com} \\
  \small\texttt{georgios.korpas@hsbc.com.sg, jluo@anyoncomputing.com, s.chen@airbnb.com}
}

\iclrfinalcopy

\hypersetup{
    pdftitle={Training Hybrid Quantum Policies with Classical Surrogates
              for Actor–Critic Reinforcement Learning},
    pdfauthor={eremy Kulcsar, Giulio Giaconi, Jie Luo, Xueyin Chen, Georgios Korpas}
}

\begin{document}
\maketitle
% Restore the template's running header outside its local title group.
% Recent fancyhdr versions scope \lhead locally; page dimensions are unchanged.

%\fancyhead[L]{Under review as a conference paper at ICLR 2027}

\fancyhead[L]{Preprint}

\begin{abstract}
Reading out a parametrized quantum circuit through finite-shot measurements introduces a discrete stochastic component into hybrid quantum-classical neural networks, making gradient-based training inefficient. In this work, we introduce a surrogate-assisted actor-critic method for training hybrid quantum policies that avoids differentiating the quantum circuit during backpropagation. A classical neural surrogate trained on the circuit's input-output pairs from a replay buffer provides a differentiable substitute for actor updates and critic-target computation, while forward interaction and evaluation retain the quantum circuit's measurements. Our experiments on standard MuJoCo continuous-control tasks demonstrate promising policy learning. Notably, on Humanoid, more runs reach high returns than with learned classical baselines. Moreover, the best average return in our ablation comes from a smaller circuit / surrogate, suggesting that strong performance does not depend on a large network. We also derive conditional upstream gradient-error bounds that separate surrogate fitting from the mean-feature relaxation. Together, the results support surrogate-assisted training as an effective and efficient approach to learning hybrid quantum policies.
\end{abstract}

\section{Introduction}
Parametrized quantum circuits (PQCs) have been used to represent policies and value functions in reinforcement learning \citep{jerbi2021parametrizedquantumpoliciesreinforcement,Skolik2022quantumagentsingym,jin2025ppoqproximalpolicyoptimization}. An $N$-qubit circuit transforms a quantum state in a $2^N$-dimensional Hilbert space, with measurements producing $N$-bit outcomes. Their probabilities depend on how the input is encoded and how the circuit transforms the state, providing a nonlinear feature map \citep{PhysRevLett.122.040504}. In a hybrid policy, classical networks learn to encode observations into circuit controls and map the measured features to actions \citep{jin2025ppoqproximalpolicyoptimization}.

The functions accessible to a quantum model depend on its input encoding, circuit structure and readout \citep{PhysRevA.103.032430}. Alongside experiments on standard RL benchmarks, \citet{jerbi2021parametrizedquantumpoliciesreinforcement} construct environments in which quantum agents have a provable learning advantage over polynomial-time classical learners, assuming the classical hardness of the discrete logarithm problem. This establishes a separation for particular problem families. Meanwhile, the practical value of quantum representations in continuous control still requires empirical study.

The practical challenge is to train the surrounding networks without repeatedly differentiating or querying the circuit on large replay minibatches. The cost can restrict the tasks that are studied: \citet{lokossou2025quantumdeepreinforcementlearning} evaluate a hybrid quantum soft actor-critic on Walker2d-v4 and identify the computational demands of Humanoid as a barrier to extending their experiments.

We study a policy whose circuit output is the empirical mode of $S$ measurement shots, that is, the complete bitstring observed most often. This bitstring provides the binary features passed to the action network. Related mode-based readouts have been used in variational quantum optimisation \citep{zhang2026resourceefficientquantumapproximateoptimization}. Ordinary backpropagation does not provide gradients through the measurement sampling and discrete mode-selection steps. Parameter-shift rules can differentiate appropriate quantum expectations \citep{Wierichs2022generalparameter}, but their application here must account for the full shot-aggregation operation.

We fit a classical neural surrogate to circuit control-output pairs stored alongside transitions in the replay buffer. Its predictions supply features for actor updates and critic-target computation without new circuit queries. Gradients train the encoder through the surrogate, whose weights are updated only by response fitting. Action selection and evaluation retain quantum measurements.

We make three contributions. First, we integrate replay-based surrogate fitting into hybrid actor-critic learning, removing circuit queries from minibatch updates. Second, we evaluate learning across four continuous-control tasks and examine training-seed variability, model capacity and deployment features. After training, replacing the circuit with the surrogate's predictions yields a purely classical actor that retains all ten high-return Humanoid seeds. It shares the architecture of the matched fully connected baseline (FCM), allowing the two training procedures to be compared at deployment (Appendix~\ref{app:deployment}). The training-time mechanism behind the observed difference remains unidentified. Third, we derive conditional bounds on the encoder's gradient error that separate surrogate-fitting error from the error of replacing sampled features by their mean.

\paragraph{Relation to prior work.}
Classical surrogates have been used for quantum-model inference \citep{PhysRevLett.131.100803,nair2025localsurrogatesquantummachine} and variational optimization \citep{PhysRevA.107.032415,O_Leary_2025}. In supervised classification, they provide differentiable proxies for quantum outputs \citep{Batista_2025,xie2025quantumsurrogatedrivenimageclassifier}. We extend this approach to off-policy actor-critic learning, fitting shot-aggregated responses from a changing replay distribution for actor updates and bootstrapped critic targets.

The method also relates to gradient estimators for stochastic neurons \citep{bengio2013estimatingpropagatinggradientsstochastic} and surrogate-gradient learning in spiking networks \citep{neftci2019surrogategradientlearningspiking}. Synthetic gradients predict backpropagated error signals directly \citep{pmlr-v70-jaderberg17a}, whereas local generative surrogates approximate stochastic simulators with differentiable models \citep{shirobokov2020blackboxoptimizationlocalgenerative}. Gumbel-Softmax and Concrete provide continuous relaxations of discrete samples \citep{jang2017categoricalreparameterizationgumbelsoftmax,maddison2017concretedistributioncontinuousrelaxation}, and RELAX combines a learned control variate with correction terms to obtain an unbiased gradient estimator \citep{grathwohl2018backpropagationvoidoptimizingcontrol}. In contrast, our surrogate predicts mean features, so differentiating through these predictions generally gives a biased update. We separate response-fitting and mean-feature relaxation errors and compare this update with a sampled-forward straight-through alternative.
\FloatBarrier
\section{Surrogate-assisted actor--critic training}
\label{sec:method}
\subsection{Hybrid quantum-classical policy}

The hybrid policy combines an $N$-qubit circuit with two classical networks: a pre-network $\operatorname{Pre}_\theta$ and a post-network $\operatorname{Post}_\psi$, parameterized by $\theta$ and $\psi$, respectively. Given a state $s$, the pre-network produces
$(u,c)=(u_\theta(s),c_\theta(s))=\operatorname{Pre}_\theta(s)$,
where $u\in[0,1]^d$ controls the circuit and $c\in[0,1]^{d_c}$ passes directly to the post-network through a bypass connection.

For circuit controls $u$, each measurement shot follows the Born rule distribution $\pi_u$ on $\{0,1\}^N$. The aggregation rule $A_S$ selects the most frequent complete bitstring from $S$ independent shots. The post-network then combines this bitstring with the bypass features to produce the action, following the interaction and evaluation path shown in Figure~\ref{fig:pipeline}:
\begin{equation}
 Z^{(1)},\ldots,Z^{(S)}\overset{\mathrm{iid}}{\sim}\pi_u,\qquad
 B=A_S(Z^{(1)},\ldots,Z^{(S)})\sim P_{u,S},\qquad
 a=\operatorname{Post}_\psi(B,c).
 \label{eq:actor}
\end{equation}

Note that the aggregated mean $\mu(u)=\E_{P_{u,S}}[B]$ generally differs from the single-shot mean $\E_{\pi_u}[Z]$. We call a tie when several bitstrings share the largest shot count, and our analysis assumes a fixed rule for selecting among them: for a measurable action set $D$, the deployed policy is the pushforward
\begin{equation}
 \Pi_{\theta,\psi}(D\mid s)
 =\sum_{b\in\{0,1\}^N}P_{u_\theta(s),S}(b)
 \mathbf 1\{\operatorname{Post}_\psi(b,c_\theta(s))\in D\}.
 \label{eq:policyclass}
\end{equation}
The parameters $\theta$ and $\psi$ therefore determine the deployed hybrid policy: the pre-network selects the circuit controls, and the post-network maps the measured bitstring and bypass features to an action. The surrogate introduced below is used to compute training updates, it is not evaluated in this hybrid action-selection path.

For a quantum circuit with $M$ layers, the pre-network supplies $2NM$ rotation controls and $2M$ routing controls, giving $d=(2N+2)M$. After rounding, the routing controls select the qubit pair connected by the entangling gate in each layer. Unless varied, $N=M=d_c=10$ and $S=1000$. Appendix~\ref{app:implementation} specifies the gate sequence and network architectures.

\begin{figure}[!htbp]
\centering\small
\begin{tabular}{@{}l@{}}
\textbf{Interaction and evaluation: retain quantum measurements}\\
$s\longrightarrow\pre\longrightarrow(u,c)\longrightarrow(B\sim P_{u,S},c)\longrightarrow\post\longrightarrow a$\\[3pt]
\textbf{Replay fitting: update only the surrogate}\\
Stored $(u,B)$ pairs $\longrightarrow$ BCE fit of $f_\phi(u)$\\[3pt]
\textbf{Actor and critic-target updates: no new circuit queries}\\
$s\longrightarrow\pre\longrightarrow(f_\phi(u),c)\longrightarrow\post\longrightarrow Q_\omega(s,a)$
\end{tabular}
\caption{Computational schematic of surrogate-assisted training. Top: the hybrid policy selects actions. Middle: stored control-output pairs train the surrogate. Bottom: surrogate predictions supply the features for actor updates and critic-target computation without new circuit queries. Critic targets use the corresponding target networks. Full network diagrams appear in Appendix~\ref{app:architecture}.}
\label{fig:pipeline}
\end{figure}

\subsection{Learning the circuit response}

A neural surrogate $f_\phi:[0,1]^d\to(0,1)^N$ that we call \texttt{qtDNN} predicts the aggregated marginal means. We fit it to stored $(u,B)$ pairs with per-bit binary cross-entropy (BCE):
\begin{equation}
 \widehat{\mathcal L}_{\mathrm{BCE}}(\phi)
 =-\frac{1}{|\mathcal M|N}\sum_{(u,B)\in\mathcal M}\sum_{k=1}^N
 [B_k\log f_{\phi,k}(u)+(1-B_k)\log(1-f_{\phi,k}(u))].
 \label{eq:surrogate}
\end{equation}
Each fit uses 32 Adam steps on size-16 resamples from the current 256-transition minibatch $\mathcal M$. The surrogate retains its parameters between fits. For the base PQC and classical stochastic channels, $\phi$ is updated only by BCE fitting: the reinforcement learning loss never updates it. During actor optimization, gradients pass through $f_\phi$ with respect to its inputs to train the encoder.

\subsection{Actor and critic updates}
Let $A_{\theta,\psi}^{\phi}(s)=\operatorname{Post}_{\psi}(f_\phi(u_\theta(s)),c_\theta(s))$ denote the deterministic actor used during training updates. An overbar (e.g., $\bar\omega$) marks target-network parameters. The critic target $y$ and actor loss $\mathcal L_A$ are
\begin{gather}
 y = r+\gamma(1-d_{\mathrm{term}})\min_{j=1,2}Q_{\bar\omega_j}
 \bigl(s',\operatorname{clip}_{\mathcal A}[A_{\bar\theta,\bar\psi}^{\bar\phi}(s')+\epsilon]\bigr),\label{eq:target}\\
 \mathcal L_A(\theta,\psi) = -\E_{s\sim\rho_{\mathcal M}}
 \!\left[Q_{\omega_1}\bigl(s,A_{\theta,\psi}^{\phi}(s)\bigr)\right].
 \label{eq:actorloss}
\end{gather}
Here $\epsilon$ is clipped target-policy smoothing noise, and $d_{\mathrm{term}}$ excludes time-limit truncations. The critics regress to the detached target $y$ using executed actions stored in replay. As in TD3 \citep{pmlr-v80-fujimoto18a}, actor and Polyak target updates occur every second critic update. The surrogate target is hard-copied immediately after fitting, so $\bar\phi=\phi$ when computing $y$ while only the pre-network, post-network and critics are Polyak-lagged. Section~\ref{sec:theory} analyzes the fixed-critic actor approximation and discusses the distinct target substitution. Algorithm~\ref{alg:training} gives the update order.
\FloatBarrier
\section{Experiments}
\label{sec:experiments}
We evaluate policy learning, outcomes across seeds, model capacity, and the update-time substitution. All circuits are simulated. The default setup uses ten qubits, ten layers, 1000 shots and a width-2048 surrogate. Evaluation retains quantum sampling but removes external exploration noise. Periodic scores average ten episodes and completed runs receive a separate terminal evaluation, normally over 30 episodes. The nominal training budget is 3M interactions. Protocols, run-completion status, run-level scores and post-hoc analyses
appear in Appendices~\ref{app:implementation}-\ref{app:results}.

\subsection{Policy learning across four environments}
\label{sec:tasks}
\begin{figure}[!htbp]
\centering
\includegraphics[width=.475\linewidth]{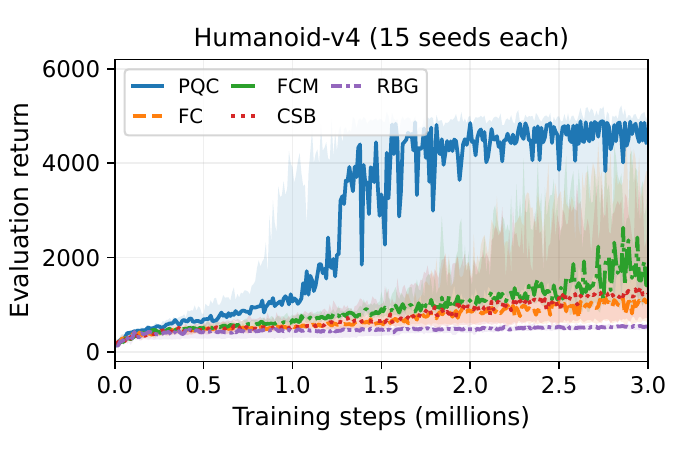}\hfill
\includegraphics[width=.475\linewidth]{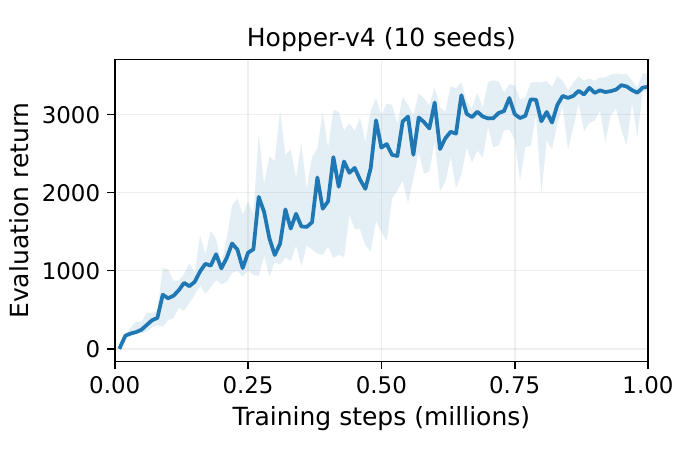}\\[2pt]
\includegraphics[width=.475\linewidth]{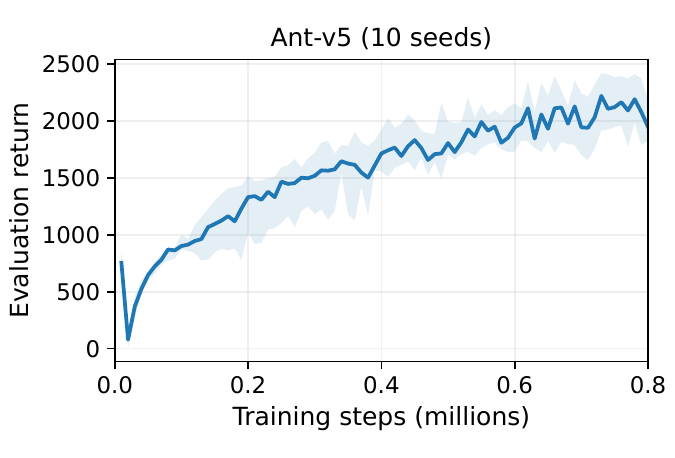}\hfill
\includegraphics[width=.475\linewidth]{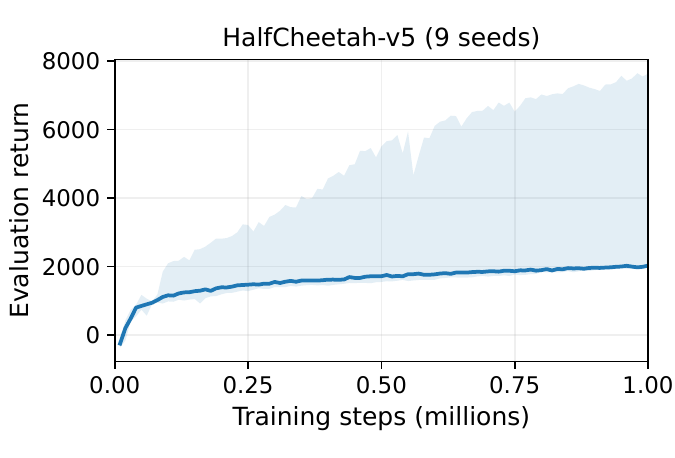}
\caption{\textbf{Surrogate-based updates train policies that retain quantum measurements.} Lines show median ten-episode returns; shading shows the interquartile range across a fixed seed cohort. Humanoid includes all five bottleneck configurations; each has 15 seeds. Appendix~\ref{app:perseed} shows individual trajectories for every task and Humanoid configuration.}
\label{fig:learning}\label{fig:maincurves}
\end{figure}

Figure~\ref{fig:learning} shows learning on Humanoid-v4, Hopper-v4, Ant-v5 and HalfCheetah-v5. The Humanoid comparison includes the quantum bottleneck (PQC), a fully connected linear-sigmoid bottleneck (FC), a fully connected bottleneck matched to the surrogate architecture (FCM), a classical stochastic bottleneck (CSB), and a random-bit generator (RBG). Section~\ref{sec:mainresults} describes how these configurations are trained. Hopper, Ant and HalfCheetah are summarized at fixed budgets using ten, ten and nine seeds, respectively (Table~\ref{tab:secondary}). Environment versions are fixed within each task and are not pooled across versions.

\begin{table}[!htb]
\centering\small\setlength{\tabcolsep}{4pt}
\begin{tabular}{lrrrr}
\toprule
Environment & Budget & Seeds & Median & IQM [95\% CI] \\
\midrule
Ant-v5 & 0.8M & 10/10 & 1,947 & 1,989 [1,540, 2,335] \\
Hopper-v4 & 1.0M & 10/10 & 3,355 & 3,394 [2,756, 3,498] \\
HalfCheetah-v5 & 1.0M & 9/10 & 2,027 & 3,133 [1,738, 6,759] \\
\bottomrule
\end{tabular}

\caption{Additional-task returns at the plotted budgets. Scores use the latest evaluation at or before each budget. Brackets give 95\% seed-bootstrap intervals for the trimmed-mean convention in Appendix~\ref{app:statistics}.}
\label{tab:secondary}
\end{table}

Hopper reaches median return 3355 and Ant 1947. HalfCheetah has a pronounced split: three runs exceed 7600 at 1M interactions while six remain below 2100, giving median 2027. The hybrid policy therefore learns on all four tasks, with consistency varying across seeds and environments. Longer-horizon summaries for nine-seed Ant and Hopper cohorts, and completed-run terminal scores, are reported separately in Appendix~\ref{app:perseed}.

\subsection{High-return outcomes and learning progress}
\label{sec:mainresults}
\subsubsection{Classical comparisons}
FC and FCM train their bottlenecks directly through the actor loss. PQC and CSB use surrogate-assisted updates, while RBG freezes the pre-network, including its bypass. The remaining actor-critic settings are shared (Appendix~\ref{app:controls}). Since PQC and CSB have different conditional output entropies, their comparison evaluates the stochastic feature map as a whole (Appendix~\ref{app:diagnostics}).

\begin{figure}[!htb]
\centering
\includegraphics[width=.48\linewidth]{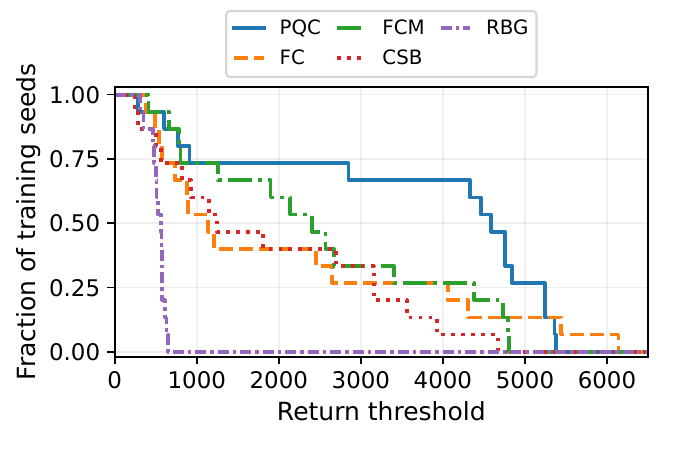}\hfill
\includegraphics[width=.48\linewidth]{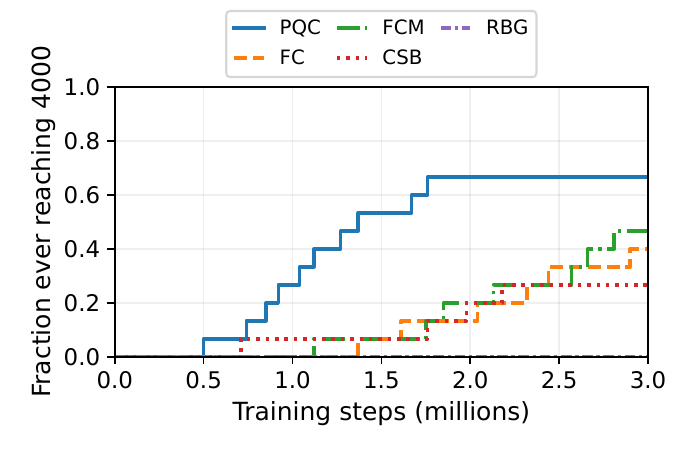}
\caption{\textbf{High-return outcomes and progress toward them.} Left: the fraction of 15 terminal scores meeting each return threshold. Right: the fraction of runs that have reached a periodic return of 4000 by each interaction budget, including transient crossings. The threshold and crossing analysis are post hoc. Pointwise intervals and run-level crossing times appear in Appendices~\ref{app:reliability} and~\ref{app:progress}.}
\label{fig:profiles}\label{fig:firstcrossing}
\end{figure}

\begin{table}[!htb]
\centering\small\setlength{\tabcolsep}{4pt}
\begin{tabular}{lrrrr}
\toprule
Bottleneck & Mean & Median & IQM [95\% CI] & $\ge4000$ \\
\midrule
PQC & 3,619 & 4,588 & 4,076 [2,513, 4,942] & 10/15 \\
FC & 2,121 & 1,138 & 1,616 [777, 3,064] & 4/15 \\
FCM & 2,509 & 2,399 & 2,387 [1,437, 3,553] & 4/15 \\
CSB & 1,911 & 1,246 & 1,722 [854, 2,766] & 1/15 \\
RBG & 521 & 558 & 537 [481, 577] & 0/15 \\
\bottomrule
\end{tabular}

\caption{Humanoid-v4 terminal returns after nominally 3M interactions, with 15 seeds and 30 terminal episodes per seed for every method. IQM intervals use 20,000 seed-bootstrap resamples. The 4000 threshold is descriptive.}
\label{tab:main}
\end{table}
The hybrid policy attains median return \MainPQCMedian\ and reaches 4000 in ten of 15 runs, compared with four for each learned classical baseline (Table~\ref{tab:main}). Figure~\ref{fig:profiles} shows the complete threshold profile. Late-training median returns also exceed 4000 in ten PQC runs, versus four FC, three FCM, two CSB and no RBG runs. Appendix~\ref{app:statistics} gives uncertainty intervals and the full-terminal and threshold tests. The PQC terminal scores in Table~\ref{tab:main} are hybrid evaluations. Appendix~\ref{app:deployment} shows that the stored surrogate actor can be deployed without the circuit while retaining the same ten high-return Humanoid seeds.

\subsubsection{Progress across training}
A complementary post-hoc measure averages each seed's evaluation returns at $0.1,0.2,\ldots,3.0$M interactions. Its median is 3144 for PQC, compared with 751, 959, 801 and 470 for FC, FCM, CSB and RBG. Exact Mann-Whitney tests favor PQC against both learned baselines after a four-comparison Holm correction ($p=0.026$ and $0.033$). Ten PQC runs first cross 4000 by 1.76M interactions, while over the full budget, FC, FCM and CSB cross in six, seven and four runs respectively (Figure~\ref{fig:firstcrossing}). The primary terminal-score tests remain inconclusive against FC and FCM (Holm-adjusted $p=0.135$ each), while four other descriptive measures (terminal high-return frequency, late-training returns, grid-averaged learning progress and first-crossing curves) favor PQC. See Appendix~\ref{app:progress} for details.

\subsection{Compact configurations and shot budget}
\label{sec:capacity}
We test widths $h\in\{64,256,1024,2048,4096\}$ at five and ten qubits, with three runs per setting. Five qubits and width 256 give the highest mean terminal return, \CompactMean, with all three scores above 4000. Its surrogate has \CompactParameters\ parameters, \CompactParameterRatio$\times$ fewer than the ten-qubit, width-2048 reference. The compact setting also exceeds 4000 in each seed's late-training median. The three-seed mean ranking is exploratory. We find that the ten qubits with width 1024 configuration has the highest median, suggesting that strong performance does not require the larger configurations tested here.

The shot sweep uses width 2048 and five runs per setting. At five qubits, reducing shots from 1000 to 100 raises mean return from 2537 to 4617, with all five lower-shot runs above 4000. At ten qubits, the corresponding means are 3301 and 4056. Changing shot count changes the aggregated output law as well as cost. A separate CPU characterisation at 200 random ten-qubit controls makes this change explicit: median aggregated entropy rises from \EntropyNTenThousand\ to \EntropyNTenHundred\ bits when reducing shots, under smallest-code tie-breaking. Five-qubit controls show the same direction of change (Appendix~\ref{app:entropy}). These measurements show how shot count changes the distribution of bitstrings returned by the circuit. However, they do not establish that higher entropy causes the improvement in policy returns. The reference re-executions share long training prefixes with primary seeds and are never pooled as extra seeds. Their later divergence is consistent with sensitivity to resumed sampling (Appendix~\ref{app:reexecutions}).

\begin{figure}[!htbp]
\centering
\includegraphics[width=.48\linewidth]{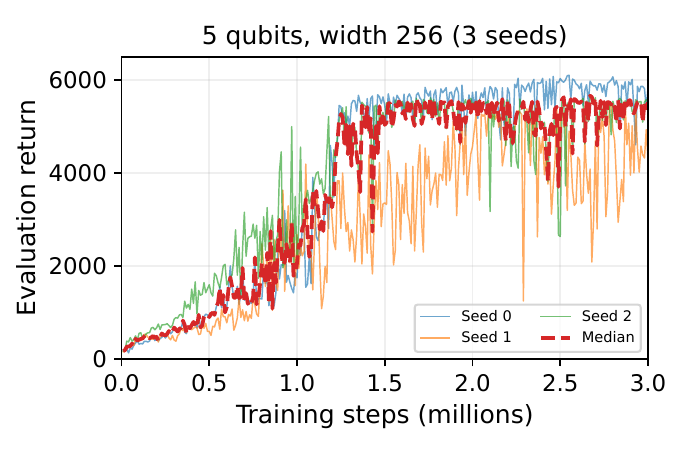}\hfill
\includegraphics[width=.48\linewidth]{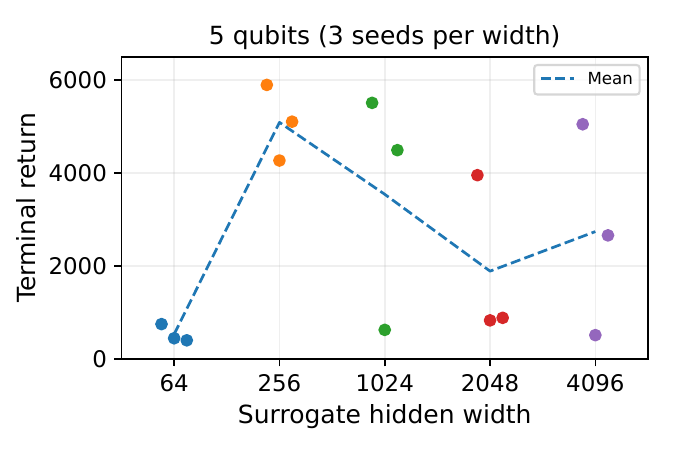}\\[2pt]
\includegraphics[width=.48\linewidth]{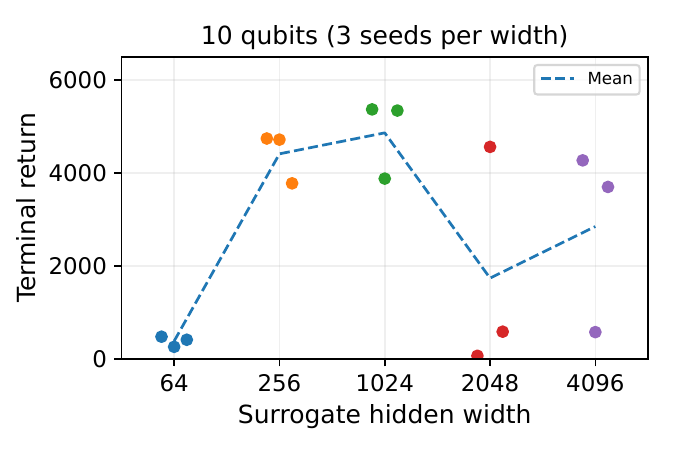}\hfill
\includegraphics[width=.48\linewidth]{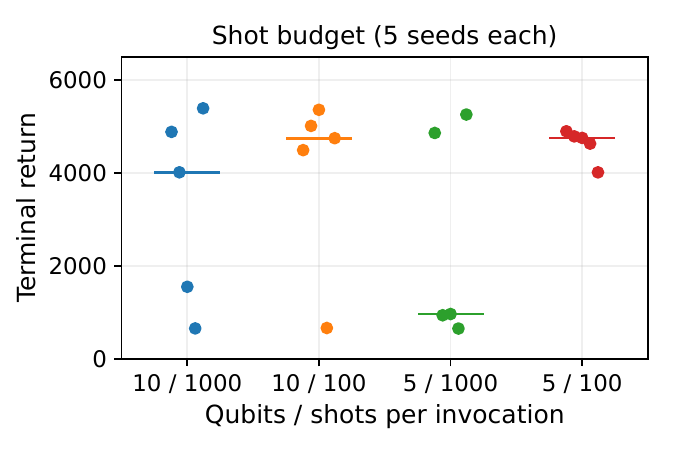}
\caption{\textbf{A compact circuit and surrogate lead the capacity sweep in mean return.} All three compact-setting trajectories and all terminal observations are shown. Dashed lines in the width panels join means, while shot-panel segments are medians. The ten-qubit, width-2048 and ten-qubit, 1000-shot reference cells are same-seed re-executions (Appendix~\ref{app:reexecutions}).}
\label{fig:compact}\label{fig:width}\label{fig:capacityshots}
\end{figure}

\subsection{Surrogate features during training updates}
\label{sec:cost}
We compare three paths for actor updates and critic-target computation. Surrogate-forward uses the continuous features $f_\phi(u)$. Hard-forward straight-through estimation (STE) uses $B+f_\phi(u)-\sg(f_\phi(u))$, where $\sg$ is the stop-gradient operator: it returns its argument unchanged but treats it as constant during backpropagation. This retains the sampled bitstring $B$ in the forward pass while using the surrogate's derivative with respect to $u$ in the backward pass. Hard-forward stop-gradient also uses sampled features but freezes the entire pre-network, including its bypass.

All three retain quantum measurements for action selection and evaluation. Comparing surrogate-forward with STE tests whether replacing sampled features with continuous surrogate predictions is an effective training choice. Comparing STE with stop-gradient tests pre-network learning as a whole, including the bypass, rather than only gradients through the surrogate.

One channel invocation evaluates a control vector using $S$ shots and returns their empirical mode. Counts include interaction, evaluation and update calls, and process time also includes surrogate fitting and other classical computation. Calls for optional diagnostics and derivative-target fitting are counted separately.

% We compare three ways of computing actor updates and critic targets. Surrogate-forward uses $f_\phi(u)$ as the feature vector during these computations. Hard-forward straight-through estimation (STE) uses $B+f_\phi(u)-\sg(f_\phi(u))$, whose forward value is the sampled bitstring $B$ and whose derivative with respect to $u$ is supplied by the surrogate. Here $\sg$ leaves its argument unchanged in the forward pass and assigns it zero derivative. The hard-forward stop-gradient configuration also uses sampled bitstrings but freezes the pre-network, including its bypass outputs.

% All three configurations use the quantum circuit when selecting actions in the environment and during evaluation. The comparison between STE and stop-gradient therefore tests learning of the entire pre-network. Since both the circuit-control and bypass branches are frozen in the stop-gradient configuration, it does not isolate the effect of gradients supplied through the surrogate.

% We report the number of calls to the quantum module and the elapsed process time. One call, termed a channel invocation, evaluates a single control vector $u$ using $S$ measurement shots and returns their most frequent bitstring. Thus a minibatch requiring quantum features at several control vectors incurs one invocation per vector. The counts include calls made during experience collection, policy evaluation and training updates. Elapsed time also includes surrogate fitting and the other classical computations. Calls used by optional diagnostics and derivative-target fitting are counted separately.
\begin{figure}[!htbp]
\centering
\includegraphics[width=.48\linewidth]{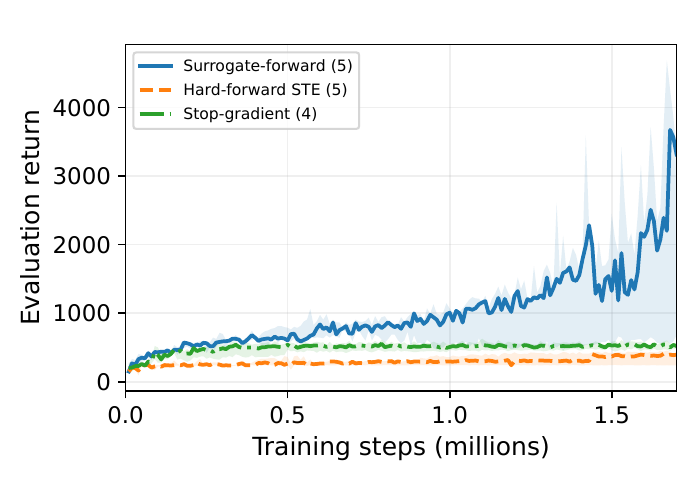}\hfill
\includegraphics[width=.48\linewidth]{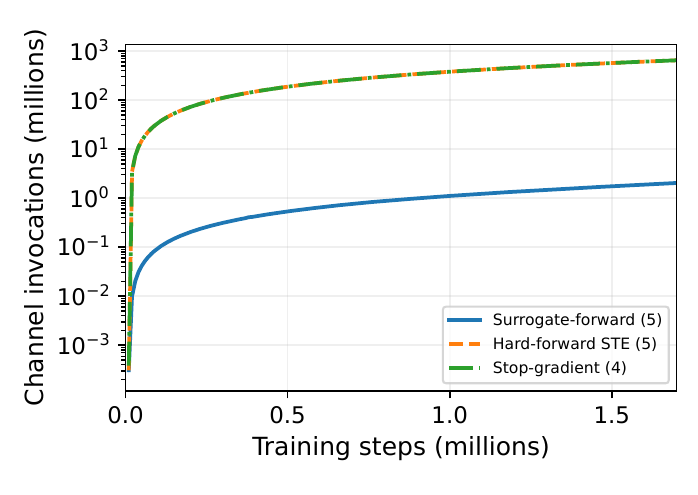}
\caption{\textbf{Surrogate-forward and sampled-forward updates.} Median evaluation returns (left) and cumulative channel invocations (right), with interquartile shading over fixed cohorts: five surrogate-forward, five hard-forward STE and four stop-gradient runs. The surrogate-forward cohort is primary PQC seeds 0-4 throughout the plotted 1.7M interactions.}
\label{fig:strategy}\label{fig:cost}
\end{figure}
\begin{table}[!htb]
\centering\small\setlength{\tabcolsep}{4pt}
\begin{tabular}{lrrrr}
\toprule
Update path & $n$ & Return & Hours & Calls \\
\midrule
Surrogate-forward & 5 & 3,316 & 19.1 & 2.05M \\
Hard-forward STE & 5 & 395 & 354.8 & 650.38M \\
Hard-forward stop-gradient & 4 & 521 & 331.6 & 650.46M \\
\bottomrule
\end{tabular}

\caption{Medians at the latest evaluation before 1.7M interactions (1.690--1.691M). Time is cumulative in-process wall time per seed, including evaluation and resumes but excluding queue waiting. This is a matched-interaction, not a time-to-target comparison. The surrogate arm is primary seeds 0--4.}
\label{tab:strategy}
\end{table}
\FloatBarrier
Surrogate-forward training reaches median return 3316, versus 395 for STE and 521 for stop-gradient (Table~\ref{tab:strategy}). It uses \STECallRatio$\times$ fewer channel invocations than STE and \STETimeRatio$\times$ less median logged time. At batch size 256 and actor delay two, both sampled-forward controls add 384 invocations per update on average; surrogate-forward adds none. The small-cohort return contrasts are uncertain: at 1.7M, raw two-sided $p$-values are $0.056$ against STE and $0.190$ against stop-gradient. Earlier checkpoints, four-seed sensitivity and all three pairwise tests appear in Appendix~\ref{app:strategydetail}. This comparison evaluates a change in the update objective as well as in query cost.

\FloatBarrier
\section{Analysis of the surrogate approximation}
\label{sec:theory}
We separate response-fitting error from the substitution of mean features in the actor objective. The analysis applies to PQC and CSB under the stated assumptions. Fix a state $s$, critic parameters $\omega$, and post-network parameters $\psi$, and define $G_s(z,c)=Q_{\omega_1}(s,\post(z,c))$. With $\mu(u)=\E_{P_{u,S}}[B]$, let
\begin{equation}
 F(u,c)=\E_{B\sim P_{u,S}}[G_s(B,c)],\quad
 F_\mu(u,c)=G_s(\mu(u),c),\quad
 F_f(u,c)=G_s(f_\phi(u),c).
 \label{eq:three}
\end{equation}
Here $F$ is the sampled fixed-critic objective, $F_\mu$ its mean-feature relaxation, and $F_f$ the surrogate update objective. The replay distribution is fixed; these are not environment-return objectives.

\subsection{The response learned from replay}
The base surrogate is fitted coordinatewise to the \emph{shot-aggregated} bitstring. Its population target is therefore the aggregated marginal mean, not the single-shot Born mean.
\begin{proposition}[BCE identifies aggregated marginals]
\label{prop:bce}
Let $u\sim\rho$ and $B\mid u\sim P_{u,S}$. For the population per-bit BCE risk $\mathcal L$ with natural logarithms and a measurable $f:[0,1]^d\to(0,1)^N$ of finite risk,
\begin{gather}
 \mathcal L(f)-\mathcal L(\mu)
 =\frac1N\E_{u\sim\rho}\!\left[\sum_{k=1}^N
 \KL\bigl(\Ber(\mu_k(u))\,\|\,\Ber(f_k(u))\bigr)\right],\nonumber\\
 \E_\rho\!\left[\norm{f-\mu}_2^2\right]\le\frac N2[\mathcal L(f)-\mathcal L(\mu)].
 \label{eq:bcebound}
\end{gather}
With the usual boundary convention for BCE, $\mu$ is the Bayes solution over the closed output cube, up to $\rho$-null sets. No independence between output bits is required.
\end{proposition}
The proof is the conditional cross-entropy identity followed by Pinsker's inequality (Appendix~\ref{app:bceproof}). The bound controls prediction error under $\rho$. A finite replay fit additionally incurs approximation, optimization, and estimation errors; derivative accuracy is a separate property.

\subsection{Error propagated to the state encoder}
Fix a state $s$ and write $u=u_\theta(s)$ and $c=c_\theta(s)$.
Let $p_\theta$ denote the number of encoder parameters. We use
Jacobians with output coordinates indexing rows and input coordinates
indexing columns:
\[
\begin{aligned}
J_f &:= \frac{\partial f_\phi(u)}{\partial u}
       \in \mathbb{R}^{N\times d},
&\qquad
J_\mu &:= \frac{\partial \mu(u)}{\partial u}
       \in \mathbb{R}^{N\times d},\\
J_\theta u &:= \frac{\partial u_\theta(s)}{\partial\theta}
       \in \mathbb{R}^{d\times p_\theta},
&\qquad
J_\theta c &:= \frac{\partial c_\theta(s)}{\partial\theta}
       \in \mathbb{R}^{d_c\times p_\theta}.
\end{aligned}
\]
All derivatives are evaluated at the current inputs and parameters.
The notation $\nabla_\theta F_\bullet$ denotes the gradient of
$F_\bullet(u_\theta(s),c_\theta(s))$ with respect to $\theta$,
holding the surrogate, post-network and critic parameters fixed.
We use Euclidean norms for vectors and the corresponding induced
operator norms for matrices.
\begin{theorem}[Conditional upstream-gradient error, including the bypass]
\label{thm:upstream}
Consider a differentiability point within a fixed routing cell. Write $f=f_\phi(u)$, $\mu=\mu(u)$, and assume
$\norm{f-\mu}_2\le\varepsilon_1$, $\norm{J_f-J_\mu}_\op\le\varepsilon_2$,
$\norm{J_\theta u}_\op\le M_u$, $\norm{J_\theta c}_\op\le M_c$, and
$\norm{J_\mu}_\op\le M_\mu$.
Suppose $G_s$ is continuously differentiable on a neighborhood of $[f,\mu]\times\{c\}$, with $\norm{\nabla_zG_s(f,c)}_2\le L_z$.
On the feature segment $[f,\mu]$, let $\nabla_zG_s(\cdot,c)$ and $\nabla_cG_s(\cdot,c)$ be $L_{zz}$- and $L_{cz}$-Lipschitz, respectively. Then
\begin{equation}
 \norm{\nabla_\theta F_f-\nabla_\theta F_\mu}_2
 \le M_uL_z\varepsilon_2+(M_uM_\mu L_{zz}+M_cL_{cz})\varepsilon_1.
 \label{eq:upstreambound}
\end{equation}
\end{theorem}
The three terms respectively capture Jacobian error, feature error through the circuit path, and feature error through the bypass (Appendix~\ref{app:upstreamproof}). An accurate BCE fit alone is insufficient. Routing boundaries and activation kinks lie outside this local result.

\subsection{Replacing sampled features by their mean}
Let $K=\operatorname{conv}(\operatorname{supp}P_{u,S})$ and $v(u)=\sum_k\mu_k(u)(1-\mu_k(u))$. Since $B$ is binary, $\E\!\left[\norm{B-\mu}_2^2\right]=v(u)\le N/4$, even with correlated bits. If $G_s(\cdot,c)$ is $L_z^{\mathrm{supp}}$-Lipschitz on $K$, then
\begin{equation}
 \abs{F-F_\mu}\le L_z^{\mathrm{supp}}\sqrt{v(u)}.
 \label{eq:valuegap}
\end{equation}
If $G_s$ is also $L_z^{\mathrm{fit}}$-Lipschitz on $[f,\mu]$, then
\begin{equation}
 \abs{F-F_f}\le L_z^{\mathrm{supp}}\sqrt{v(u)}+L_z^{\mathrm{fit}}\varepsilon_1.
 \label{eq:totalvaluegap}
\end{equation}
This value bound needs no Jacobian approximation. Live channel diagnostics give mean $\sqrt{v(u)}$ estimates of $1.40$--$1.47$ for PQC and $0.17$--$0.55$ for CSB, compared with the ten-bit maximum $1.58$ (Appendix~\ref{app:diagnostics}). They measure the variance factor, not the downstream Lipschitz constants. With a continuous Hessian bounded by $H$ on a neighborhood of $K$, Taylor expansion replaces the first term by $Hv(u)/2$.

For a channel law with fixed finite support and positive differentiable probabilities near $u$, define $\ell_u(B)=\nabla_u\log P_{u,S}(B)$. Holding $c$ fixed gives the exact identity
\begin{equation}
 \nabla_uF-\nabla_uF_\mu
 =\E\!\left[\left\{G_s(B,c)-G_s(\mu,c)
 -\nabla_zG_s(\mu,c)^\top(B-\mu)\right\}\ell_u(B)\right].
 \label{eq:scoreidentity}
\end{equation}
Under the Hessian condition its norm is at most
$\frac H2\E\!\left[\norm{B-\mu}_2^2\norm{\ell_u(B)}_2\right]$.
Extending this comparison to $\theta$ also requires a bound on the bypass derivative over the \emph{sampled support hull}: $\nabla_cG_s(\cdot,c)$ must be $L_{cz}^{\mathrm{supp}}$-Lipschitz on $K$. This adds $M_cL_{cz}^{\mathrm{supp}}\sqrt{v(u)}$ to the relaxation bound. The fitting-segment assumption in Theorem~\ref{thm:upstream} alone does not imply this support-wide condition. Corollary~\ref{cor:sampled} combines the fitting and relaxation contributions and states a sufficient local ascent condition.

For an affine map in the sampled features, the mean-feature relaxation is exact. For nonlinear maps, even a perfect surrogate can reverse the gradient (Appendix~\ref{app:example}). The source-generated tanh-map probes give median mean-feature/reference cosines of $0.994$--$1.000$, while fitted-surrogate directions are mixed. Matched probes with random LeakyReLU and post-network--critic maps also have high median alignment. By contrast, two oscillatory maps with identical values on every binary output give eight-qubit median cosines of $0.752$ and $0.294$, and opposing directions occur at individual anchors (Appendix~\ref{app:gmapfamilies}). Thus the off-vertex geometry matters: neither a benign relaxation nor derivative fitting follows from BCE alone.

\subsection{The straight-through direction and critic targets}
Let $A=J_\theta u$ and $C=J_\theta c$. Assuming $G_s$ is differentiable at the sampled feature points, the expected upstream STE update is
\begin{equation}
 \E[\widehat g^{\mathrm{STE}}_\theta]
 = A^\top J_f^\top\E[\nabla_zG_s(B,c)]
   +C^\top\E[\nabla_cG_s(B,c)].
 \label{eq:stegradient}
\end{equation}
It is generally not $\nabla_\theta F$, $\nabla_\theta F_\mu$, or $\nabla_\theta F_f$; replacing only the backward pass is a different approximation (Appendix~\ref{app:ste}). The critic is fitted on executed sampled actions, with exploration noise, but its target uses the relaxed actor. A value-gap bound also applies to this target under a Lipschitz assumption on the clipped, smoothed minimum of target critics. It does not establish Bellman consistency, critic accuracy, or convergence of the coupled training process (Appendix~\ref{app:targetgap}).

\section{Discussion and conclusion}
Surrogate-assisted training produces policies that learn on four continuous-control tasks while using quantum measurements to select actions. On Humanoid, the hybrid evaluations show more high-return outcomes than the learned classical baselines, and the post-hoc learning-progress summaries show a consistent pattern. A compact circuit and surrogate also achieve the highest average return in the capacity sweep. The approximation analysis distinguishes errors in fitting the circuit response from those introduced by mean-feature and straight-through updates.

The trained networks also support deployment without the circuit. Using the stored surrogate to supply the features retains the same ten high-return Humanoid seeds without further fitting (Appendix~\ref{app:deployment}). This classical actor has the FCM architecture, so the follow-up compares two training procedures for a shared final architecture; the primary tables continue to report hybrid evaluations. Constant-half and fair-bit features retain nine high-return seeds, while zeros retain six. These substitutions measure dependence on the deployed features, but do not explain why surrogate-assisted hybrid training reaches high returns more often in these runs than end-to-end FCM training.

\paragraph{Limitations.}
The experiments use simulated circuits and finite seed cohorts. Capacity rankings are exploratory; reference re-executions share primary-run prefixes. PQC and CSB are not entropy-matched, and HalfCheetah shows substantial seed variability. The directional probes use fixed synthetic maps, not trained critics. The relation between estimated channel entropy and outcome is correlational: the controls were collected under the surrogate actor, and no training-time mechanism is identified. The learning results show that surrogate-based updates can train useful hybrid policies without requiring the larger configurations tested here.

\label{lastmainpage}
\section*{Reproducibility statement}
Appendix~\ref{app:architecture} illustrates the network and training flow; Appendix~\ref{app:implementation} documents the implementation, experimental protocols and data coverage; and Appendix~\ref{app:statistics} specifies the statistical analysis. The theoretical assumptions and proofs are provided in Appendix~\ref{app:proofs}. Appendix~\ref{app:gradientdiagnostics} describes the synthetic probes, for which a complete repeat of all seven configurations agrees in every non-timing numerical field. The channel-characterization and downstream-map protocols are also detailed in the appendices.

We will make the code and analysis package available upon acceptance. The package will include run configurations, source hashes, per-seed scores, raw terminal evaluations, periodic training histories, control-level and probe-level records, and scripts for reproducing the tables and figures. Analysis of the recorded results requires no quantum simulator, and rerunning policy training requires the simulator and environment dependencies specified in Appendix~\ref{app:implementation}.

\section*{AI use statement}
The authors used generative AI to revise and polish the manuscript text, to assist with code review and debugging for the experiments and with producing code for selected supplementary studies in the appendix, and to help develop and check the mathematical arguments in the main text and appendices.

The authors have reviewed, verified and improved all AI-assisted work. The authors take full responsibility for the final content of this work, including the mathematical claims, the theoretical and experimental results, their verification and final presentation.

\section*{Ethics statement}
This work does not involve human subjects or personally identifiable data. Potential dual–use considerations arise from quantum computing research in general, but the presented method is limited to reinforcement–learning benchmarks and carries no direct security or safety risks. We note that training with large quantum simulations consumes substantial compute, and future work should continue to assess energy efficiency.

\section*{Disclaimer}
This paper was prepared for information purposes and is not a product of HSBC Bank Plc. or its affiliates. Neither HSBC Bank Plc. nor any of its affiliates make any explicit or implied representation or warranty and none of them accept any liability in connection with this paper, including, but not limited to, the completeness, accuracy, reliability of information contained herein and the potential legal, compliance, tax or accounting effects thereof. Copyright HSBC Group 2026.

\bibliography{references}

@InProceedings{pmlr-v80-fujimoto18a,
  title = 	 {Addressing Function Approximation Error in Actor-Critic Methods},
  author =       {Fujimoto, Scott and van Hoof, Herke and Meger, David},
  booktitle = 	 {Proceedings of the 35th International Conference on Machine Learning},
  pages = 	 {1587--1596},
  year = 	 {2018},
  editor = 	 {Dy, Jennifer and Krause, Andreas},
  volume = 	 {80},
  series = 	 {Proceedings of Machine Learning Research},
  month = 	 {10--15 Jul},
  publisher =    {PMLR},
  url = 	 {https://proceedings.mlr.press/v80/fujimoto18a.html}
}

@misc{zhang2026resourceefficientquantumapproximateoptimization,
      title={Resource-efficient quantum approximate optimization algorithm via Bayesian optimization and maximum-probability evaluation}, 
      author={Siran Zhang and Shuming Cheng},
      year={2026},
      eprint={2603.28413},
      archivePrefix={arXiv},
      primaryClass={quant-ph},
      url={https://arxiv.org/abs/2603.28413}, 
}

@misc{agarwal2022deepreinforcementlearningedge,
      title={Deep Reinforcement Learning at the Edge of the Statistical Precipice}, 
      author={Rishabh Agarwal and Max Schwarzer and Pablo Samuel Castro and Aaron Courville and Marc G. Bellemare},
      year={2022},
      eprint={2108.13264},
      archivePrefix={arXiv},
      primaryClass={cs.LG},
      url={https://arxiv.org/abs/2108.13264}, 
}

@misc{jerbi2021parametrizedquantumpoliciesreinforcement,
      title={Parametrized quantum policies for reinforcement learning}, 
      author={Sofiene Jerbi and Casper Gyurik and Simon C. Marshall and Hans J. Briegel and Vedran Dunjko},
      year={2021},
      eprint={2103.05577},
      archivePrefix={arXiv},
      primaryClass={quant-ph},
      url={https://arxiv.org/abs/2103.05577}, 
}

@article{Skolik2022quantumagentsingym,
  doi = {10.22331/q-2022-05-24-720},
  url = {https://doi.org/10.22331/q-2022-05-24-720},
  title = {Quantum agents in the {G}ym: a variational quantum algorithm for deep {Q}-learning},
  author = {Skolik, Andrea and Jerbi, Sofiene and Dunjko, Vedran},
  journal = {{Quantum}},
  issn = {2521-327X},
  publisher = {{Verein zur F{\"{o}}rderung des Open Access Publizierens in den Quantenwissenschaften}},
  volume = {6},
  pages = {720},
  month = may,
  year = {2022}
}

@misc{jin2025ppoqproximalpolicyoptimization,
      title={PPO-Q: Proximal Policy Optimization with Parametrized Quantum Policies or Values}, 
      author={Yu-Xin Jin and Zi-Wei Wang and Hong-Ze Xu and Wei-Feng Zhuang and Meng-Jun Hu and Dong E. Liu},
      year={2025},
      eprint={2501.07085},
      archivePrefix={arXiv},
      primaryClass={quant-ph},
      url={https://arxiv.org/abs/2501.07085}, 
}

@article{PhysRevLett.131.100803,
  title = {Classical Surrogates for Quantum Learning Models},
  author = {Schreiber, Franz J. and Eisert, Jens and Meyer, Johannes Jakob},
  journal = {Phys. Rev. Lett.},
  volume = {131},
  issue = {10},
  pages = {100803},
  numpages = {6},
  year = {2023},
  month = {Sep},
  publisher = {American Physical Society},
  doi = {10.1103/PhysRevLett.131.100803},
  url = {https://link.aps.org/doi/10.1103/PhysRevLett.131.100803}
}

@misc{nair2025localsurrogatesquantummachine,
      title={Local surrogates for quantum machine learning}, 
      author={Sreeraj Rajindran Nair and Christopher Ferrie},
      year={2025},
      eprint={2506.09425},
      archivePrefix={arXiv},
      primaryClass={quant-ph},
      url={https://arxiv.org/abs/2506.09425}, 
}

@article{PhysRevA.107.032415,
  title = {Surrogate-based optimization for variational quantum algorithms},
  author = {Shaffer, Ryan and Kocia, Lucas and Sarovar, Mohan},
  journal = {Phys. Rev. A},
  volume = {107},
  issue = {3},
  pages = {032415},
  numpages = {10},
  year = {2023},
  month = {Mar},
  publisher = {American Physical Society},
  doi = {10.1103/PhysRevA.107.032415},
  url = {https://link.aps.org/doi/10.1103/PhysRevA.107.032415}
}

@article{O_Leary_2025,
   title={Efficient online quantum circuit learning with no upfront training},
   volume={8},
   ISSN={2399-3650},
   url={http://dx.doi.org/10.1038/s42005-025-02423-4},
   DOI={10.1038/s42005-025-02423-4},
   number={1},
   journal={Communications Physics},
   publisher={Springer Science and Business Media LLC},
   author={O’Leary, Tom and Czarnik, Piotr and Pelofske, Elijah and Sornborger, Andrew T. and McKerns, Michael and Cincio, Lukasz},
   year={2025},
   month=Nov }

@misc{bengio2013estimatingpropagatinggradientsstochastic,
      title={Estimating or Propagating Gradients Through Stochastic Neurons for Conditional Computation}, 
      author={Yoshua Bengio and Nicholas Léonard and Aaron Courville},
      year={2013},
      eprint={1308.3432},
      archivePrefix={arXiv},
      primaryClass={cs.LG},
      url={https://arxiv.org/abs/1308.3432}, 
}

@article{Wierichs2022generalparameter,
  doi = {10.22331/q-2022-03-30-677},
  url = {https://doi.org/10.22331/q-2022-03-30-677},
  title = {General parameter-shift rules for quantum gradients},
  author = {Wierichs, David and Izaac, Josh and Wang, Cody and Lin, Cedric Yen-Yu},
  journal = {{Quantum}},
  issn = {2521-327X},
  publisher = {{Verein zur F{\"{o}}rderung des Open Access Publizierens in den Quantenwissenschaften}},
  volume = {6},
  pages = {677},
  month = mar,
  year = {2022}
}

@inproceedings{Batista_2025,
   title={Medical Imaging Classification with Cold-Atom Reservoir Computing Using Auto-Encoders and Surrogate-Driven Training},
   url={http://dx.doi.org/10.1109/QAI63978.2025.00064},
   DOI={10.1109/qai63978.2025.00064},
   booktitle={2025 IEEE International Conference on Quantum Artificial Intelligence (QAI)},
   publisher={IEEE},
   author={Batista, Nuno and Morgado, Ana and Ferraz, Oscar and Pratapsi, Sagar Silva and Lobo, Jorge and Falcao, Gabriel},
   year={2025},
   month=Nov, pages={373–380} }

@misc{grathwohl2018backpropagationvoidoptimizingcontrol,
      title={Backpropagation through the Void: Optimizing control variates for black-box gradient estimation}, 
      author={Will Grathwohl and Dami Choi and Yuhuai Wu and Geoffrey Roeder and David Duvenaud},
      year={2018},
      eprint={1711.00123},
      archivePrefix={arXiv},
      primaryClass={cs.LG},
      url={https://arxiv.org/abs/1711.00123}, 
}

@InProceedings{pmlr-v70-jaderberg17a,
  title = 	 {Decoupled Neural Interfaces using Synthetic Gradients},
  author =       {Max Jaderberg and Wojciech Marian Czarnecki and Simon Osindero and Oriol Vinyals and Alex Graves and David Silver and Koray Kavukcuoglu},
  booktitle = 	 {Proceedings of the 34th International Conference on Machine Learning},
  pages = 	 {1627--1635},
  year = 	 {2017},
  editor = 	 {Precup, Doina and Teh, Yee Whye},
  volume = 	 {70},
  series = 	 {Proceedings of Machine Learning Research},
  month = 	 {06--11 Aug},
  publisher =    {PMLR},
  url = 	 {https://proceedings.mlr.press/v70/jaderberg17a.html}
}

@misc{shirobokov2020blackboxoptimizationlocalgenerative,
      title={Black-Box Optimization with Local Generative Surrogates}, 
      author={Sergey Shirobokov and Vladislav Belavin and Michael Kagan and Andrey Ustyuzhanin and Atılım Güneş Baydin},
      year={2020},
      eprint={2002.04632},
      archivePrefix={arXiv},
      primaryClass={cs.LG},
      url={https://arxiv.org/abs/2002.04632}, 
}

@misc{neftci2019surrogategradientlearningspiking,
      title={Surrogate Gradient Learning in Spiking Neural Networks}, 
      author={Emre O. Neftci and Hesham Mostafa and Friedemann Zenke},
      year={2019},
      eprint={1901.09948},
      archivePrefix={arXiv},
      primaryClass={cs.NE},
      url={https://arxiv.org/abs/1901.09948}, 
}

@misc{jang2017categoricalreparameterizationgumbelsoftmax,
      title={Categorical Reparameterization with Gumbel-Softmax}, 
      author={Eric Jang and Shixiang Gu and Ben Poole},
      year={2017},
      eprint={1611.01144},
      archivePrefix={arXiv},
      primaryClass={stat.ML},
      url={https://arxiv.org/abs/1611.01144}, 
}

@misc{maddison2017concretedistributioncontinuousrelaxation,
      title={The Concrete Distribution: A Continuous Relaxation of Discrete Random Variables}, 
      author={Chris J. Maddison and Andriy Mnih and Yee Whye Teh},
      year={2017},
      eprint={1611.00712},
      archivePrefix={arXiv},
      primaryClass={cs.LG},
      url={https://arxiv.org/abs/1611.00712}, 
}

@misc{lokossou2025quantumdeepreinforcementlearning,
      title={Quantum deep reinforcement learning for humanoid robot navigation task}, 
      author={Romerik Lokossou and Birhanu Shimelis Girma and Ozan K. Tonguz and Ahmed Biyabani},
      year={2025},
      eprint={2509.11388},
      archivePrefix={arXiv},
      primaryClass={cs.RO},
      url={https://arxiv.org/abs/2509.11388}, 
}

@misc{xie2025quantumsurrogatedrivenimageclassifier,
      title={Quantum Surrogate-Driven Image Classifier: A Gradient-Free Approach to Avoid Barren Plateaus}, 
      author={Yichen Xie},
      year={2025},
      eprint={2505.05249},
      archivePrefix={arXiv},
      primaryClass={quant-ph},
      url={https://arxiv.org/abs/2505.05249}, 
}

@misc{gymnasiumdocs,
  title  = {Humanoid - {Gymnasium} Documentation},
  author = {{Farama Foundation}},
  year   = {2026},
  url    = {https://gymnasium.farama.org/environments/mujoco/humanoid/},
}

@article{PhysRevLett.122.040504,
  title = {Quantum Machine Learning in Feature Hilbert Spaces},
  author = {Schuld, Maria and Killoran, Nathan},
  journal = {Phys. Rev. Lett.},
  volume = {122},
  issue = {4},
  pages = {040504},
  numpages = {6},
  year = {2019},
  month = {Feb},
  publisher = {American Physical Society},
  doi = {10.1103/PhysRevLett.122.040504},
  url = {https://link.aps.org/doi/10.1103/PhysRevLett.122.040504}
}

@article{PhysRevA.103.032430,
  title = {Effect of data encoding on the expressive power of variational quantum-machine-learning models},
  author = {Schuld, Maria and Sweke, Ryan and Meyer, Johannes Jakob},
  journal = {Phys. Rev. A},
  volume = {103},
  issue = {3},
  pages = {032430},
  numpages = {12},
  year = {2021},
  month = {Mar},
  publisher = {American Physical Society},
  doi = {10.1103/PhysRevA.103.032430},
  url = {https://link.aps.org/doi/10.1103/PhysRevA.103.032430}
}
\bibliographystyle{iclr2027_conference}
\clearpage
\appendix
\section{Network architecture and training flow}
\label{app:architecture}
The diagrams below show the complete hybrid actor and its surrogate-assisted training procedure. They use $\mathbf q_{\mathrm i}$ for the circuit controls $u$, $\mathbf q_{\mathrm o}$ for the aggregated bitstring $B$, and $\mathbf q_{\mathrm o,\mathrm{pred}}$ for the surrogate prediction $f_\phi(u)$. The direct connections from PreDNN to PostDNN form the classical bypass $c$.

\begin{figure}[!ht]
\centering
\includegraphics[width=\linewidth]{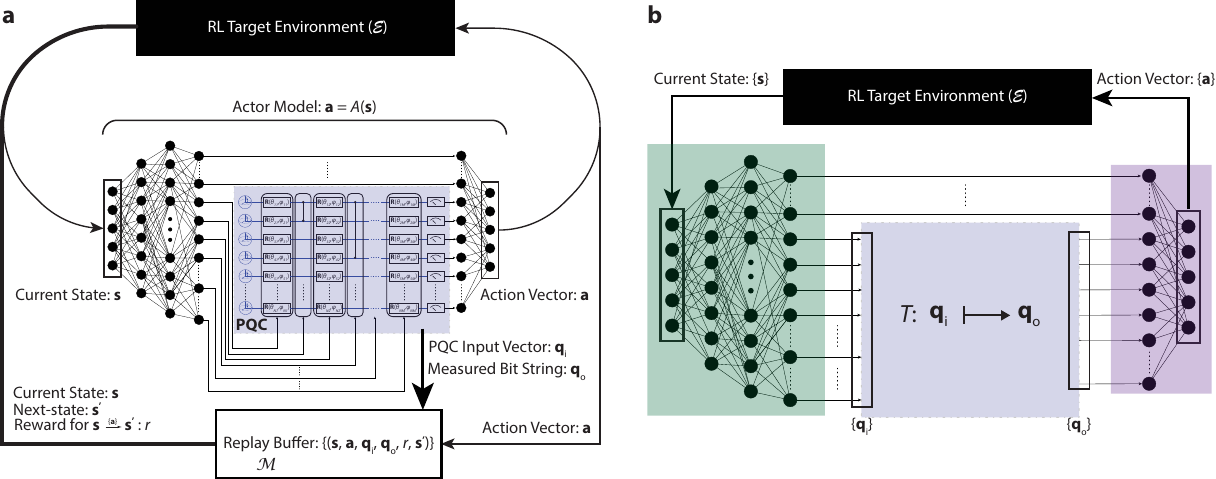}
\caption{\textbf{Hybrid actor and interaction with the environment.} (a) PreDNN encodes the state, the circuit supplies measured features, and PostDNN produces the action. Replay stores the transition together with the circuit controls and aggregated output. (b) The modular view separates the classical networks from the middle module $T$. For the hybrid policy, $T$ is the quantum channel during interaction and evaluation, and the surrogate during minibatch updates. Circuit connectivity is schematic; Appendix~\ref{app:implementation} specifies the implemented gate sequence.}
\label{fig:architecture_actor}
\end{figure}

\begin{figure}[!ht]
\centering
\includegraphics[width=\linewidth]{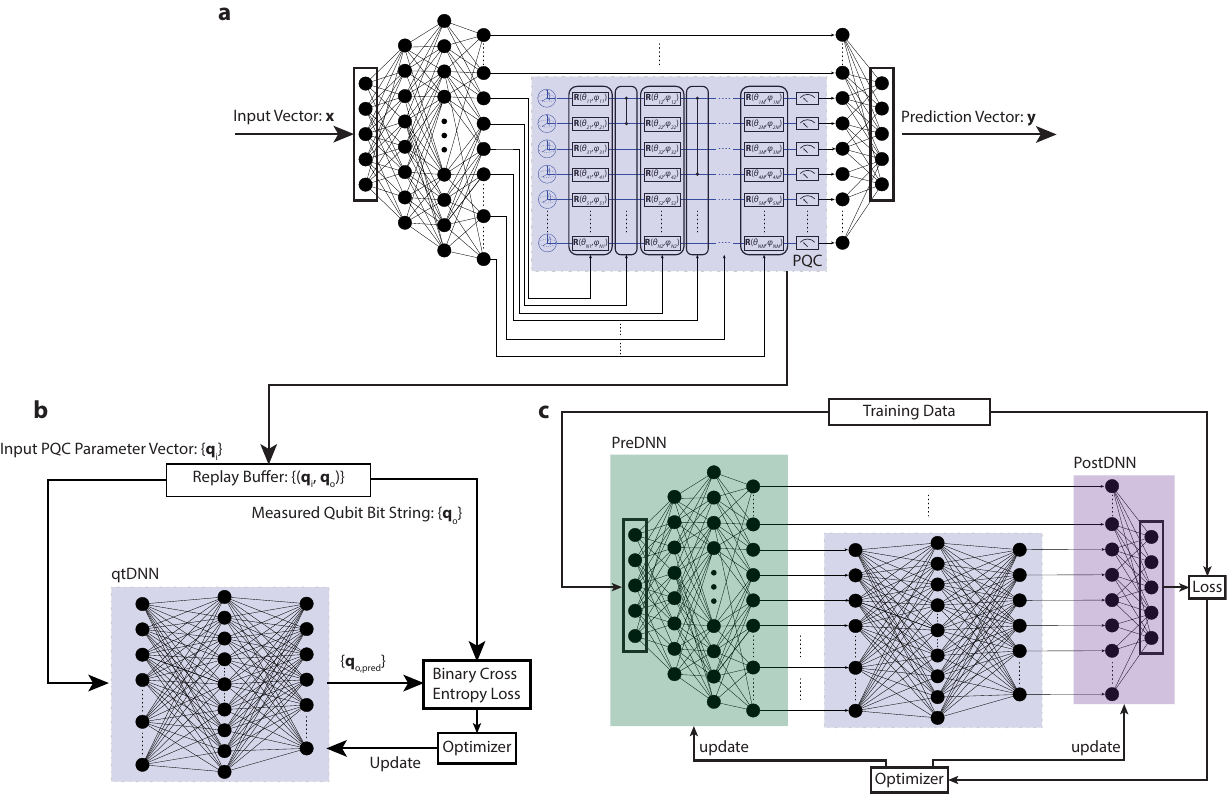}
\caption{\textbf{Surrogate-assisted training within the full network.} (a) The hybrid network evaluates the circuit and records its response. (b) The surrogate learns from stored control--output pairs by binary cross-entropy. (c) The fitted surrogate is held fixed while gradients pass through it to update PreDNN and PostDNN. For the RL actor, the input $x$ is the state $s$, the output $y$ is the action $a$, and the outer loss is Eq.~\eqref{eq:actorloss}, evaluated on replay states. The bypass is used in both network paths.}
\label{fig:architecture_surrogate}
\end{figure}
\clearpage

\begin{figure}[!ht]
\centering
\includegraphics[width=\linewidth,height=.83\textheight,keepaspectratio]{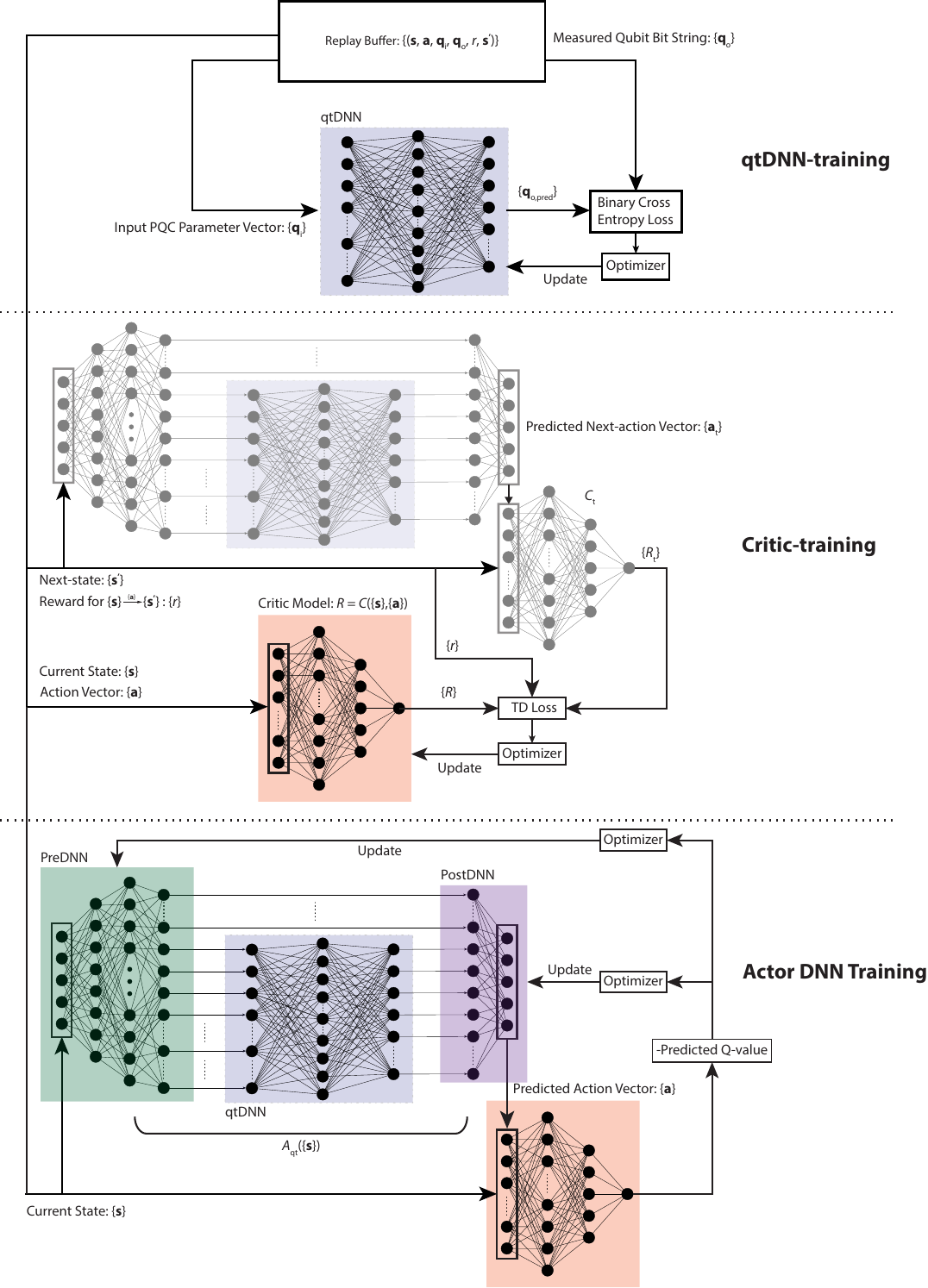}
\caption{\textbf{Replay-based surrogate, critic, and actor updates.} Top: fit \qt\ to recorded circuit responses. Middle: use the target surrogate actor to construct critic targets, then update the critics from replayed transitions. Bottom: hold the surrogate and critic fixed and differentiate the actor objective through them to update PreDNN and PostDNN. The labels $R$ and $R_t$ denote critic estimates, not the reward $r$. The middle panel represents TD3's twin critics and their targets; the actor update uses $Q_{\omega_1}$. Clipped-double targets, smoothing, terminal masks, and delayed updates follow Eqs.~\eqref{eq:target}--\eqref{eq:actorloss} and Algorithm~\ref{alg:training}.}
\label{fig:architecture_updates}
\end{figure}
\clearpage

\section{Proofs and additional analysis}
\label{app:proofs}
We use Euclidean norms for vectors and induced operator norms for Jacobians. The state $s$, critic parameters $\omega$, and post-network parameters $\psi$ are fixed when differentiating with respect to the pre-network parameters $\theta$, and we suppress $s$ in $G_s$. Derivative statements apply within a fixed routing cell and away from network kinks. The second-order results additionally assume the stated Hessian bound on the relevant convex hull.

\subsection{Population target of the binary cross-entropy fit}\label{app:bceproof}
For fixed $u$ and coordinate $k$, let $p=\mu_k(u)$ and $q=f_k(u)$. Conditioning on $u$ gives
\begin{align}
 \E[-B_k\log q-(1-B_k)\log(1-q)\mid u]
 &= -p\log q-(1-p)\log(1-q)\\
 &= h(p)+\KL(\Ber(p)\|\Ber(q)),
\end{align}
where $h(p)=-p\log p-(1-p)\log(1-p)$ uses natural logarithms and $0\log0=0$. Summing over $k$, dividing by $N$, and integrating over $\rho$ proves the identity in Proposition~\ref{prop:bce}. Because BCE is applied coordinatewise, the calculation depends only on the marginals and allows arbitrary correlations between bits.

For Bernoulli laws their total-variation distance is $|\mu_k-f_k|$. Pinsker's inequality therefore gives
\begin{equation}
 2(\mu_k(u)-f_k(u))^2
 \le\KL(\Ber(\mu_k(u))\|\Ber(f_k(u))).
\end{equation}
Summing over coordinates and integrating over $\rho$ gives the norm bound. A finite neural network can also incur approximation and optimization error, while a replay sample introduces estimation error and possible distribution shift. The proposition assumes genuine channel observations. The warm-up placeholders described in Appendix~\ref{app:reprolimits} fall outside that assumption.

\paragraph{Small prediction error does not control derivatives.}
To see why the gradient bound needs a separate Jacobian assumption, take the constant marginal $\mu(u)=1/2$ on $[0,1]$ and
\begin{equation}
 f_\varepsilon(u)=\frac12+\varepsilon\sin\!\left(\frac{u-1/2}{\varepsilon^2}\right),
 \qquad 0<\varepsilon\le\frac14.
\end{equation}
The uniform value error is at most $\varepsilon$, and the pointwise excess BCE is
$-\frac12\log(1-4\varepsilon^2\sin^2((u-1/2)/\varepsilon^2))=O(\varepsilon^2)$ uniformly. Nevertheless, $f_\varepsilon'(1/2)=1/\varepsilon$, while $\mu'(1/2)=0$. Thus even uniformly vanishing excess BCE cannot replace a Jacobian-error hypothesis.

\subsection{Proof of the upstream-gradient bound}\label{app:upstreamproof}
Write $A=J_\theta u$ and $C=J_\theta c$. The chain rule, retaining both outputs of the pre-network, yields
\begin{align}
 \nabla_\theta F_f
 &=A^\top J_f^\top\nabla_zG(f,c)+C^\top\nabla_cG(f,c),\\
 \nabla_\theta F_\mu
 &=A^\top J_\mu^\top\nabla_zG(\mu,c)+C^\top\nabla_cG(\mu,c).
\end{align}
Add and subtract $A^\top J_\mu^\top\nabla_zG(f,c)$. Their difference is exactly
\begin{align}
 \nabla_\theta F_f-\nabla_\theta F_\mu
 &= A^\top(J_f-J_\mu)^\top\nabla_zG(f,c)\nonumber\\
 &\quad+A^\top J_\mu^\top[\nabla_zG(f,c)-\nabla_zG(\mu,c)]\nonumber\\
 &\quad+C^\top[\nabla_cG(f,c)-\nabla_cG(\mu,c)].
 \label{eq:appendixchainsplit}
\end{align}
Apply the triangle inequality and submultiplicativity. The three terms have norms at most $M_u\varepsilon_2L_z$, $M_uM_\mu L_{zz}\varepsilon_1$, and $M_cL_{cz}\varepsilon_1$, respectively, proving Theorem~\ref{thm:upstream}.

For the post-network parameters, let $G_\psi$ denote the same composition with its $\psi$-dependence explicit. If $\nabla_\psi G_\psi(z,c)$ is $L_{\psi z}$-Lipschitz in $z$, then, with $u,c$ fixed,
\begin{equation}
 \norm{\nabla_\psi G_\psi(f,c)-\nabla_\psi G_\psi(\mu,c)}_2
 \le L_{\psi z}\varepsilon_1.
\end{equation}
The post-network bound has no Jacobian-of-$f$ term because $u$ and $c$ are fixed for this parameter block. For a finite replay distribution, differentiating its finite sum and averaging the pointwise bounds is immediate. For a general fixed distribution, the same conclusion holds under a locally integrable dominating envelope for the relevant derivatives, which justifies exchanging differentiation and expectation. A policy-dependent state-occupancy measure would introduce additional derivatives that are outside this fixed-critic analysis.

\subsection{Value bias caused by the mean-feature relaxation}
Set $X=B-\mu$. Because $B_k\in\{0,1\}$,
\begin{equation}
 \E\!\left[\norm{X}_2^2\right]=\sum_k\E\!\left[(B_k-\mu_k)^2\right]
 =\sum_k\mu_k(1-\mu_k)=v(u)\le N/4.
 \label{eq:varianceidentity}
\end{equation}
No off-diagonal covariance terms occur in a squared Euclidean norm. If $G(\cdot,c)$ is $L_z^{\mathrm{supp}}$-Lipschitz on $K=\operatorname{conv}(\operatorname{supp}P_{u,S})$, then
\begin{equation}
 |\E[G(B,c)]-G(\mu,c)|
 \le L_z^{\mathrm{supp}}\E\!\left[\norm{X}_2\right]\le L_z^{\mathrm{supp}}\sqrt{v(u)}.
\end{equation}
For the second-order result, assume $G(\cdot,c)$ is twice continuously differentiable on an open neighborhood of $K$ and $\sup_{z\in K}\norm{\nabla_z^2G(z,c)}_\op\le H$. Taylor's theorem on every segment from $\mu$ to a support point gives
\begin{equation}
 G(B,c)=G(\mu,c)+\nabla_zG(\mu,c)^\top X+r_G(B,\mu),
 \qquad |r_G(B,\mu)|\le\frac H2\norm{X}_2^2.
 \label{eq:taylorremainder}
\end{equation}
The linear term has zero expectation. Taking expectations and using Eq.~\eqref{eq:varianceidentity} proves $|F-F_\mu|\le Hv(u)/2$.

For Eq.~\eqref{eq:totalvaluegap}, use a separate Lipschitz constant $L_z^{\mathrm{fit}}$ on $[f,\mu]$. Then
\begin{align}
 |F-F_f|&\le |F-F_\mu|+|G(\mu,c)-G(f,c)|\nonumber\\
 &\le L_z^{\mathrm{supp}}\sqrt{v(u)}+L_z^{\mathrm{fit}}\norm{f-\mu}_2.
 \label{eq:appendixvaluegap}
\end{align}
The pointwise bound on $\norm{\nabla_zG(f,c)}_2$ in Theorem~\ref{thm:upstream} alone does not supply this segment-wide constant. No Jacobian error enters the value comparison.

The actual bias depends on the downstream function and the full output law; the bound alone does not imply that it grows monotonically with $N$. For example, when $G(z,c)$ is affine in $z$, the value bias and fixed-$c$ gradient bias both vanish wherever the derivatives exist.

\subsection{Score identity and a sufficient ascent condition}
\label{app:score}
Suppose the channel law has fixed finite support with positive probabilities in an open neighborhood of $u$, and these probabilities are differentiable. Finite summation justifies differentiation under the expectation. With $\ell_u(b)=\nabla_u\log P_{u,S}(b)$,
\begin{equation}
 \E[\ell_u(B)]=0,\qquad
 J_\mu=\E[B\ell_u(B)^\top]=\E[(B-\mu)\ell_u(B)^\top],\qquad
 \nabla_uF=\E[G(B,c)\ell_u(B)].
\end{equation}
Subtract the constant $G(\mu,c)\E[\ell_u(B)]$ and the linear contribution
$J_\mu^\top\nabla_zG(\mu,c)$ to obtain Eq.~\eqref{eq:scoreidentity}. Equation~\eqref{eq:taylorremainder} then implies
\begin{equation}
 \norm{\nabla_uF-\nabla_uF_\mu}_2
 \le\frac H2\E\!\left[\norm{B-\mu}_2^2\norm{\ell_u(B)}_2\right]
 \le\frac H2\sqrt{\E\!\left[\norm{B-\mu}_2^4\right]\,\E\!\left[\norm{\ell_u(B)}_2^2\right]}.
 \label{eq:scoremomentbound}
\end{equation}
The score moment measures an additional sensitivity of the sampled law. It can be large even when BCE prediction error is small.

The bypass also contributes to the difference between the mean-feature and sampled objectives. Assume, in addition, that $G$ is continuously differentiable in $(z,c)$ on a neighborhood of $K\times\{c\}$ and that $\nabla_cG(\cdot,c)$ is $L_{cz}^{\mathrm{supp}}$-Lipschitz on $K$. This is a condition on the sampled support hull, not just on the fitting segment $[f,\mu]$. Then
\begin{equation}
 \norm{\E[\nabla_cG(B,c)]-\nabla_cG(\mu,c)}_2
 \le L_{cz}^{\mathrm{supp}}\E\!\left[\norm{B-\mu}_2\right]\le L_{cz}^{\mathrm{supp}}\sqrt{v(u)}.
\end{equation}
Combining this with the pre-network chain rule and Theorem~\ref{thm:upstream} gives the following conditional extension.

\begin{corollary}[From surrogate fitting to the sampled critic objective]
\label{cor:sampled}
Assume Theorem~\ref{thm:upstream}; a fixed finite-support channel law with positive differentiable probabilities on an open neighborhood of $u$; and the Hessian and bypass-Lipschitz conditions on the support hull $K$ stated above. Let
\begin{align}
 \Delta_{\mathrm{fit}}
 &=M_uL_z\varepsilon_2+(M_uM_\mu L_{zz}+M_cL_{cz})\varepsilon_1,\\
 \Delta_{\mathrm{relax}}
 &=\frac{M_uH}{2}\E\!\left[\norm{B-\mu}_2^2\norm{\ell_u(B)}_2\right]
   +M_cL_{cz}^{\mathrm{supp}}\sqrt{v(u)}.
\end{align}
Then $\norm{\nabla_\theta F_f-\nabla_\theta F}_2\le
\Delta_{\mathrm{fit}}+\Delta_{\mathrm{relax}}$.
If $g=\nabla_\theta F\ne0$ and the right-hand side is strictly smaller than $\norm g_2$, the surrogate gradient is a strict local ascent direction for $F$.
\end{corollary}
\begin{proof}
The first assertion follows by inserting $\nabla_\theta F_\mu$ and applying the bounds already proved. If $\delta$ is the combined bound, Cauchy--Schwarz gives
$\langle g,\nabla_\theta F_f\rangle\ge\norm g_2^2-\norm g_2\delta>0$.
Differentiability therefore implies improvement along this direction for sufficiently small positive step sizes.
\end{proof}
The strict inner product implies ascent of the fixed-critic objective for a sufficiently small Euclidean step that remains inside the differentiability neighborhood. The fitting and support-wide constants are assumptions of this local result, rather than quantities certified by the experiments. Adam updates, critic evolution, and changes in the state distribution are outside its scope.

\subsection{Exact one-qubit counterexample and aggregation-aware differentiation}
\label{app:example}
\begin{example}[Zero value error with an opposing gradient]
\label{ex:opposing}
Prepare one qubit as $R_y(\vartheta)|0\rangle$, retain the majority of five shots, and let $G(z)=z+\frac14\sin(2\pi z)$. At $\vartheta=\pi/2$ the exact aggregated marginal is $\mu=1/2$. Hence $F=F_\mu=1/2$, but
\begin{equation}
 F'(\pi/2)=15/16>0,\qquad
 F_\mu'(\pi/2)=(1-\pi/2)\,15/16<0.
 \label{eq:counter}
\end{equation}
This failure occurs even with a perfect surrogate, $f=\mu$ and $J_f=J_\mu$.
\end{example}

In Example~\ref{ex:opposing}, the probability of measuring one in a single shot is $p(\vartheta)=\sin^2(\vartheta/2)$. The five-shot majority is one when at least three shots return one, giving
\begin{equation}
 \mu(\vartheta)
 =10p^3(1-p)^2+5p^4(1-p)+p^5
 =10p^3-15p^4+6p^5,
 \qquad \frac{\mathrm d\mu}{\mathrm dp}=30p^2(1-p)^2.
\end{equation}
At $\vartheta=\pi/2$, $p=1/2$, $p'=1/2$, and $\mu'=15/16$. Since $G(0)=0$ and $G(1)=1$, the exact sampled objective is $F=\mu$. In contrast, $F_\mu=\mu+\frac14\sin(2\pi\mu)$ has derivative
\begin{equation}
 F_\mu'=\left[1+\frac\pi2\cos(2\pi\mu)\right]\mu'.
\end{equation}
Evaluated at the center, these derivatives are $0.9375$ and approximately $-0.53512$. The scalar gradients have cosine $-1$, although the two objective values coincide and the surrogate is exact.

\begin{figure}[H]
\centering
\includegraphics[width=\linewidth]{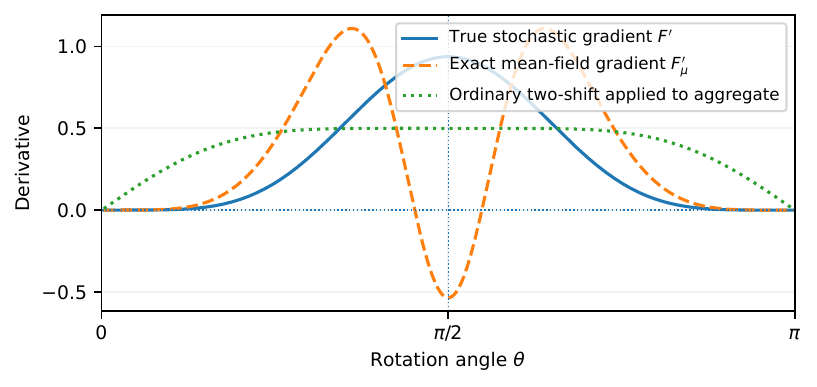}
\caption{Derivatives in the one-qubit example. The sampled objective and its exact mean-feature relaxation have opposite gradients at $\vartheta=\pi/2$. Applying an ordinary two-shift expression to the aggregated objective gives a third, incorrect derivative. The curves are evaluated from the exact binomial polynomial.}
\label{fig:counterexample}
\end{figure}

\paragraph{Differentiating the shot-aggregated objective.}
For this example,
$\frac12[F(\pi/2+\pi/2)-F(\pi/2-\pi/2)]=1/2$, not $15/16$.
Nonlinear shot aggregation changes the objective to which a differentiation rule must apply. The ordinary two-shift rule for a single-parameter Born expectation is therefore insufficient here. Generalized shift rules can handle richer Fourier spectra \citep{Wierichs2022generalparameter}; alternatively, an estimator can be derived directly for the full shot-bag objective.

More generally, enumerate the $K=2^N$ bitstrings. Under a fixed deterministic tie rule, the probability that mode aggregation returns $b$ is
\begin{equation}
 P_{u,S}(b)=\sum_{\substack{n_1+\cdots+n_K=S\\ A_S(n_1,\ldots,n_K)=b}}
 \frac{S!}{n_1!\cdots n_K!}\prod_{j=1}^K\pi_u(j)^{n_j}.
 \label{eq:multinomialmode}
\end{equation}
Within a fixed routing cell, Eq.~\eqref{eq:multinomialmode} is a homogeneous polynomial of degree $S$ in the single-shot probabilities, possibly of lower degree on the probability simplex. For the implemented circuit, each rotation control $u_j$ occurs once, as one angle of $U_3(2\pi u_j,\cdot,0)$ or $U_3(\cdot,2\pi u_j,0)$. Holding other controls fixed, the single-shot probability of an outcome has the form
\begin{equation}
 \pi_u(k)=a_k+b_k\cos(2\pi u_j)+c_k\sin(2\pi u_j).
 \label{eq:bornfourier}
\end{equation}
For the first angle, the gate is a fixed unitary times $R_y(2\pi u_j)$; conjugating a density matrix produces only constant, sine and cosine terms. For the second, the gate contains a diagonal relative phase with the same frequency. Fixed surrounding gates and parameter-independent noise are linear maps on the density matrix, so they preserve this coordinatewise degree. Consequently $P_{u,S}(b)$ has Fourier degree at most $S$ in each rotation control. For fixed $G$ and $c$, $F(u,c)$ inherits this degree; $G(\mu(u),c)$ need not. Aggregation can therefore invalidate the ordinary two-shift rule even though it is valid for the single-shot probabilities. Generalized shift rules can exploit the larger spectrum \citep{Wierichs2022generalparameter}; this is not a lower bound on estimation cost.

The channel score can also be related directly to circuit probabilities. On an interior region where all $\pi_u(k)>0$, let $w_u(n)$ be the multinomial weight in Eq.~\eqref{eq:multinomialmode}. Finite differentiation gives
\begin{equation}
 \nabla_u P_{u,S}(b)
 =\sum_{n:A_S(n)=b}w_u(n)\sum_{k=1}^{2^N}
 n_k\frac{\nabla_u\pi_u(k)}{\pi_u(k)},\qquad
 \ell_u(b)=\frac{\nabla_u P_{u,S}(b)}{P_{u,S}(b)}.
 \label{eq:circuitscore}
\end{equation}
Each rotation derivative of $\pi_u$ is available by a suitable shift rule; routing derivatives vanish locally. At zero probabilities, one can differentiate the polynomial directly rather than use the displayed divisions. This is an in-principle expression, not a proposal to enumerate all shot-count vectors during RL training. It identifies aggregation sensitivity absent from the single-shot marginal Jacobian.

We evaluated the closed-form one-qubit expressions at 801 angles. Central finite differences with step $10^{-6}$ agree with the analytic derivative to below $2.5\times10^{-10}$. We also checked the finite-support value and gradient identities, including the bypass, and verified that five-shot aggregation can generate frequency five. These numerical checks test the analytic expressions; they do not establish the assumptions along a trained RL policy.

\subsection{Computational complexity}
The surrogate removes update-time quantum queries but adds classical computation. A single-hidden-layer surrogate with width $h$, input dimension $d$, and output dimension $N$ has $(d+N+1)h+N$ parameters and costs order $h(d+N)$ per dense evaluation. Fitting it for $K_{\mathrm{fit}}$ inner steps with sub-batch size $b$ costs order $K_{\mathrm{fit}}b h(d+N)$ per outer update, in addition to the actor and critic computations. These operation counts do not predict end-to-end runtime. The primary configuration uses $h=2048$ at $N=10$; a width tied to $2^{N+1}$ is a possible design choice rather than a requirement. The capacity sweep treats width and qubit count as independent choices.

For an appropriate unaggregated expectation, a naive two-shift estimate with $I=2NM$ continuous gate angles, batch size $b_{\mathrm{RL}}$, and $S$ shots per circuit would use $2Ib_{\mathrm{RL}}S$ shots. Setting $I=200$, $b_{\mathrm{RL}}=256$, and $S=1000$ gives 102.4 million shots. This illustrates the cost of that particular estimator, not a lower bound or a measured baseline. Its validity for a Born expectation also does not make it valid for mode aggregation. Our measured comparison is with hard-forward STE (Section~\ref{sec:cost}).

\subsection{Synthetic probes of fitting and relaxation}
\label{app:gradientdiagnostics}\label{sec:gradientdiagnostics}
This study uses exact CPU simulation of four-, six- and eight-qubit circuits and a fixed random smooth map $G$, not a trained RL critic. The map is a width-64 tanh network with scalar output scale three. It is fixed within each qubit count. At each synthetic control anchor, we compare the sampled objective $F(u)=\E[G(B)]$, the aggregated-mean relaxation $F_\mu(u)=G(\mu(u))$, and the Born-mean substitute $G(p(u))$, where $p(u)=\E_{\pi_u}[Z]$. Tables~\ref{tab:alignmentlatest}-\ref{tab:alignmentfits} summarize the seven synthetic configurations described below. We repeated the complete study with the same implementation and settings; all non-timing probe values agree between executions. The software environment and source hashes were recorded for both executions.

\paragraph{Protocols and reference directions.}
Cold fits use 256 local control--output pairs, with Gaussian control perturbations of standard deviation $0.05$, clipped to $[0,1]^d$, and one shot-aggregated output per control. A model is evaluated after 32 optimization steps and again after 500 cumulative steps. The latter continues the same fit. The product surrogate uses width $2^{N+1}$ or 64, and the autoregressive surrogate uses width $2^{N+1}$. The persistent protocol uses six qubits and width 128, retaining parameters over 12 drifting batches. Each batch adds 64 pairs and performs 32 fitting steps, for 768 pairs and 384 optimization steps in total. It is a growing synthetic replay, not the full RL replay process.

Each anchor uses eight normalized directions in the rotation-control subspace, displacement $h=0.05$, and 4000 common-random-number (CRN) shot-bag replicates at each endpoint. For direction $v_j$, the endpoints are $u_j^\pm=\operatorname{clip}_{[0,1]^d}(u\pm hv_j)$. Let $\hat\mu_\pm$ and $\hat F_\pm$ denote the empirical aggregated means and sampled-objective values at these endpoints. The compared reference vectors have components
\begin{align}
 a^{F}_j&=\frac{\widehat F_+-\widehat F_-}{2h}, &
 a^{\mu}_j&=\frac{G(\widehat\mu_+)-G(\widehat\mu_-)}{2h},\\
 a^{\mathrm{Born}}_j&=v_j^\top J_p(u)^\top\nabla_zG(p(u)).
\end{align}
The Born Jacobian is computed by an exact ordinary parameter-shift rule on the CPU simulator. The fitted direction uses coordinate finite differences of $G(f(u))$ with step $10^{-3}$, with the same endpoint clipping, and is projected onto the eight directions. Both finite-difference calculations retain their nominal denominators after clipping. When clipping is active, these differences compare projected endpoint values rather than ordinary unprojected directional derivatives.

\begin{table}[!htb]
\centering\small\setlength{\tabcolsep}{3pt}
\begin{tabular}{lrrrrrr}
\toprule
$N/M/S$ & Protocol & Defined & $F_\mu$ FD & Born PS & Fit 32/384 & Fit 500 \\
\midrule
4/6/1000 & Cold & 7/10 & 0.998 & 0.838 & 0.092 & 0.090 \\
6/10/1000 & Cold & 10/10 & 0.998 & 0.196 & -0.079 & -0.062 \\
6/6/100 & Cold & 10/10 & 0.994 & 0.808 & -0.056 & -0.086 \\
6/6/1000 & Cold & 9/10 & 0.995 & 0.386 & -0.232 & -0.159 \\
6/6/1000 & Persistent & 10/10 & 0.999 & 0.642 & -0.081 & --- \\
6/6/100 & Persistent & 10/10 & 0.998 & 0.886 & 0.269 & --- \\
8/6/1000 & Cold & 5/5 & 1.000 & 0.557 & 0.224 & 0.113 \\
\bottomrule
\end{tabular}

\caption{Median directional cosines against the CRN sampled-objective reference $a^F$. $F_\mu$ FD uses the aggregated-mean reference, and Born PS uses the single-shot mean and its exact shift Jacobian. Fit columns use the product surrogate with width $2^{N+1}$. The first fit column is 32 steps for cold fits and 384 for persistent fits. ``Defined'' counts distinct anchors, not duplicated reference entries across model variants.}
\label{tab:alignmentlatest}
\end{table}

Estimated aggregate-mean directions align closely in these fixed-$G$ probes: median cosines range from $0.994$ to $1.000$ across the seven settings. Ordinary Born-mean directions have lower medians, from $0.196$ to $0.886$, with negative individual cosines in several settings. The single-shot Jacobian is exact for its own target but does not automatically differentiate the shot-aggregated objective. This is the circuit-specific distinction illustrated analytically in Appendix~\ref{app:example}.

Fitted-surrogate directions are more variable. At four qubits, product width 32 gives medians $0.092$ and $0.090$ after 32 and 500 steps; product width 64 gives $-0.048$ and $0.162$, and autoregressive width 32 gives $-0.032$ and $0.155$. Six-qubit persistent fits give $-0.081$ at 1000 shots and $0.269$ at 100 shots. These observations distinguish benign relaxation for this particular $G$ from successful derivative fitting. They neither certify alignment of the trained RL policies nor contradict their measured learning curves.

\begin{table}[!htb]
\centering\small\setlength{\tabcolsep}{4pt}
\begin{tabular}{lrrrrr}
\toprule
$N/M/S$ & Surrogate & Width & Defined & 32/384 steps & 500 steps \\
\midrule
4/6/1000 & ar & 32 & 7/10 & -0.032 & 0.155 \\
4/6/1000 & product & 32 & 7/10 & 0.092 & 0.090 \\
4/6/1000 & product & 64 & 7/10 & -0.048 & 0.162 \\
6/10/1000 & ar & 128 & 10/10 & 0.047 & -0.014 \\
6/10/1000 & product & 64 & 10/10 & 0.097 & -0.025 \\
6/10/1000 & product & 128 & 10/10 & -0.079 & -0.062 \\
6/6/100 & ar & 128 & 10/10 & 0.056 & 0.287 \\
6/6/100 & product & 64 & 10/10 & 0.081 & 0.238 \\
6/6/100 & product & 128 & 10/10 & -0.056 & -0.086 \\
6/6/1000 & ar & 128 & 9/10 & -0.114 & 0.092 \\
6/6/1000 & product & 64 & 9/10 & -0.023 & 0.131 \\
6/6/1000 & product & 128 & 9/10 & -0.232 & -0.159 \\
6/6/1000 & product (persistent) & 128 & 10/10 & -0.081 & --- \\
6/6/100 & product (persistent) & 128 & 10/10 & 0.269 & --- \\
8/6/1000 & ar & 512 & 5/5 & 0.038 & 0.009 \\
8/6/1000 & product & 64 & 5/5 & 0.027 & 0.068 \\
8/6/1000 & product & 512 & 5/5 & 0.224 & 0.113 \\
\bottomrule
\end{tabular}

\caption{Every fitted surrogate setting in the synthetic study. Entries are median cosines with $a^F$. The first budget is 32 steps for cold fits or 384 for persistent fits. Three four-qubit reference vectors and one six-qubit, six-layer, 1000-shot cold reference vector vanish, making 18 and six model/budget rows undefined. The coverage denominators include these undefined cases; the medians use only defined cosines. Other configurations have ten defined anchors, except the five-anchor eight-qubit study.}
\label{tab:alignmentfits}
\end{table}

Each cold-protocol control anchor is shared by six model-budget combinations. Reference-direction summaries count each anchor once, so a ten-anchor study contributes ten reference probes rather than 60. Shared CRN noise also couples the estimated mean and sampled directions. High cosine does not imply matching gradient magnitudes or an accurate uniform bound. The counterexample and the probe observations concern different downstream maps: together they show why the relaxation must be assessed for the map actually used.

\begin{table}[!htb]
\centering\small\setlength{\tabcolsep}{3pt}
\begin{tabular}{lrrr}
\toprule
$N/M/S$ & Born PS evaluations & PS seconds / control & 32-step fit seconds \\
\midrule
4/6/1000 & 96 & 0.184 & 0.004 \\
6/10/1000 & 240 & 3.640 & 0.007 \\
6/6/100 & 144 & 1.014 & 0.006 \\
6/6/1000 & 144 & 0.985 & 0.006 \\
8/6/1000 & 192 & 25.900 & 0.018 \\
\bottomrule
\end{tabular}

\caption{Recorded CPU microbenchmarks from the cold synthetic studies. The shift computation is an exact single-shot marginal Jacobian for one control vector; the fit uses 256 stored pairs and 32 optimizer steps. They have different outputs and accuracies. These are not RL runtimes or timings of an aggregation-corrected gradient estimator.}
\label{tab:gradienttiming}
\end{table}
The single-shot marginal Jacobian computation uses 240 probability evaluations and takes a median 3.640 seconds per control for six qubits and ten layers; the corresponding values for eight qubits and six layers are 192 evaluations and 25.900 seconds. Corresponding short fits take 0.007 and 0.018 seconds. The timings use single-threaded numerical kernels in a shared four-vCPU environment and include ordinary process contention. They are descriptive measurements, not isolated hardware benchmarks, and are not transferred to the cluster runtime comparisons.

\subsubsection{Dependence on the downstream map}
\label{app:gmapfamilies}
For each qubit count, the controls, shot draws, replay data, fitting budgets and projection directions are the same as in the corresponding tanh experiment. The fitted surrogates are also held fixed across downstream maps.

The first alternative has a width-256 LeakyReLU hidden layer, slope $0.01$, and a linear scalar head. The second uses the paper's post-network--critic composition: a width-256 LeakyReLU post-network maps the bit features and a fixed bypass to 17 tanh actions in $[-0.4,0.4]$; a critic combines a width-256 state encoding with the actions, applies another width-256 LeakyReLU layer, and outputs a scalar. Xavier-uniform weights and zero biases match the initialization convention. The 376-dimensional standard-normal state and uniform bypass vector are fixed synthetic draws. This reproduces the architecture, not learned weights or a trained policy's state distribution. The NumPy composition is checked against the actual PyTorch building blocks in \texttt{hdqnn/networks.py}.

The remaining alternatives are controlled oscillatory maps,
\begin{equation}
 G_{a,k}(z)=\frac3N\sum_{j=1}^N\bigl[z_j+a\sin(2\pi k z_j)\bigr],
 \qquad (a,k)\in\{(1/4,1),(1/2,2)\}.
 \label{eq:sinemaps}
\end{equation}
For integer $k$, both maps agree on every binary feature vector: $G_{a,k}(B)=3N^{-1}\sum_j B_j$. Their sampled objective is therefore unchanged, whereas the mean-feature objective depends on $a$ and $k$. The Hessian norm is bounded by $3a(2\pi k)^2/N$. This tests the effect of off-vertex curvature without changing the map's values on deployed features, generalizing the mechanism in Example~\ref{ex:opposing}.

\begin{table}[!htb]
\centering\small\setlength{\tabcolsep}{3pt}
\begin{tabular}{lrrrrr}
\toprule
$N$ & Downstream map & Defined & Median [IQR] & Minimum & $<0$ \\
\midrule
4 & Tanh & 7/10 & 0.998 [0.982, 0.999] & 0.913 & 0 \\
4 & LeakyReLU & 7/10 & 0.999 [0.995, 1.000] & 0.961 & 0 \\
4 & Post--critic & 7/10 & 0.997 [0.996, 0.998] & 0.995 & 0 \\
4 & Sine $(a=1/4,k=1)$ & 7/10 & 0.972 [0.964, 0.975] & 0.950 & 0 \\
4 & Sine $(a=1/2,k=2)$ & 7/10 & 0.837 [0.620, 0.881] & -0.539 & 1 \\
6 & Tanh & 9/10 & 0.995 [0.992, 1.000] & 0.749 & 0 \\
6 & LeakyReLU & 9/10 & 0.996 [0.985, 1.000] & 0.811 & 0 \\
6 & Post--critic & 9/10 & 0.993 [0.979, 0.999] & 0.949 & 0 \\
6 & Sine $(a=1/4,k=1)$ & 9/10 & 0.910 [0.906, 0.997] & 0.774 & 0 \\
6 & Sine $(a=1/2,k=2)$ & 9/10 & 0.656 [0.503, 0.788] & -0.984 & 1 \\
8 & Tanh & 5/5 & 1.000 [1.000, 1.000] & 0.998 & 0 \\
8 & LeakyReLU & 5/5 & 0.993 [0.989, 0.994] & 0.942 & 0 \\
8 & Post--critic & 5/5 & 0.968 [0.965, 0.993] & 0.176 & 0 \\
8 & Sine $(a=1/4,k=1)$ & 5/5 & 0.752 [0.728, 0.860] & -0.775 & 1 \\
8 & Sine $(a=1/2,k=2)$ & 5/5 & 0.294 [0.169, 0.681] & 0.025 & 0 \\
\bottomrule
\end{tabular}

\caption{Dependence of the mean-feature relaxation on the downstream map. Each row summarizes distinct matched control anchors, comparing the finite-difference vectors $a^\mu$ and $a^F$. ``$<0$'' counts opposing estimated directions among defined cosines. The reported coverage includes undefined cases; cosine summaries use only defined directions. These are synthetic fixed-map probes, not measurements of trained RL gradients.}
\label{tab:gmapfamilies}
\end{table}

Across the nine random-map/qubit settings, median cosines range from $0.968$ to $1.000$. The stronger oscillatory map has medians $0.837$, $0.656$ and $0.294$ at four, six and eight qubits, respectively, with negative individual cosines at four and six qubits. The lower-frequency sine map also has an opposing eight-qubit estimate. The random-map probes and the oscillatory probes test different downstream geometries. The latter can produce opposing directions despite using the exact same sampled objective. Because this comparison uses aggregated-marginal references rather than fitted surrogates, its discrepancy concerns the relaxation and reference estimation, not surrogate fitting. Conversely, high alignment for one fixed map does not establish a bound for a learned actor and critic. All reference directions retain the original 4000-bag Monte Carlo and finite-difference errors, including shared CRN noise.

\subsection{Straight-through updates and relaxed critic targets}
\label{app:ste}
Assume $G$ is differentiable at each sampled feature point and that $f$, $u_\theta$, and $c_\theta$ are differentiable at the inputs under consideration. For a sampled feature $B$, the automatic-differentiation expression $z=B+f_\phi(u)-\sg(f_\phi(u))$ has forward value $B$ and assigned input derivative $J_f$. Holding $\phi$, $\psi$, the critic and the replay state fixed gives
\begin{equation}
 \widehat g^{\mathrm{STE}}_\theta
 =A^\top J_f^\top\nabla_zG(B,c)+C^\top\nabla_cG(B,c).
\end{equation}
Taking the finite-support expectation proves Eq.~\eqref{eq:stegradient}. Under the fixed-support and differentiability assumptions of Appendix~\ref{app:score}, the corresponding objective gradients are,
\begin{align}
 \nabla_\theta F&=A^\top\E[G(B,c)\ell_u(B)]+C^\top\E[\nabla_cG(B,c)],\\
 \nabla_\theta F_\mu&=A^\top J_\mu^\top\nabla_zG(\mu,c)+C^\top\nabla_cG(\mu,c).
\end{align}
Even $J_f=J_\mu$ does not make $J_\mu^\top\E[\nabla_zG(B,c)]$ equal to either fixed-$c$ contribution in general. In Example~\ref{ex:opposing}, exact-mean STE gives $(1+\pi/2)15/16$ at $\vartheta=\pi/2$, whereas $F'=15/16$ and $F_\mu'=(1-\pi/2)15/16$. The implemented stop-gradient control detaches both feature paths and never steps the pre-network optimizer. Its comparison with STE concerns encoder and bypass learning together.

\paragraph{The target substitution.}\label{app:targetgap}
For fixed next state, target networks and smoothing draw, write
\begin{equation}
 H_{s',\epsilon}(z,c)=\min_{j=1,2}Q_{\bar\omega_j}
 \bigl(s',\operatorname{clip}_{\mathcal A}[\operatorname{Post}_{\bar\psi}(z,c)+\epsilon]\bigr).
\end{equation}
Here $u=u_{\bar\theta}(s')$, $c=c_{\bar\theta}(s')$, $B\sim P_{u,S}$, and $f=f_{\bar\phi}$; all target-network parameters are held fixed.

If this function is Lipschitz in $z$ with constants $L_H^{\mathrm{supp}}$ and $L_H^{\mathrm{fit}}$ on the sampled support hull and fitting segment, respectively, then
\begin{equation}
 |\E_B[H_{s',\epsilon}(B,c)]-H_{s',\epsilon}(f(u),c)|
 \le L_H^{\mathrm{supp}}\sqrt{v(u)}+L_H^{\mathrm{fit}}\norm{f(u)-\mu(u)}_2.
\end{equation}
Multiplication by $\gamma(1-d_{\mathrm{term}})$ gives the conditional target-value gap. An integrable dominating bound permits averaging over smoothing noise and replay. A finite minimum and action clipping preserve suitable Lipschitz bounds, although they can introduce kinks that preclude a Hessian argument. Critic regressors are trained on executed sampled actions, with exploration noise, whereas the target action is relaxed. This local value statement neither removes that distribution mismatch nor proves that either critic equals the deployed policy's action value.

\FloatBarrier
\section{Implementation and experimental protocol}
\label{app:implementation}
This appendix specifies the implementation used for the reported experiments, including its evaluation schedule and checkpoint behavior. The network and training diagrams are in Appendix~\ref{app:architecture}.

\subsection{Circuit and neural-network architecture}
The quantum circuit starts in $|+\rangle^{\otimes N}$. The first $2NM$ sigmoid outputs are multiplied by $2\pi$ and used as pairs of angles for $U_3(\alpha,\beta,0)$. At each layer, these gates are applied to all qubits before a single controlled-$Z$ gate. The last $2M$ controls select its two endpoints by $\operatorname{rint}((N-1)u_j)$; NumPy's nearest-even rounding applies at exact half-integers. If the two endpoints coincide, the controlled-$Z$ gate is omitted. Both bit-flip and phase-flip noise with probability $10^{-3}$ are attached to each $U_3$ gate. Computational-basis measurement is followed by full-bitstring mode aggregation through CUDA-Q's \texttt{counts.most\_probable}. The backend's resolution of tied modes is not recorded in the run manifests. The returned string is converted to an $N$-dimensional vector of zeros and ones.

The pre-network has one width-256 LeakyReLU hidden layer and a sigmoid output of dimension $d+N$. Its last $N$ outputs form the bypass. The post-network concatenates the $N$ channel features with the $N$ bypass features, uses a width-256 LeakyReLU layer, and produces a tanh action mapped to the environment's action bounds. Each of the two critics first transforms the state through a width-256 layer, concatenates the action, and applies another width-256 layer before its scalar output. The hidden LeakyReLU slope is 0.01. Linear layers use Xavier-uniform weight initialization and zero biases.

The default surrogate has a width-2048 LeakyReLU hidden layer and $N$ sigmoid outputs. With $N=M=10$, $d=220$ and it has $(220+10+1)2048+10=473{,}098$ parameters. In the base hybrid deployment, the surrogate is used only during training; Appendix~\ref{app:deployment} evaluates it as an alternative readout. We count its parameters separately from those of the deployed actor, since it contributes to training cost but not to action selection.

\begin{table}[H]
\centering\small
\begin{tabular}{ll}
\toprule
Setting & Value or convention \\
\midrule
Primary task / nominal budget & Humanoid-v4 / 3,000,000 environment steps \\
Training seeds & 15 per primary bottleneck; 3 per width; 5 per ablation \\
Quantum default & $N=10$, $M=10$, $S=1000$ \\
Replay / minibatch & Capacity $10^6$ / 256, uniform sampling \\
Surrogate fit & 32 Adam steps, sub-batches of 16 with replacement \\
Learning rates & $3\times10^{-4}$ for pre, post, surrogate, and critics \\
Gradient clipping & Disabled (all configured clipping values are $+\infty$) \\
Discount / target rate / actor delay & $0.99$ / $0.005$ / 2 critic updates \\
Warm-up & 1001 random actions, due to the inclusive index condition \\
Exploration noise & Standard deviation $0.05$ times the full action range \\
Target smoothing & Standard deviation $0.2$ times first-coordinate half-range \\
Target noise clipping & $[-0.5,0.5]$ in absolute action units \\
Periodic / terminal evaluation & 10 / 30 episodes, no external exploration noise \\
Maximum episode length & 1000 environment steps \\
\bottomrule
\end{tabular}
\caption{Hyperparameters and implementation conventions for the primary experiments. Environment-step budgets count training interactions only.}
\label{tab:hyperparameters}
\end{table}

\subsection{Training order and target construction}
\begin{algorithm}[H]
\caption{Replay-trained surrogate-forward TD3 (base quantum configuration)}
\label{alg:training}
\begin{algorithmic}[1]
\STATE Initialize pre-network $\theta$, post-network $\psi$, surrogate $\phi$, twin critics $\omega_1,\omega_2$, their targets, optimizers, and replay $\mathcal D$.
\FOR{each environment interaction}
\IF{inside random-action warm-up}
\STATE Draw a random action; record zero placeholders for $u$ and $B$.
\ELSE
\STATE Compute $(u,c)=\pre(s)$; query $B\sim P_{u,S}$.
\STATE Act with $\operatorname{clip}_{\mathcal A}(\post(B,c)+\xi)$.
\ENDIF
\STATE Observe $(r,s',d_{\mathrm{term}})$ and store $(s,u,B,a,r,s',d_{\mathrm{term}})$ in $\mathcal D$.
\IF{warm-up is over and replay has more than 256 entries}
\STATE Draw a uniform minibatch $\mathcal M$ of 256 transitions.
\STATE Fit $\phi$ by 32 BCE steps on size-16 resamples of $(u,B)$ in $\mathcal M$; set $\bar\phi\gets\phi$.
\STATE Construct the surrogate-forward target $y$ in Eq.~\eqref{eq:target}, without channel queries.
\STATE Update both critics by squared-error regression to the detached target.
\IF{this is every second critic update}
\STATE Update $\theta,\psi$ using Eq.~\eqref{eq:actorloss}, keeping $\phi$ fixed but differentiating through its inputs.
\STATE Polyak-update pre, post, and both critic targets with rate $\tau$.
\ENDIF
\ENDIF
\STATE Evaluate at the scheduled crossings using the sampled deployed channel, without exploration noise.
\ENDFOR
\end{algorithmic}
\end{algorithm}
After warm-up, each interaction is followed by an update. The bootstrap multiplier is zero for terminal transitions but remains active at time-limit truncations. We hard-copy the surrogate target after every fit and Polyak-update the pre-network, post-network, and critic targets on the delayed actor schedule. Autograd can accumulate gradients for the surrogate parameters during the actor backward pass; its optimizer is not stepped at that stage, and the buffers are cleared before the next fit.

The exploration noise standard deviation is $0.05(a_{\max}-a_{\min})$ componentwise. Target smoothing instead uses the scalar $0.2(a_{\max,1}-a_{\min,1})/2$ for all action coordinates, then clips the noise in absolute action units. All compared configurations use these same noise conventions.

\subsection{Control configurations}\label{app:controls}
\paragraph{FC and FCM.}
For FC the bottleneck is a learned linear map followed by a sigmoid. FCM adds the same width-2048 hidden layer as the quantum surrogate. Both use the deterministic module for interaction and updates, training it directly with the actor objective and Polyak-averaging its target parameters. FCM matches the surrogate's architecture and parameter count; its action-selection cost differs because it executes a classical network rather than a sampled circuit. Neither control requires an auxiliary BCE fit.

\paragraph{CSB.}
The classical stochastic bottleneck uses fixed random matrices in a $d\to256\to N$ network, with tanh hidden units and sigmoid output logits scaled by gain 3. The channel-construction seed is fixed at 20260611 across training seeds. Its private shot generator is also initialized with the same constant, 20260612, in every training seed, and its state is not restored from checkpoints. Each shot draws independent Bernoulli coordinates conditional on the controls; the same full-bitstring mode operation is applied to $S$ shots. Mode aggregation can introduce dependencies between output bits even though the individual shots have a product distribution. CSB uses the same surrogate-forward protocol as PQC. Its output law is not entropy-matched: the conditional entropy estimates and variance coefficients in Appendix~\ref{app:diagnostics} are substantially smaller than for PQC. This is a comparison of different stochastic feature maps, not a controlled intervention on quantum correlations alone.

\paragraph{RBG and stop-gradient.}
RBG returns control-independent random bits and disables upstream learning, including gradients through the bypass. Its fixed random pre-network still provides state-dependent bypass features to the trainable post-network. It is therefore a control for this full frozen-encoder architecture, rather than an isolated test of quantum randomness or entanglement. The hard-forward stop-gradient variant samples the control-dependent quantum channel but detaches both the sampled features and the bypass; the pre-network optimizer is not stepped. Four runs supply the matched-budget comparison in Section~\ref{sec:cost}. The surrogate is still fitted from replay, but it supplies no policy-update gradient in this control.

\paragraph{Hard-forward STE.}
The actor update and critic-target computation both request sampled channel features. The expression $B+f_\phi(u)-\sg(f_\phi(u))$ then supplies the surrogate derivative during backpropagation. Every minibatch element requires a channel call, giving 384 extra calls per update on average at batch size 256 and actor delay two. Switching to surrogate-forward changes the forward objective as well as the query count, so return differences reflect both changes. Batching and parallel execution can change runtime but leave this control-setting count unchanged.

\subsection{Evaluation, termination, and checkpoints}
The training loop stops at the first episode boundary after reaching 3M environment steps. The overshoot is at most one episode. A step budget keeps sample counts comparable even when policy quality changes episode length. Evaluation interactions are counted separately.

A ten-episode evaluation occurs on the first completed episode and when the cumulative training count crosses a multiple of $10\times$ the maximum episode length (10,000 steps). A dedicated 30-episode terminal evaluation is scheduled at the end of each completed run. One scalar-sensitivity-regularized run has only ten recorded terminal episodes; Appendix~\ref{app:statistics} identifies this exception. Environment reset seeds are generated by
$\mathrm{episode\_seed}(b,e)=(1000003b+e)\bmod(2^{31}-1)$; different evaluation episodes use different values. Evaluation disables external exploration noise while retaining the stochastic finite-shot channel.

Elapsed-time counters include simulation, learning, and periodic evaluation, but exclude scheduler waiting. Resumed runs restore the previously logged elapsed time. The last metrics row precedes the separate terminal evaluation, whereas the final channel counter includes it. We therefore use time and call counts from the same periodic checkpoint when comparing update strategies. Matching integer seeds alone does not ensure matched hardware.

\subsection{Reproducibility and data availability}
\label{app:reprolimits}
\paragraph{Warm-up observations.}
Initial random actions are stored with $(u,B)=(0,0)$ placeholders. The surrogate fit does not mask these records, so early BCE updates can be biased by labels that are not genuine channel observations. There are 1001 such records, occupying approximately 1\% of replay at 100,000 interactions. With capacity $10^6$, they remain until individually overwritten after replay fills. All reported training runs use this unmasked warm-up procedure.

\paragraph{Randomness and restarts.}
Environment, PyTorch, NumPy, and replay random streams are seeded and checkpointed. CUDA-Q's shot generator is not explicitly seeded or checkpointed, and the private CSB/RBG generator state is absent from model checkpoints. The reported seed labels therefore identify repeated training runs without guaranteeing bitwise reproducibility, especially after a restart.

\paragraph{Hardware and software records.}
Training uses PyTorch, Gymnasium/MuJoCo, and CUDA-Q. The target-selection code prefers the NVIDIA simulator when a GPU is available; strict mode stops on failure rather than falling back to another backend. The run manifests do not record every run's GPU model and software versions. Runtime comparisons are therefore reported as observed process-time ratios, without device normalization.

\paragraph{Experimental records.}
Our experimental records contain 159 terminal evaluations. All 75 primary runs have 30 distinct episode returns and complete periodic evaluation histories. Across all experiments, 158 terminal evaluations contain 30 episodes; scalar-sensitivity-regularization seed 1 contains ten. For every run, the mean calculated from the episode returns agrees with the recorded summary. We distinguish completed terminal evaluations, partial histories and same-seed re-executions, and do not count re-executions as additional independent primary runs. Missing evaluations are not assigned zero returns. Appendix~\ref{app:reexecutions} details the shared training histories.

\paragraph{Run completion.}
The primary Humanoid comparison is complete. Some additional-task and update-path runs were still in progress when we finalized the results for this submission. We report these experiments at the interaction budgets reached by the stated seed cohorts, using only completed evaluations. Each aggregate learning curve uses a fixed cohort, and individual trajectories end at their last recorded evaluation. Runs that have not reached a comparison budget do not contribute to its statistics.

\paragraph{Synthetic-study reproducibility.}
We executed all seven small-circuit configurations twice with the same implementation and settings. Every non-timing probe value agrees between executions. Reference-direction summaries count each control anchor once, and undefined cosines are excluded from the summaries with their coverage reported explicitly. The channel-characterisation and downstream-map studies have separate protocols and control-level or probe-level results. These CPU studies are distinct from the MuJoCo policy-training experiments. The code and analysis package will be released upon acceptance, including the implementations, configurations, recorded software information and source hashes, both synthetic-study executions, and their numerical comparison.

\paragraph{Environment versions.}
Run configurations fix Humanoid-v4 and Hopper-v4, and Ant-v5 and HalfCheetah-v5. These IDs specify separate task implementations, not interchangeable labels. In particular, Gymnasium's version history documents changes to observations, reward handling and contact costs between Humanoid-v4 and v5 \citep{gymnasiumdocs}. All within-task comparisons use the same ID; we neither pool versions nor treat the additional tasks as a matched cross-version benchmark.

\FloatBarrier
\section{Additional experiments and statistical analysis}
\label{app:results}
\subsection{Scores, coverage and uncertainty estimates}
\label{app:statistics}
For run $i$ with $n_i$ terminal evaluation episodes, the score is
$X_i=n_i^{-1}\sum_{j=1}^{n_i}R_{i,j}$.
All 75 primary runs have $n_i=30$ with distinct episode scores. Across the 159 terminal evaluations, 158 contain 30 scores and scalar-sensitivity-regularization seed 1 contains ten. Training runs, not individual evaluation episodes, are the statistical units.

The data release will include per-episode terminal returns, periodic histories, run configurations, and an inventory distinguishing completed runs, partial histories and same-seed re-executions.

For ordered scores $X_{(1)}\le\cdots\le X_{(n)}$, let $k=\lfloor n/4\rfloor$. Our finite-sample interquartile-mean convention is
\begin{equation}
 \widehat{\mathrm{IQM}}=\frac{1}{n-2k}\sum_{i=k+1}^{n-k}X_{(i)}.
 \label{eq:iqm}
\end{equation}
It retains nine central scores when $n=15$. Pointwise 95\% percentile intervals use 20,000 resamples of training runs, with analysis RNG seed 270913 for each statistic. They quantify between-run variation on one task; they are neither simultaneous nor multitask intervals. We follow the emphasis on uncertainty and seed-level distributions in \citet{agarwal2022deepreinforcementlearningedge}.

The probability-of-improvement estimate between configurations $A$ and $B$ is
\begin{equation}
 \widehat p_{A>B}=\frac{1}{n_An_B}\sum_{i,j}
 \left[\mathbf1\{X_i^A>X_j^B\}+\frac12\mathbf1\{X_i^A=X_j^B\}\right].
\end{equation}
We bootstrap the two groups separately. Matching seed labels do not establish a paired common-random-number design across different architectures. Terminal-score tests are exact, two-sided Mann--Whitney tests; the data have no ties. Holm adjustment is applied to the four comparisons against PQC. Welch tests and Fisher exact tests at 1500 and 4000 form separate four-comparison families. Both thresholds are descriptive; no pre-registration is claimed. Width, shot and surrogate-family comparisons are exploratory.

\begin{table}[!htb]
\centering\small\setlength{\tabcolsep}{3pt}
\begin{tabular}{lrrrrr}
\toprule
PQC vs. & PoI [95\% CI] & MW & Welch & Fisher 1500 & Fisher 4000 \\
\midrule
FC & 0.70 [0.49, 0.89] & 0.135 & 0.091 & 0.418 & 0.131 \\
FCM & 0.68 [0.47, 0.87] & 0.135 & 0.101 & 1.000 & 0.131 \\
CSB & 0.76 [0.56, 0.92] & 0.044 & 0.037 & 0.527 & 0.005 \\
RBG & 0.92 [0.78, 1.00] & $<10^{-3}$ & $<10^{-3}$ & $<10^{-3}$ & $<10^{-3}$ \\
\bottomrule
\end{tabular}

\caption{Primary terminal-score comparisons against PQC. Probability-of-improvement intervals are pointwise seed-bootstrap intervals. Each test column is Holm-adjusted within its own four-comparison family. Fisher columns use the named thresholds. The analysis package to be released upon acceptance will include both raw and adjusted $p$-values.}
\label{tab:tests}
\end{table}
\begin{table}[!htb]
\centering\small\setlength{\tabcolsep}{5pt}
\begin{tabular}{lrrrrr}
\toprule
Seed & PQC & FC & FCM & CSB & RBG \\
\midrule
0 & 4,453.2 & 885.4 & 654.1 & 2,695.9 & 564.6 \\
1 & 4,326.2 & 722.7 & 1,891.2 & 551.1 & 571.2 \\
2 & 901.1 & 6,135.8 & 2,664.8 & 275.9 & 471.5 \\
3 & 270.3 & 1,137.6 & 797.1 & 926.3 & 629.6 \\
4 & 5,375.9 & 1,201.7 & 400.8 & 3,926.4 & 575.0 \\
5 & 596.5 & 4,304.2 & 2,399.1 & 816.2 & 305.9 \\
6 & 5,355.3 & 5,437.9 & 4,379.9 & 1,802.2 & 601.7 \\
7 & 5,237.6 & 531.5 & 4,806.9 & 4,669.3 & 502.8 \\
8 & 4,751.5 & 2,649.4 & 1,253.5 & 1,147.1 & 344.3 \\
9 & 4,750.4 & 4,054.6 & 777.1 & 236.4 & 572.1 \\
10 & 4,834.9 & 570.6 & 3,394.4 & 3,150.3 & 526.6 \\
11 & 768.6 & 2,448.5 & 4,787.3 & 3,159.0 & 462.9 \\
12 & 5,231.1 & 480.7 & 2,563.4 & 3,558.6 & 641.9 \\
13 & 2,845.9 & 379.3 & 2,137.9 & 1,245.7 & 557.7 \\
14 & 4,588.0 & 873.8 & 4,720.5 & 498.6 & 493.0 \\
\bottomrule
\end{tabular}

\caption{All primary terminal scores, recomputed from raw 30-episode evaluations. Rounded values are shown; all analyses use full precision.}
\label{tab:seeds}
\end{table}

\paragraph{Learning-curve extraction.}
The metrics log repeats the most recent evaluation between evaluation events. We extract the first completed episode and rows crossing a 10,000-interaction boundary, as specified by the evaluation code. Zero-valued reset markers do not represent evaluations and are excluded; none occurs among the selected evaluation records. A run must reach the stated interaction budget, and the statistic uses its latest genuine evaluation at or before that budget. This past-only rule prevents future-checkpoint leakage. Main curves use fixed cohorts: 15 for each Humanoid configuration, ten for Ant and Hopper, nine for HalfCheetah, and five/five/four for surrogate-forward/STE/stop-gradient. Lines are medians and bands are interquartile ranges across those runs, not confidence intervals. The compact panel additionally shows individual trajectories. Appendix per-seed lines stop at their own last evaluation, and each plotting axis stops at the last observation in that panel.

\subsection{High-return frequencies and late-training behavior}
\label{app:reliability}
For terminal scores $X_1,\ldots,X_n$, the observed high-return frequency is
\begin{equation}
 \widehat p_\tau=\frac1n\sum_{i=1}^n\mathbf1\{X_i\ge\tau\}.
\end{equation}
We selected $\tau=4000$ after inspecting the Humanoid results. It denotes a high-return regime at a finite training budget, not an official success criterion or a mathematical equilibrium. Figure~\ref{fig:profiles} shows the full threshold profile. The exact binomial intervals in Table~\ref{tab:reliability} are conditional on a fixed configuration and threshold; they do not account for exploratory selection. For $k$ successes in $n$ runs, their endpoints are
\begin{equation}
 L=\begin{cases}0,&k=0,\\ \operatorname{Beta}^{-1}_{k,n-k+1}(0.025),&k>0,\end{cases}
 \qquad
 U=\begin{cases}1,&k=n,\\ \operatorname{Beta}^{-1}_{k+1,n-k}(0.975),&k<n.\end{cases}
\end{equation}
The compact setting's three successes in three runs yield $[29.2\%,100\%]$; this width illustrates the uncertainty of a three-run frequency estimate.

\begin{figure}[!htb]
\centering\includegraphics[width=.95\linewidth]{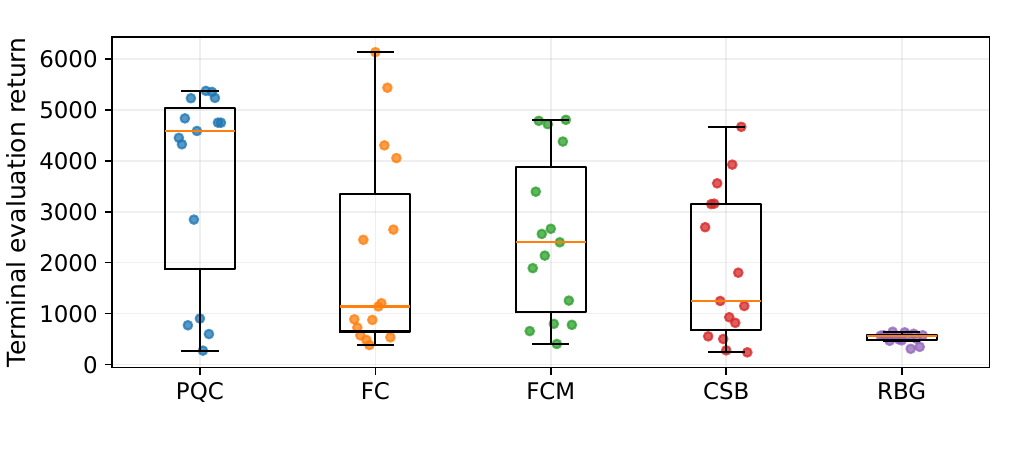}
\caption{Humanoid terminal distributions. Dots are training-run scores; boxes show medians and interquartile ranges, with 1.5-IQR whiskers. Each primary configuration has 15 raw terminal evaluations.}
\label{fig:terminal}
\end{figure}
\begin{table}[!htb]
\centering\small\setlength{\tabcolsep}{3pt}
\begin{tabular}{lrrrrr}
\toprule
Configuration & $n$ & Terminal & 95\% interval & Late median & Ever \\
\midrule
PQC & 15 & 10/15 & [38.4, 88.2]\% & 10/15 & 10/15 \\
FC & 15 & 4/15 & [7.8, 55.1]\% & 4/15 & 6/15 \\
FCM & 15 & 4/15 & [7.8, 55.1]\% & 3/15 & 7/15 \\
CSB & 15 & 1/15 & [0.2, 31.9]\% & 2/15 & 4/15 \\
RBG & 15 & 0/15 & [0.0, 21.8]\% & 0/15 & 0/15 \\
$N=5,h=256,S=1000$ & 3 & 3/3 & [29.2, 100.0]\% & 3/3 & 3/3 \\
$N=5,h=2048,S=100$ & 5 & 5/5 & [47.8, 100.0]\% & 5/5 & 5/5 \\
\bottomrule
\end{tabular}

\caption{Terminal and late-training high-return frequencies at threshold 4000. ``Late median'' counts runs whose median periodic return over 2.5--3M interactions exceeds the threshold. ``Ever'' counts any periodic evaluation reaching it by 3M. The last two settings are exploratory configurations, not extra primary seeds.}
\label{tab:reliability}
\end{table}
Late-training summaries use one median per seed from actual evaluations in $[2.5\mathrm M,3\mathrm M]$, without smoothing or extrapolation. The counts are 10/15 for PQC, 4/15 for FC, 3/15 for FCM, 2/15 for CSB and 0/15 for RBG. For the compact configuration, the late medians are approximately 5890, 4333 and 5525. Its least stable seed exceeds 4000 at only 62\% of these checkpoints, so a high late median does not imply a flat learning curve.

\subsection{Capacity, shot and surrogate-family comparisons}
\label{app:ablations}
\begin{table}[!htb]
\centering\small\setlength{\tabcolsep}{4pt}
\begin{tabular}{lrrrrr}
\toprule
$N$ & Width & Parameters & Mean & Median & Range \\
\midrule
5 & 64 & 8,069 & 535 & 448 & [403, 752] \\
5 & 256 & 32,261 & 5,088 & 5,103 & [4,268, 5,893] \\
5 & 1024 & 129,029 & 3,542 & 4,491 & [628, 5,508] \\
5 & 2048 & 258,053 & 1,891 & 887 & [832, 3,953] \\
5 & 4096 & 516,101 & 2,741 & 2,660 & [515, 5,048] \\
10 & 64 & 14,794 & 387 & 416 & [264, 481] \\
10 & 256 & 59,146 & 4,412 & 4,717 & [3,778, 4,742] \\
10 & 1024 & 236,554 & 4,863 & 5,342 & [3,880, 5,367] \\
10 & 2048 & 473,098 & 1,741 & 589 & [71, 4,561] \\
10 & 4096 & 946,186 & 2,851 & 3,700 & [580, 4,272] \\
\bottomrule
\end{tabular}

\caption{The full capacity sweep, with three runs per setting. Ranges are minima and maxima, not confidence intervals. The ten-qubit, width-2048 cell contains same-seed re-executions of primary seeds 0-2, not independent replications.}
\label{tab:capacity}
\end{table}
The five-qubit, width-256 setting has the largest observed mean and minimum in the sweep; ten qubits with width 1024 have the largest median. Parameter counts are $(d+N+1)h+N$, with $d=(2N+2)M$ and $M=10$. Circuit controls and bypass outputs also change with qubit count, so changing $N$ is not an intervention on surrogate width alone. The comparison concerns complete configurations.

\begin{table}[!htb]
\centering\small\setlength{\tabcolsep}{3pt}
\begin{tabular}{lrrrr}
\toprule
Configuration ($n=5$) & Mean & Median & Range & $\ge4000$ \\
\midrule
$N=10,S=1000$ (re-execution) & 3,301 & 4,015 & [660, 5,391] & 3/5 \\
$N=10,S=100$ & 4,056 & 4,748 & [668, 5,359] & 4/5 \\
$N=5,S=1000$ & 2,537 & 969 & [657, 5,257] & 2/5 \\
$N=5,S=100$ & 4,617 & 4,754 & [4,014, 4,897] & 5/5 \\
Autoregressive soft rollout & 3,092 & 4,042 & [461, 5,187] & 3/5 \\
Scalar sensitivity penalty & 4,120 & 4,956 & [687, 5,130] & 4/5 \\
Directional derivative targets & 4,171 & 4,917 & [606, 5,366] & 4/5 \\
Forward-path re-execution & 3,212 & 4,675 & [242, 5,344] & 3/5 \\
\bottomrule
\end{tabular}

\caption{Five-run exploratory configurations. The ten-qubit, 1000-shot reference and forward-path repeat are same-seed re-executions of primary seeds 0--4. They are displayed separately and never pooled as extra primary seeds. Scalar-sensitivity seed 1 has ten terminal episodes; every other row uses 30 episodes per run.}
\label{tab:ablations}
\end{table}
At ten qubits, reducing shots from 1000 to 100 changes mean return from 3301 to 4056 and median return from 4015 to 4748. At five qubits, the corresponding means are 2537 and 4617. These observations concern different mode-aggregated laws, not only different precision of a fixed expectation estimate. The capacity and shot sweeps test separate configurations.

\paragraph{Autoregressive surrogate.}
The autoregressive variant is trained with teacher-forced conditional Bernoulli likelihoods and a masked lower-triangular coupling matrix, adding 45 effective coupling parameters at ten qubits. At update time, previous bits are replaced recursively by soft probabilities. This differentiable rollout generally differs from the exact autoregressive marginal means; Proposition~\ref{prop:bce} concerns the base coordinatewise BCE model. The variant changes both fitting and downstream relaxation, so its returns do not isolate a benefit from modeling output correlations.

\paragraph{Scalar sensitivity regularization.}
The source option \texttt{jac\_reg} differentiates the sum of surrogate outputs and penalizes its projection onto a random unit input direction $v$. Its penalty obeys
\begin{equation}
 \lambda[v^\top J_f(u)^\top\mathbf1]^2,\qquad
 \E_v[vv^\top]=I/d
 \quad\Longrightarrow\quad
 \E_v[\mathrm{penalty}]=\frac\lambda d\norm{J_f(u)^\top\mathbf1}_2^2.
\end{equation}
Output gradients may cancel, so this is neither a Frobenius-Jacobian penalty nor a penalty on $J_f-J_\mu$. The logged fitting loss excludes this extra penalty.

\paragraph{Directional derivative targets.}
The Sobolev variant adds squared error between surrogate Jacobian--vector products and cached channel finite differences. These labels require additional calls, include finite-shot and finite-difference error, and can cross routing boundaries. The supervised directions are not exact parameter-shift derivatives. These runs are kept distinct from diagnostic-only runs in Table~\ref{tab:diag}.

\subsection{Channel entropy and response-fitting diagnostics}
\label{app:diagnostics}
The live diagnostic records cover three PQC runs, three CSB runs and five PQC runs with derivative-target fitting. Table~\ref{tab:diag} reports each run's last diagnostic event, not a uniform bound over the channel. Marginal estimation uses eight replay-control probes with 64 aggregated samples each. Four normalized directions and displacement $\delta=0.02$ are used for central differences. The table's $\widehat e_1$ is a mean marginal discrepancy and $\widehat D_\delta$ a mean directional discrepancy. The field \texttt{eps2\_fro} rescales random-direction measurements; it is not a measured full Jacobian norm or an operator-norm bound.

\begin{table}[!htb]
\centering\small\setlength{\tabcolsep}{3pt}
\begin{tabular}{lrrrrrrr}
\toprule
Channel & Seed & Steps (M) & $\widehat e_1$ & $\widehat D_\delta$ & $\widehat H(B\mid u)$ & $\widehat{\sqrt v}$ & BCE \\
\midrule
CSB & 0 & 2.993 & 0.213 & 2.97 & 0.65 & 0.40 & 0.066 \\
CSB & 1 & 2.973 & 0.051 & 1.01 & 0.41 & 0.17 & 0.019 \\
CSB & 2 & 2.953 & 0.186 & 4.05 & 1.53 & 0.55 & 0.072 \\
PQC & 0 & 2.974 & 0.689 & 15.16 & 6.66 & 1.47 & 0.664 \\
PQC & 1 & 2.968 & 0.809 & 13.40 & 5.90 & 1.40 & 0.528 \\
PQC & 2 & 2.924 & 0.693 & 14.83 & 7.01 & 1.44 & 0.587 \\
PQC + derivatives & 0 & 2.893 & 0.485 & 21.13 & 5.42 & 1.22 & 0.533 \\
PQC + derivatives & 1 & 2.970 & 0.933 & 18.96 & 6.79 & 1.34 & 0.695 \\
PQC + derivatives & 2 & 2.995 & 1.024 & 38.54 & 5.24 & 1.09 & 0.407 \\
PQC + derivatives & 3 & 2.971 & 1.099 & 18.96 & 6.68 & 1.43 & 0.615 \\
PQC + derivatives & 4 & 2.781 & 0.735 & 16.23 & 6.40 & 1.45 & 0.663 \\
\bottomrule
\end{tabular}

\caption{All eleven live diagnostic runs, separated by channel and fitting protocol. Entropy is in bits. $\widehat{\sqrt v}$ is the mean empirical variance coefficient across the marginal probes. BCE is the median logged episode-end fitting loss over 2.5--3M interactions (nats per bit), not the loss at the single diagnostic event.}
\label{tab:diag}
\end{table}
PQC diagnostic runs have conditional entropy estimates between 5.90 and 7.01 bits, versus 0.41--1.53 for CSB. Their empirical variance coefficients are 1.40--1.47 and 0.17--0.55, respectively, compared with the binary worst-case bound $\sqrt{10/4}=1.581$. The mean-feature value bound in Eq.~\eqref{eq:valuegap} can therefore be substantially looser for the PQC channel. The actual bias still depends on the downstream map and is not determined by this coefficient alone. The BCE and marginal discrepancies likewise measure different channel predictability, not a causal effect of quantum correlations on return.

\paragraph{Entropy estimator limits.}
Entropy uses 16 replay-control probes, 128 aggregated samples at each, and the plug-in estimate with Miller--Madow correction. If $k_{\mathrm{obs}}$ outcomes are observed among $R$ samples, the estimator in bits is
\begin{equation}
 \widehat H=-\sum_b\widehat p_b\log_2\widehat p_b
 +\frac{k_{\mathrm{obs}}-1}{2R\ln2}
 \le\log_2 R+\frac{R-1}{2R\ln2}.
\end{equation}
At $R=128$ the ceiling is \EntropyCeiling\ bits, below the ten-bit maximum possible true entropy. High estimates near this ceiling do not establish maximal true entropy; rare outcomes can remain unseen. The logged total-correlation estimate subtracts this undersampled joint entropy from summed marginal entropies and can consequently be biased upward. We therefore do not interpret this total-correlation estimate as evidence of entanglement or as an explanation of learning performance.

\paragraph{Directional and Lipschitz proxies.}
For independent Bernoulli replicate estimates at two unclipped endpoints, the RMS sampling error of the central difference is at most $\sqrt{N/(8R\delta^2)}\simeq6.99$ for $N=10,R=64,\delta=0.02$. The logger's ``noise floor'' is this worst-case upper bound, not a lower bound. Actual perturbations include routing coordinates and are clipped; smaller endpoint displacement can increase the bound, and crossing a routing cell invalidates a local smooth-derivative interpretation. Finite-difference bias is separate. The maximum observed downstream gradient and secant norms are empirical maxima, not certified Lipschitz constants on a support hull. The diagnostic cost is $8\cdot64(1+2\cdot4)+16\cdot128=6656$ channel invocations per event.

\subsubsection{Shot-count and control dependence}
\label{app:entropy}
We isolate the readout law using the same ten-layer circuit and $10^{-3}$ bit-flip and phase-flip channels as in training. For each of $N=5$ and $N=10$, we evaluate 200 controls drawn uniformly from $[0,1]^d$. Density-matrix simulation supplies the Born probability vector without sampling error, up to floating-point arithmetic. For each vector and $S\in\{100,1000\}$, 20,000 independent multinomial shot-count bags estimate the mode-output law. The Born probabilities are exact simulator outputs; the aggregated probabilities, entropy and marginal variance are Monte Carlo estimates.

The main calculation breaks mode ties by the smallest bitstring code, as in the small-circuit study. We also choose uniformly among tied modes using the same shot bags. The latter is a sensitivity analysis, not a claim about CUDA-Q's unrecorded tie convention. Entropy uses the plug-in estimate with Miller--Madow correction; brackets in Table~\ref{tab:randomentropy} describe variation across controls, not confidence intervals. The 20,000-bag calculation is not subject to the live diagnostic's 128-sample ceiling, but retains finite-sample error.
\begin{table}[!htb]
\centering\small\setlength{\tabcolsep}{3pt}
\begin{tabular}{lrrrrrr}
\toprule
$N$ & $S$ & Tie rule & $H(\pi_u)$ & $\widehat H(P_{u,S})$ [IQR] & $\widehat{\sqrt v}$ & Ties (\%) \\
\midrule
5 & 1000 & Smallest & 4.27 & 0.16 [0.00, 0.79] & 0.21 & 0.4 \\
5 & 1000 & Uniform & 4.27 & 0.17 [0.00, 0.80] & 0.20 & 0.4 \\
5 & 100 & Smallest & 4.27 & 1.28 [0.83, 1.87] & 0.65 & 8.4 \\
5 & 100 & Uniform & 4.27 & 1.25 [0.83, 1.90] & 0.64 & 8.4 \\
10 & 1000 & Smallest & 8.35 & 1.92 [1.14, 2.79] & 0.81 & 9.3 \\
10 & 1000 & Uniform & 8.35 & 1.92 [1.12, 2.82] & 0.81 & 9.3 \\
10 & 100 & Smallest & 8.35 & 4.62 [3.73, 5.63] & 1.15 & 40.8 \\
10 & 100 & Uniform & 8.35 & 4.59 [3.72, 5.78] & 1.18 & 40.8 \\
\bottomrule
\end{tabular}

\caption{Channel characterisation at 200 random controls per qubit count. Entropies are in bits; entries are medians across controls. Both shot counts and tie rules use the same controls. ``Ties'' is the percentage of bags with more than one modal bitstring. $\widehat{\sqrt v}$ uses the estimated aggregated marginals.}
\label{tab:randomentropy}
\end{table}
For ten qubits and smallest-code ties, reducing shots raises median aggregated entropy from \EntropyNTenThousand\ to \EntropyNTenHundred\ bits. For five qubits the corresponding medians are \EntropyNFiveThousand\ and \EntropyNFiveHundred. The alternative tie rule gives the same qualitative random-control comparison. A large Born entropy need not imply a similarly random mode output: repeated shots can concentrate the output on a few modal bitstrings. These are conditional readout measurements, not additional policy-training outcomes or a causal explanation of the shot-sweep returns.

\begin{figure}[!htb]
\centering
\includegraphics[width=.48\linewidth]{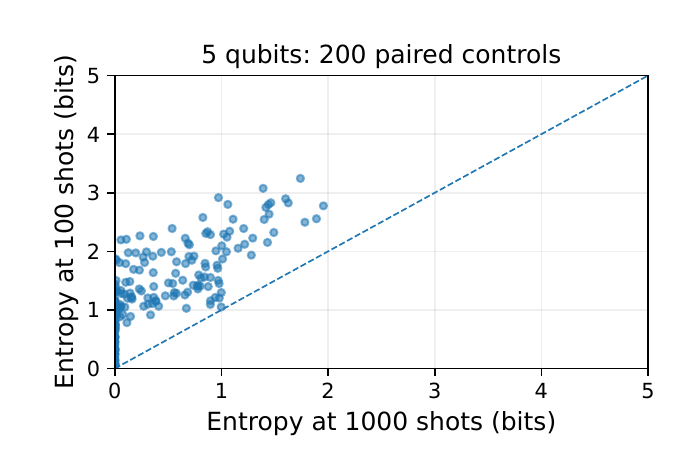}\hfill
\includegraphics[width=.48\linewidth]{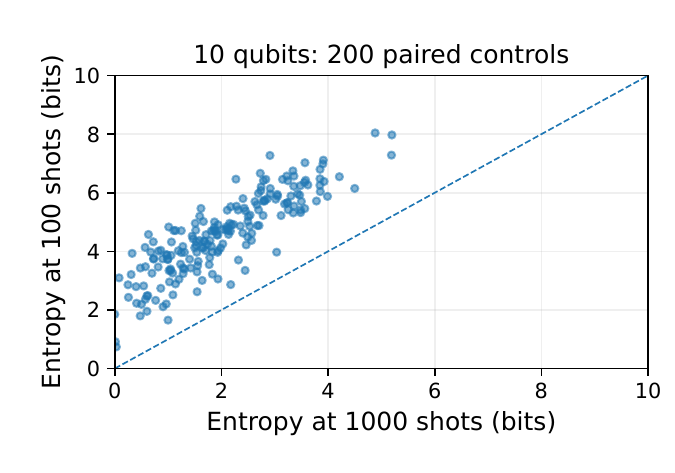}
\caption{Paired change in aggregated entropy when reducing the shot count. Each point is one random control, using smallest-code ties. The diagonal denotes equal entropy at the two shot counts. Tables~\ref{tab:randomentropy}--\ref{tab:saturationentropy} retain both tie conventions.}
\label{fig:randomentropy}
\end{figure}

\paragraph{Rotation saturation.}
A separate sweep holds the routing controls fixed while varying rotation controls as $u^{\mathrm{rot}}=\sigma(gz)$, with $z$ standard Gaussian and $g\in\{0.5,1,2,4,8,32,128\}$. The same 12 Gaussian vectors and routing choices are used at every scale, giving 84 controls per qubit count. This changes rotation saturation without simultaneously changing the routed gate endpoints. Four further controls place all rotation controls exactly at zero or one, with random fixed routing. The full study therefore contains 288 controls per qubit count.

\begin{table}[!htb]
\centering\small\setlength{\tabcolsep}{3pt}
\begin{tabular}{lrrrrrrr}
\toprule
$N$ & Logit scale & $n$ & $H(\pi_u)$ & $S=1000$, min & Uniform & $S=100$, min & Uniform \\
\midrule
5 & $0.5$ & 12 & 4.16 & 0.78 & 0.78 & 1.65 & 1.71 \\
5 & $1$ & 12 & 4.41 & 0.91 & 0.90 & 1.71 & 1.66 \\
5 & $2$ & 12 & 4.48 & 0.20 & 0.20 & 1.85 & 1.85 \\
5 & $4$ & 12 & 4.00 & 0.42 & 0.43 & 1.09 & 1.06 \\
5 & $8$ & 12 & 4.25 & 0.23 & 0.23 & 1.38 & 1.40 \\
5 & $32$ & 12 & 4.26 & 1.84 & 1.84 & 2.52 & 2.52 \\
5 & $128$ & 12 & 4.94 & 4.00 & 4.00 & 4.19 & 4.29 \\
5 & Endpoints & 4 & 5.00 & 4.99 & 5.00 & 4.95 & 5.00 \\
10 & $0.5$ & 12 & 8.30 & 1.56 & 1.58 & 4.28 & 4.26 \\
10 & $1$ & 12 & 8.60 & 2.16 & 2.18 & 4.84 & 5.15 \\
10 & $2$ & 12 & 8.51 & 2.02 & 2.06 & 4.94 & 4.99 \\
10 & $4$ & 12 & 7.87 & 2.17 & 2.13 & 3.89 & 3.93 \\
10 & $8$ & 12 & 7.78 & 1.47 & 1.43 & 3.79 & 3.70 \\
10 & $32$ & 12 & 8.37 & 5.02 & 4.99 & 6.43 & 6.49 \\
10 & $128$ & 12 & 9.75 & 8.89 & 9.02 & 8.82 & 9.25 \\
10 & Endpoints & 4 & 10.00 & 9.81 & 10.00 & 9.38 & 10.00 \\
\bottomrule
\end{tabular}

\caption{Median entropies in the matched saturation sweep, in bits. ``Min'' breaks ties by the smallest code; ``Uniform'' samples among tied modes. The endpoint rows use exactly zero-or-one rotation controls, not an extrapolated finite scale. All values for the aggregated law are Monte Carlo estimates; the true entropy is at most $N$.}
\label{tab:saturationentropy}
\end{table}
\begin{figure}[!htb]
\centering
\includegraphics[width=.48\linewidth]{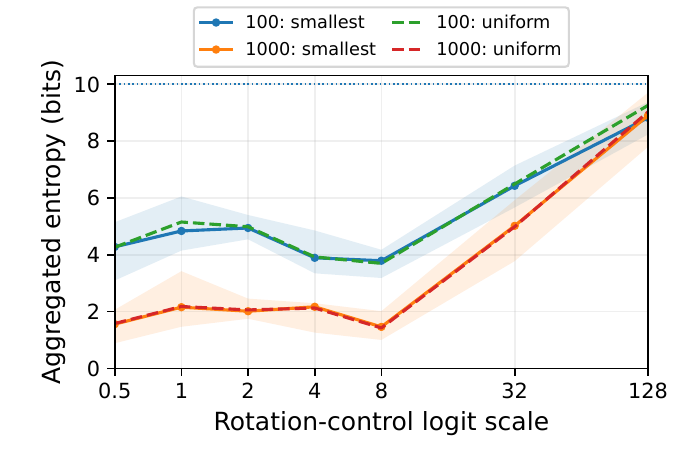}\hfill
\includegraphics[width=.48\linewidth]{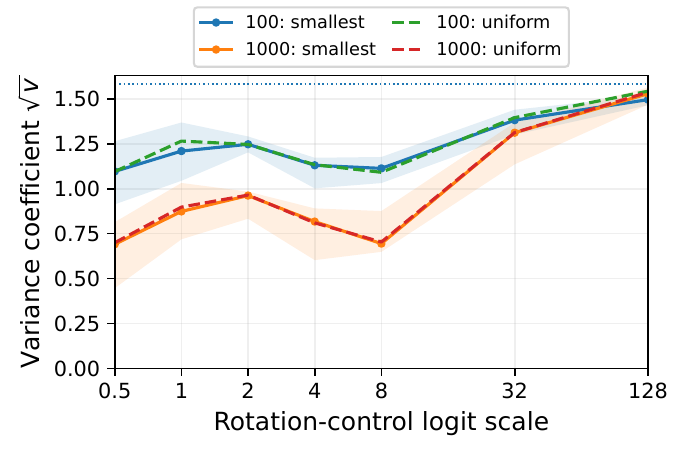}
\caption{Ten-qubit rotation-saturation sweep. Solid curves and interquartile shading use smallest-code ties; dashed curves use uniform tie-breaking on the same bags. Each point summarizes the same 12 underlying Gaussian controls and fixed routing choices at its logit scale. The horizontal dotted lines are the ten-bit entropy and $\sqrt{10/4}$ variance-coefficient maxima.}
\label{fig:saturationentropy}
\end{figure}
At exact rotation endpoints, every $U_3$ is the identity up to a global phase. The intervening CZ gates can change phases, so the state need not return to $|+\rangle^{\otimes N}$; nevertheless its computational-basis probabilities remain uniform. The Pauli noise also preserves this uniform diagonal. With uniform tie-breaking, symmetry then makes the mode-output law exactly uniform for either shot count. Smallest-code ties break that symmetry and can reduce its entropy, particularly when many modes tie. Thus shot count and tie handling both form part of the readout law; an entropy increase with fewer shots is not a universal monotonicity result.

The random-control and saturation sweeps characterize prescribed circuit inputs. Table~\ref{tab:behavior} and Appendix~\ref{app:visitedcontrols} instead characterize controls produced by the stored surrogate actors during their own rollouts, not hybrid-policy rollouts. Their seed-level median entropies reach 8.92 bits. The live estimates in Table~\ref{tab:diag} use 128 samples per control and are constrained by the estimator ceiling. Since control selection, estimation budget and potentially tie handling differ, these studies do not provide a comparison at matched circuit inputs. A higher-replication check on 11 fixed controls compares 20,000 with 80,000 bags across 44 shot/tie cells. The largest observed changes are 0.025 bits in corrected entropy and 0.0028 in $\widehat{\sqrt v}$. This is a sensitivity check on selected controls, not a uniform accuracy guarantee.

\subsection{Individual histories and additional-task coverage}
\label{app:perseed}
The main Ant, Hopper and HalfCheetah summaries use ten, ten and nine seeds. Their actual evaluations occur at 790,046--790,726, 990,036--990,911 and exactly 1,000,000 steps, respectively. HalfCheetah seed 3 has two logged episodes through 2000 interactions and one evaluation at 1000; it terminated early and cannot contribute to the 1M cohort. Some additional-task runs were still in progress when we finalized the results, so each summary uses a fixed interaction budget reached by every run in its stated cohort. Table~\ref{tab:additionalfull} also reports nine-seed cohorts at longer horizons: Ant through 1M interactions and Hopper through 2.1M. Each aggregate learning curve uses a fixed cohort throughout.
\begin{table}[!htb]
\centering\small\setlength{\tabcolsep}{4pt}
\begin{tabular}{lrrrr}
\toprule
Environment & Budget & Seeds & Median & IQM [95\% CI] \\
\midrule
Ant-v5 & 0.8M & 10/10 & 1,947 & 1,989 [1,540, 2,335] \\
Hopper-v4 & 1.0M & 10/10 & 3,355 & 3,394 [2,756, 3,498] \\
HalfCheetah-v5 & 1.0M & 9/10 & 2,027 & 3,133 [1,738, 6,759] \\
Ant-v5 & 1.0M & 9/10 & 2,014 & 2,143 [1,466, 2,585] \\
Hopper-v4 & 2.1M & 9/10 & 3,394 & 3,403 [2,695, 3,507] \\
\bottomrule
\end{tabular}

\caption{Additional-task summaries, with coverage shown for every horizon. Longer-horizon subsets are not replacements for the larger short-horizon cohorts.}
\label{tab:additionalfull}
\end{table}
\begin{table}[!htb]
\centering\small\setlength{\tabcolsep}{5pt}
\begin{tabular}{lrrrr}
\toprule
Environment & Seed & Steps (M) & Episodes & Terminal mean \\
\midrule
HalfCheetah-v5 & 4 & 3.000 & 30 & 2,542 \\
HalfCheetah-v5 & 5 & 3.000 & 30 & 11,302 \\
HalfCheetah-v5 & 6 & 3.000 & 30 & 2,540 \\
HalfCheetah-v5 & 7 & 3.000 & 30 & 2,547 \\
HalfCheetah-v5 & 8 & 3.000 & 30 & 1,116 \\
HalfCheetah-v5 & 9 & 3.000 & 30 & 2,416 \\
Hopper-v4 & 7 & 3.001 & 30 & 3,476 \\
Hopper-v4 & 8 & 3.000 & 30 & 3,001 \\
\bottomrule
\end{tabular}

\caption{All completed additional-task runs with separate terminal evaluations. These are a completion-selected subset, not a full-cohort terminal comparison. In particular, the strong HalfCheetah terminal score coexists with several lower-return completed runs.}
\label{tab:additionalterminal}
\end{table}
We plot all recorded periodic evaluations for the five primary Humanoid configurations and the three additional tasks. Each trajectory ends at its last recorded evaluation, without extrapolation or carrying the last value forward. Each horizontal axis ends at the last observation in its panel; Ant therefore ends near 2.14M rather than at 3M.
\begin{figure}[!htbp]
\centering\includegraphics[width=.98\linewidth]{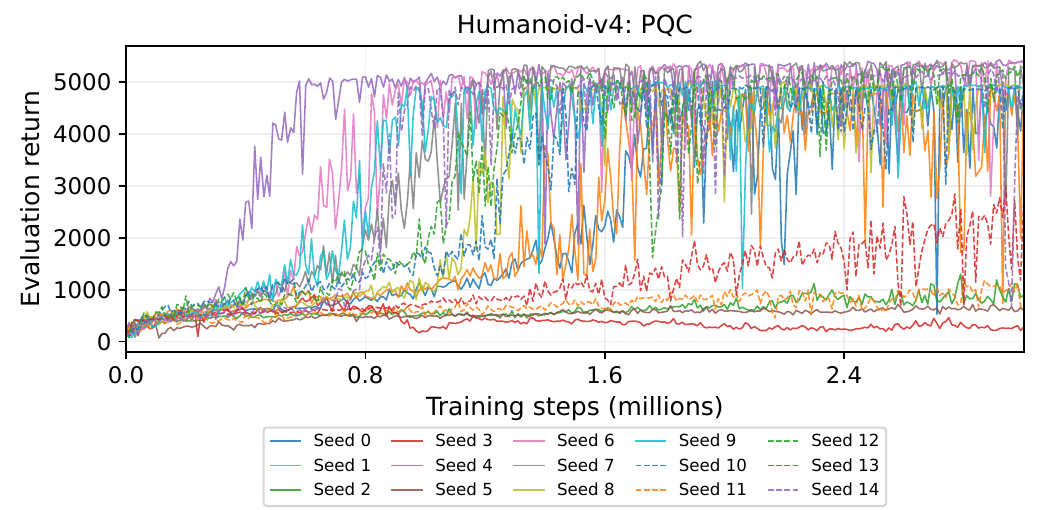}
\caption{Humanoid-v4, PQC per-seed periodic evaluation histories. Every line uses actual observations through that run's final evaluation.}
\label{fig:pqc}
\end{figure}
\begin{figure}[!htbp]
\centering\includegraphics[width=.98\linewidth]{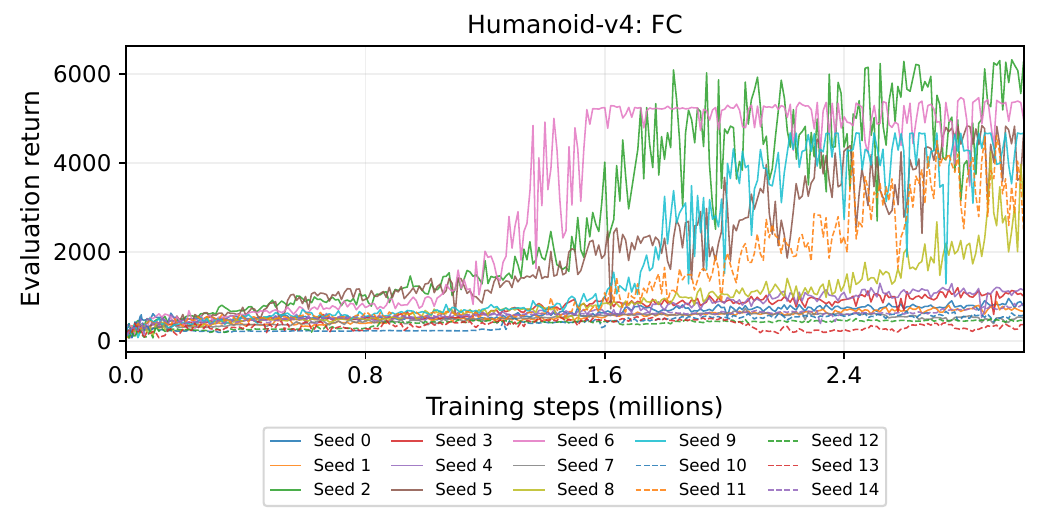}
\caption{Humanoid-v4, FC per-seed periodic evaluation histories. Every line uses actual observations through that run's final evaluation.}
\label{fig:fc}
\end{figure}
\begin{figure}[!htbp]
\centering\includegraphics[width=.98\linewidth]{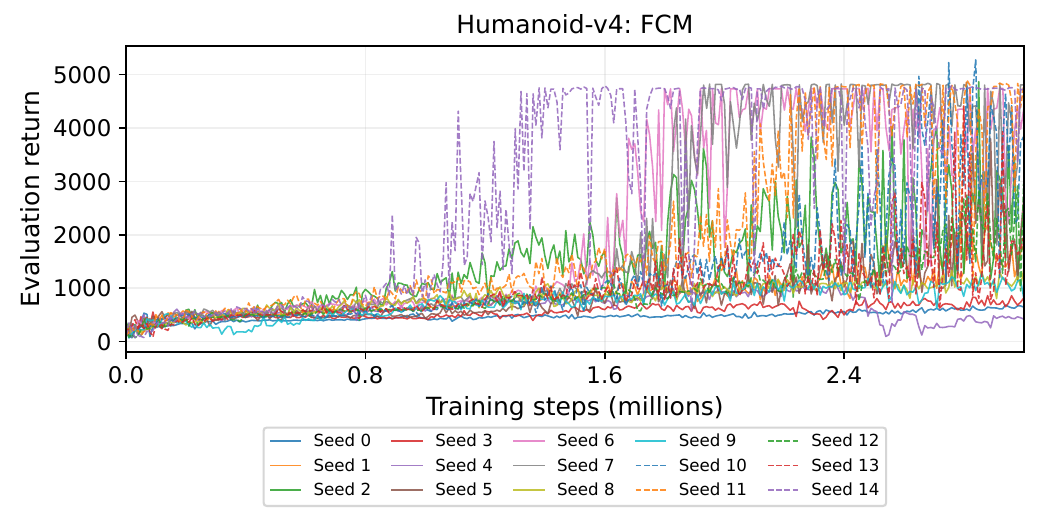}
\caption{Humanoid-v4, FCM per-seed periodic evaluation histories. Every line uses actual observations through that run's final evaluation.}
\label{fig:fcm}
\end{figure}
\begin{figure}[!htbp]
\centering\includegraphics[width=.98\linewidth]{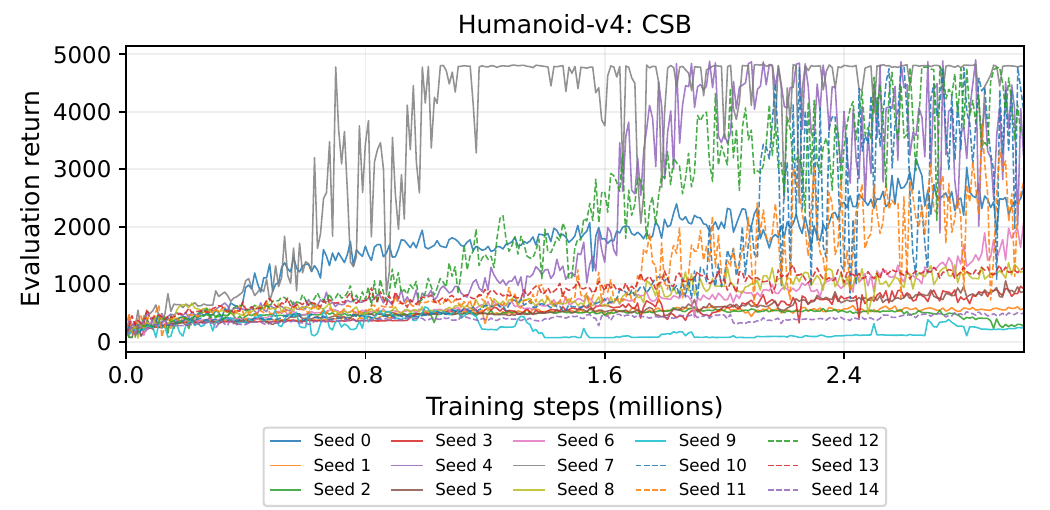}
\caption{Humanoid-v4, CSB per-seed periodic evaluation histories. Every line uses actual observations through that run's final evaluation.}
\label{fig:csb}
\end{figure}
\begin{figure}[!htbp]
\centering\includegraphics[width=.98\linewidth]{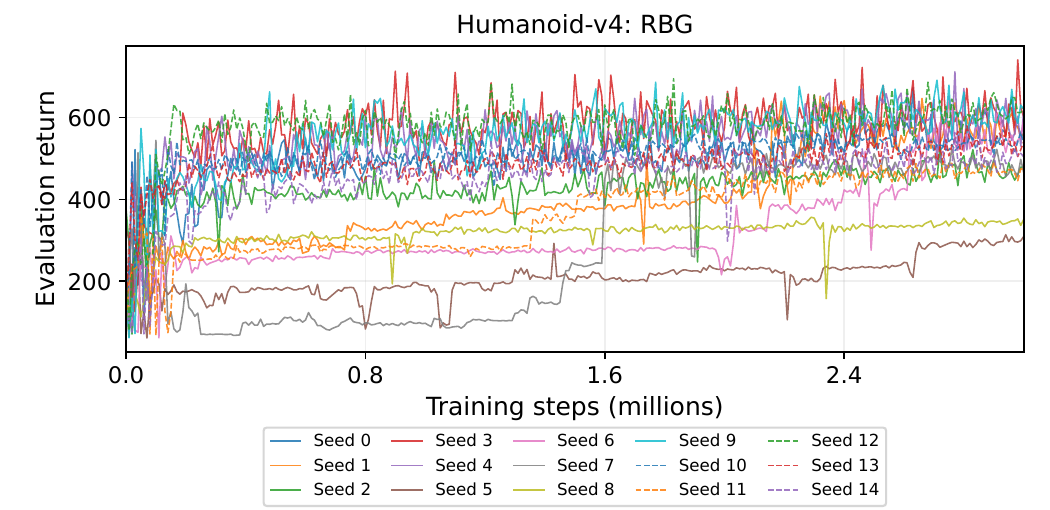}
\caption{Humanoid-v4, RBG per-seed periodic evaluation histories. Every line uses actual observations through that run's final evaluation.}
\label{fig:rbg}
\end{figure}
\begin{figure}[!htbp]
\centering\includegraphics[width=.98\linewidth]{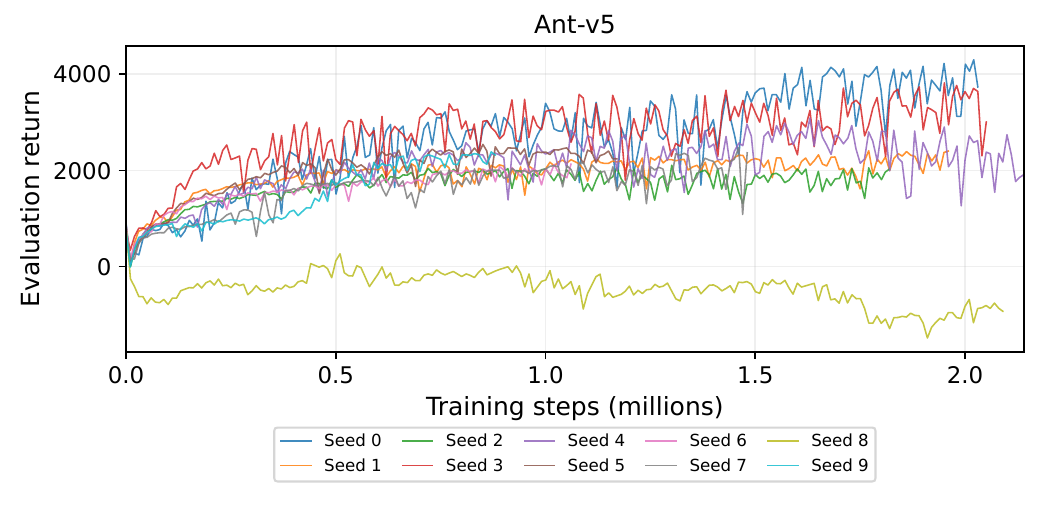}
\caption{Ant-v5 per-seed periodic evaluation histories. Every line uses actual observations through that run's final evaluation.}
\label{fig:ant}
\end{figure}
\begin{figure}[!htbp]
\centering\includegraphics[width=.98\linewidth]{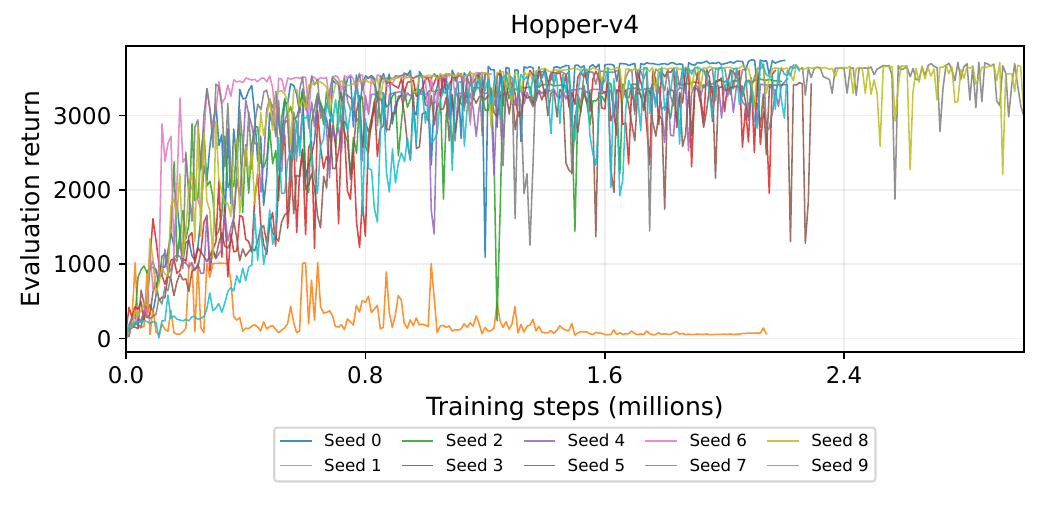}
\caption{Hopper-v4 per-seed periodic evaluation histories. Every line uses actual observations through that run's final evaluation.}
\label{fig:hopper}
\end{figure}
\begin{figure}[!htbp]
\centering\includegraphics[width=.98\linewidth]{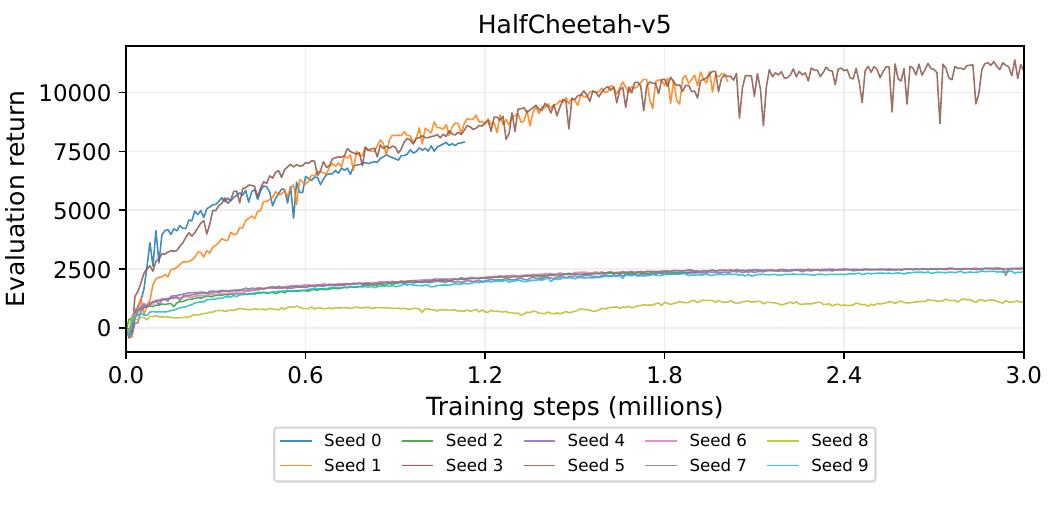}
\caption{HalfCheetah-v5 per-seed periodic evaluation histories. Every line uses actual observations through that run's final evaluation.}
\label{fig:halfcheetah}
\end{figure}
\FloatBarrier
\subsection{Update paths at matched interaction budgets}
\label{app:strategydetail}
Return, process time and channel count in Table~\ref{tab:strategyfull} come from the same latest eligible evaluation at each budget. The common 1.7M horizon is the largest 0.1M grid point covered by all five STE and four stop-gradient runs, chosen from step coverage rather than returns. The straight-through and stop-gradient runs were still in progress when we finalized the reported results. The surrogate-forward histories equal primary seeds 0-4 over these budgets. At 1.4M, their median is 1557, whereas the full 15-seed PQC cohort has median 3967. The medians differ because the update-path comparison uses a five-seed subset of the 15-seed primary cohort, not additional independent surrogate-forward runs.
\begin{table}[!htb]
\centering\small\setlength{\tabcolsep}{3pt}
\begin{tabular}{lrrrrr}
\toprule
Budget & Update path & $n$ & Return & Hours & Calls \\
\midrule
1.0M & Surrogate-forward & 5 & 1,007 & 10.9 & 1.11M \\
1.0M & Hard-forward STE & 5 & 294 & 205.9 & 380.83M \\
1.0M & Hard-forward stop-gradient & 4 & 506 & 194.0 & 380.89M \\
1.4M & Surrogate-forward & 5 & 1,557 & 15.6 & 1.61M \\
1.4M & Hard-forward STE & 5 & 308 & 290.9 & 534.86M \\
1.4M & Hard-forward stop-gradient & 4 & 533 & 273.1 & 534.91M \\
1.7M & Surrogate-forward & 5 & 3,316 & 19.1 & 2.05M \\
1.7M & Hard-forward STE & 5 & 395 & 354.8 & 650.38M \\
1.7M & Hard-forward stop-gradient & 4 & 521 & 331.6 & 650.46M \\
\bottomrule
\end{tabular}

\caption{Three update paths at common budgets. All horizons reuse the same runs. Times accumulate in-process wall time across resumes, include periodic evaluation, and exclude scheduler waiting. Channel counts also include evaluation.}
\label{tab:strategyfull}
\end{table}
\begin{table}[!htb]
\centering\small\setlength{\tabcolsep}{3pt}
\begin{tabular}{lrrrrr}
\toprule
Budget & Comparison & $n_A/n_B$ & PoI & Raw $p$ & Holm $p$ \\
\midrule
1.0M & Surrogate vs. STE & 5/5 & 0.80 & 0.151 & 0.302 \\
1.0M & Surrogate vs. Stop-gradient & 5/4 & 0.75 & 0.286 & 0.302 \\
1.0M & STE vs. Stop-gradient & 5/4 & 0.10 & 0.063 & 0.190 \\
1.4M & Surrogate vs. STE & 5/5 & 0.92 & 0.032 & 0.095 \\
1.4M & Surrogate vs. Stop-gradient & 5/4 & 0.75 & 0.286 & 0.286 \\
1.4M & STE vs. Stop-gradient & 5/4 & 0.10 & 0.063 & 0.127 \\
1.7M & Surrogate vs. STE & 5/5 & 0.88 & 0.056 & 0.167 \\
1.7M & Surrogate vs. Stop-gradient & 5/4 & 0.80 & 0.190 & 0.222 \\
1.7M & STE vs. Stop-gradient & 5/4 & 0.15 & 0.111 & 0.222 \\
\bottomrule
\end{tabular}

\caption{Exploratory exact two-sided Mann--Whitney comparisons. Each budget forms a three-comparison Holm family; corrections do not make the repeated horizons independent or account for post-hoc budget selection. PoI gives the observed probability of improvement.}
\label{tab:strategytests}
\end{table}
The five-versus-four comparison uses all available seeds. Table~\ref{tab:fourstrategy} additionally restricts all paths to labels 0--3, without claiming a paired experimental design. The surrogate-forward median remains higher, but the small-sample tests remain uncertain. The raw $p=0.032$ surrogate-versus-STE contrast at 1.4M is not a multiplicity-adjusted finding.
\begin{table}[!htb]
\centering\small\setlength{\tabcolsep}{4pt}
\begin{tabular}{lrrrr}
\toprule
Budget & Surrogate & STE & Stop-gradient & Surrogate/stop $p$ \\
\midrule
1.0M & 779 & 300 & 506 & 0.486 \\
1.4M & 1,040 & 336 & 533 & 0.486 \\
1.7M & 1,997 & 326 & 521 & 0.343 \\
\bottomrule
\end{tabular}

\caption{Four-seed subset sensitivity. The last column is an unadjusted exploratory test. These are the same runs as the full-cohort table, not a new experiment.}
\label{tab:fourstrategy}
\end{table}

\subsection{Same-seed re-executions and resumed sampling}
\label{app:reexecutions}
The forward-path reference (five runs), ten-qubit/1000-shot ablation reference (five), and ten-qubit/width-2048 reference (three) reuse primary seed labels and have long identical histories. We compare episode indices, interaction counts, training returns, episode lengths, periodic scores, losses and channel counters at tolerance $10^{-6}$, excluding elapsed time. Table~\ref{tab:reexecutions} gives the final identical point and separately evaluated terminal scores. For the forward-path comparison, all five histories remain identical beyond 1.7M, justifying their identification as primary seeds 0--4 within that budget.
\begin{table}[!htb]
\centering\small\setlength{\tabcolsep}{3pt}
\begin{tabular}{lrrrr}
\toprule
Re-execution & Seed & Last identical (M) & Primary final & Re-execution final \\
\midrule
Forward & 0 & 2.098 & 4,453 & 4,675 \\
Forward & 1 & 2.154 & 4,326 & 4,839 \\
Forward & 2 & 2.101 & 901 & 958 \\
Forward & 3 & 1.917 & 270 & 242 \\
Forward & 4 & 2.065 & 5,376 & 5,344 \\
Shot reference & 0 & 2.098 & 4,453 & 4,884 \\
Shot reference & 1 & 2.154 & 4,326 & 4,015 \\
Shot reference & 2 & 1.931 & 901 & 1,554 \\
Shot reference & 3 & 1.881 & 270 & 660 \\
Shot reference & 4 & 2.089 & 5,376 & 5,391 \\
Width reference & 0 & 1.806 & 4,453 & 71 \\
Width reference & 1 & 2.154 & 4,326 & 4,561 \\
Width reference & 2 & 1.580 & 901 & 589 \\
\bottomrule
\end{tabular}

\caption{Thirteen same-seed re-executions with shared training prefixes. Scores after divergence are not extra independent primary seeds. ``Last identical'' refers to the episode-level log comparison, not only to smoothed or periodic curves.}
\label{tab:reexecutions}
\end{table}
The first discrepancies occur between approximately 1.58M and 2.15M interactions, around the first process-restart regime. The CUDA-Q shot stream is neither explicitly seeded nor checkpointed, so resumed executions can follow different subsequent sample paths. The observed divergence is consistent with restart-sensitive sampling, but the logs do not separate that effect from other execution differences. We report the reference re-executions separately and do not treat them as additional independent primary runs or as a controlled estimate of a restart effect.

\subsection{Post-hoc progress, threshold crossings and response fit}
\label{app:progress}
To summarize learning across the full interaction budget, define one value per seed:
\begin{equation}
 A_i=\frac1{30}\sum_{j=1}^{30}\widehat R_i(10^5j),
\end{equation}
where $\widehat R_i(t)$ is the latest genuine ten-episode evaluation at or before $t$. This is a grid-averaged return, not an exact integral of a continuous learning curve. It uses the same fixed 15-seed cohort per primary configuration and was selected after observing the trajectories. Tables~\ref{tab:progress}--\ref{tab:progresstests} report its distribution and four-comparison inference. The correction addresses the four method contrasts, not selection of this statistic among possible learning summaries.
\begin{table}[!htb]
\centering\small\setlength{\tabcolsep}{6pt}
\begin{tabular}{lrr}
\toprule
Method & Median grid average & IQM [95\% CI] \\
\midrule
PQC & 3,144 & 2,638 [1,579, 3,444] \\
FC & 751 & 913 [560, 1,590] \\
FCM & 959 & 1,130 [809, 1,648] \\
CSB & 801 & 924 [598, 1,498] \\
RBG & 470 & 452 [376, 517] \\
\bottomrule
\end{tabular}

\caption{Post-hoc grid-averaged returns across 0.1--3M interactions. Each entry aggregates 15 seed-level values.}
\label{tab:progress}
\end{table}
\begin{table}[!htb]
\centering\small\setlength{\tabcolsep}{5pt}
\begin{tabular}{lrrr}
\toprule
PQC versus & PoI [95\% CI] & Exact MW $p$ & Holm $p$ \\
\midrule
FC & 0.78 [0.60, 0.93] & 0.009 & 0.026 \\
FCM & 0.73 [0.52, 0.92] & 0.033 & 0.033 \\
CSB & 0.76 [0.57, 0.93] & 0.013 & 0.026 \\
RBG & 0.94 [0.83, 1.00] & $<10^{-3}$ & $<10^{-3}$ \\
\bottomrule
\end{tabular}

\caption{Post-hoc learning-progress comparisons against PQC. Confidence intervals resample training runs; p-values use exact two-sided tests and a four-comparison Holm adjustment.}
\label{tab:progresstests}
\end{table}

\paragraph{First threshold crossing.}
The first genuine periodic evaluation at or above 4000 is recorded for each seed, with non-crossing runs right-censored at 3M. A transient crossing and a high terminal score are different events: FC crosses in six runs and FCM in seven, but each has four terminal scores above 4000. PQC crosses in ten runs, CSB in four and RBG in none. Figure~\ref{fig:firstcrossing} uses all 15 runs at every plotted step, rather than averaging crossing times over successful runs alone.
\begin{table}[!htb]
\centering\small\setlength{\tabcolsep}{6pt}
\begin{tabular}{lrrrrr}
\toprule
Seed & PQC & FC & FCM & CSB & RBG \\
\midrule
0 & 1.751 & $>3$ & $>3$ & $>3$ & $>3$ \\
1 & 1.660 & $>3$ & $>3$ & $>3$ & $>3$ \\
2 & $>3$ & 1.600 & 2.561 & $>3$ & $>3$ \\
3 & $>3$ & $>3$ & $>3$ & $>3$ & $>3$ \\
4 & 0.491 & $>3$ & $>3$ & 1.750 & $>3$ \\
5 & $>3$ & 2.310 & $>3$ & $>3$ & $>3$ \\
6 & 0.731 & 1.360 & 1.740 & $>3$ & $>3$ \\
7 & 1.031 & $>3$ & 1.840 & 0.701 & $>3$ \\
8 & 1.260 & 2.890 & $>3$ & $>3$ & $>3$ \\
9 & 0.840 & 2.030 & $>3$ & $>3$ & $>3$ \\
10 & 1.361 & $>3$ & 2.650 & 2.170 & $>3$ \\
11 & $>3$ & 2.431 & 2.120 & $>3$ & $>3$ \\
12 & 1.111 & $>3$ & $>3$ & 1.970 & $>3$ \\
13 & $>3$ & $>3$ & 2.800 & $>3$ & $>3$ \\
14 & 0.911 & $>3$ & 1.111 & $>3$ & $>3$ \\
\bottomrule
\end{tabular}

\caption{First threshold crossing in millions of interactions. $>3$ denotes no observed crossing by the budget, not a predicted future crossing.}
\label{tab:crossings}
\end{table}

\paragraph{Behavior associated with high returns.}
For each primary PQC run, Table~\ref{tab:behavior} summarizes training episodes ending in $[2.5\mathrm M,3\mathrm M]$. Among the ten runs with terminal scores above 4000, 69.8--96.7\% of these episodes last the full 1000 steps, and median reward per step is 4.82--5.39. Humanoid-v4's reward includes a healthy reward of five, forward reward weight 1.25 and action-cost weight 0.1 \citep{gymnasiumdocs}. These logs support sustained episode survival; total reward alone does not recover forward speed or separate the reward components. They should not be read as evidence of optimal or fast locomotion.
\begin{table}[!htb]
\centering\small\setlength{\tabcolsep}{5pt}
\begin{tabular}{lrrrrrr}
\toprule
Seed & Terminal return & Full episodes (\%) & Reward / step & Late BCE & $\widehat H_{1000}$ & $\widehat{\sqrt v}_{1000}$ \\
\midrule
0 & 4,453 & 87.1 & 4.88 & 0.240 & 4.54 & 1.03 \\
1 & 4,326 & 76.7 & 4.82 & 0.265 & 4.29 & 1.01 \\
2 & 901 & 0.0 & 4.89 & 0.661 & 8.86 & 1.49 \\
3 & 270 & 0.0 & 5.35 & 0.662 & 8.92 & 1.49 \\
4 & 5,376 & 96.7 & 5.33 & 0.296 & 5.39 & 1.14 \\
5 & 596 & 0.0 & 4.99 & 0.653 & 8.59 & 1.47 \\
6 & 5,355 & 88.0 & 5.39 & 0.307 & 5.43 & 1.14 \\
7 & 5,238 & 95.3 & 5.32 & 0.232 & 3.78 & 0.95 \\
8 & 4,751 & 95.7 & 4.94 & 0.183 & 2.75 & 0.79 \\
9 & 4,750 & 94.4 & 4.93 & 0.466 & 6.52 & 1.26 \\
10 & 4,835 & 87.2 & 4.90 & 0.163 & 2.47 & 0.74 \\
11 & 769 & 0.0 & 5.00 & 0.557 & 7.41 & 1.35 \\
12 & 5,231 & 92.3 & 5.24 & 0.228 & 3.37 & 0.90 \\
13 & 2,846 & 0.1 & 4.93 & 0.417 & 4.11 & 0.99 \\
14 & 4,588 & 69.8 & 5.16 & 0.582 & 8.45 & 1.45 \\
\bottomrule
\end{tabular}

\caption{Primary PQC behavior, fitting loss and visited-control estimates. Terminal returns come from the primary hybrid evaluations; episode completion, reward-per-step and BCE summaries concern late training. The last two columns are median estimated conditional entropy (bits) and variance coefficient over 64 controls from surrogate-actor rollouts at $S=1000$ (Appendix~\ref{app:visitedcontrols}); they do not use hybrid-policy rollouts.}
\label{tab:behavior}
\end{table}

\paragraph{Response fit and training outcome.}
Proposition~\ref{prop:bce} separates response-fitting error from the stochasticity inherent in the labels. The irreducible population risk is
\begin{equation}
 \mathcal L(\mu)=\frac1N\E_{u\sim\rho}\!\left[\sum_{k=1}^N h_{\mathrm{nats}}(\mu_k(u))\right],
 \qquad h_{\mathrm{nats}}(p)=-p\log p-(1-p)\log(1-p).
\end{equation}
Only the excess risk $\mathcal L(f)-\mathcal L(\mu)$ measures fitting error. A change in visited controls or shot count can change the irreducible term as well as the fitted model. The channel studies in Appendices~\ref{app:entropy} and~\ref{app:visitedcontrols} characterize changes in the output law. Their joint entropies are not the sum of marginal entropies, and neither the random controls nor the surrogate-policy visits estimate this risk on the original replay distribution.

\begin{figure}[!htb]
\centering\includegraphics[width=.94\linewidth]{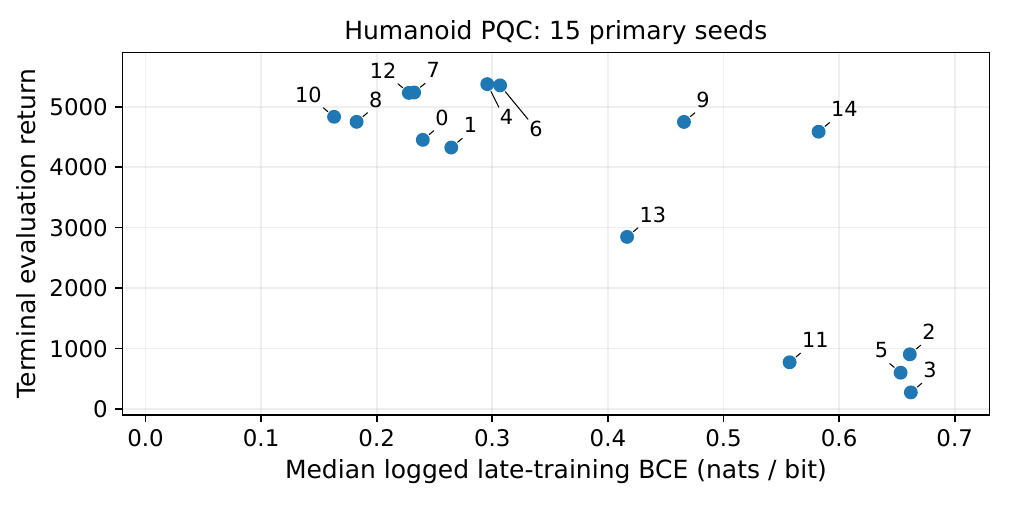}
\caption{Late fitting loss and terminal return for the 15 primary PQC seeds. Labels identify seeds. This association is post hoc and does not identify causal direction or Jacobian accuracy.}
\label{fig:bcereturn}
\end{figure}

The logged \texttt{q\_value\_loss} is the last surrogate fitting minibatch loss in an update, recorded at episode end; it is not a held-out population BCE. Taking its median over 2.5--3M interactions yields a post-hoc primary-seed Spearman correlation of $-0.686$ with terminal return ($n=15$, unadjusted $p=0.0048$). The median of these losses is 0.252 for the ten high-return runs and 0.653 for the other five (Figure~\ref{fig:bcereturn}).
Across 88 completed PQC surrogate-forward Humanoid executions, the corresponding descriptive correlation is $-0.637$. This pool includes 13 shared-prefix re-executions and different configurations; it is not an 88-independent-seed test. Excluding those re-executions leaves 75 executions with correlation $-0.644$, still confounded by architecture and fitting choices. For the autoregressive setting the logged loss is a conditional negative log-likelihood rather than the base per-bit BCE. We therefore make no pooled causal or confirmatory significance claim. Across the 15 primary checkpoints, late BCE is positively associated with estimated aggregated entropy (Spearman coefficient $+0.961$, unadjusted $p=1.30\times10^{-8}$; Appendix~\ref{app:visitedcontrols}). At 1000 shots, median entropy across the ten high-return seeds is 4.414 bits, compared with 8.589 for the other five; the corresponding variance coefficients are 1.018 and 1.472. Entropy correlates with recorded terminal return at $-0.604$ (unadjusted $p=0.0172$), with an exact two-sided group comparison of $p=0.0280$. These measurements use controls from the classical surrogate actor, not the hybrid policy or training replay, and do not identify causality or isolate excess BCE. The high-entropy, high-return seed 14 and the lower-entropy, lower-return seed 13 also rule out a deterministic separation by entropy.

\subsection{Process time and evaluation-query allocation}
\label{app:wallallocation}
\begin{table}[!htb]
\centering\small\setlength{\tabcolsep}{12pt}
\begin{tabular}{lr}
\toprule
Method & Median process hours \\
\midrule
PQC & 35.1 \\
FC & 8.8 \\
FCM & 8.2 \\
CSB & 41.4 \\
RBG & 8.9 \\
\bottomrule
\end{tabular}

\caption{Median cumulative logged process time at the end of the full 3M primary campaigns, over 15 runs each. These totals precede the separate terminal evaluation and are not normalized across machines.}
\label{tab:wallfull}
\end{table}
In a base PQC run, interaction calls equal the logged training count minus 1001 warm-up actions. The remaining logged calls are periodic evaluations; the difference between the final JSON counter and the final metrics counter is the separate terminal evaluation. Across primary runs, total calls including terminal evaluation range from 3.24M to 5.55M, and evaluation contributes 7.3--46.0\%. Longer evaluation episodes naturally require more circuit calls. This allocation explains why an end-to-end invocation ratio is not simply 384, the average additional calls per update in sampled-forward training.
\begin{table}[!htb]
\centering\small\setlength{\tabcolsep}{6pt}
\begin{tabular}{lrrr}
\toprule
Seed & Interactions (M) & All calls (M) & Evaluation share (\%) \\
\midrule
0 & 3.001 & 4.521 & 33.7 \\
1 & 3.001 & 4.556 & 34.1 \\
2 & 3.000 & 3.385 & 11.4 \\
3 & 3.000 & 3.236 & 7.3 \\
4 & 3.001 & 5.551 & 46.0 \\
5 & 3.000 & 3.312 & 9.5 \\
6 & 3.000 & 5.328 & 43.7 \\
7 & 3.000 & 5.137 & 41.6 \\
8 & 3.001 & 4.941 & 39.3 \\
9 & 3.000 & 5.238 & 42.7 \\
10 & 3.000 & 4.934 & 39.2 \\
11 & 3.000 & 3.431 & 12.6 \\
12 & 3.000 & 5.002 & 40.0 \\
13 & 3.000 & 3.720 & 19.4 \\
14 & 3.000 & 5.169 & 42.0 \\
\bottomrule
\end{tabular}

\caption{Primary PQC invocation counts, including periodic and separate terminal evaluation. Environment-step budgets count training interactions only.}
\label{tab:queryallocation}
\end{table}

\FloatBarrier
\clearpage
\section{Deployment substitutions and channel laws at visited controls}
\label{app:deployment}
This appendix evaluates the stored policies without further optimization. It separates the role of sampled features at deployment from the training procedure that produced the networks. The hybrid reference is the recorded terminal evaluation used in the primary comparison.

\subsection{Evaluation protocol and classical calibration}
\label{app:deploymentcalibration}
For each of the 15 primary PQC checkpoints, we retain the learned pre-network, post-network and surrogate, and replace only the features passed to the post-network. The bypass $c_\theta(s)$ is unchanged. Each substitute receives 30 evaluation episodes without external exploration noise, using the same environment-reset seeds as the recorded terminal evaluation (Appendix~\ref{app:implementation}). Episode index $e\in\{0,\ldots,29\}$ for training seed $i$ uses
\begin{equation*}
 r_{i,e}=\bigl(1{,}000{,}003(i+777{,}001)+999{,}999\cdot1000+e\bigr)\bmod(2^{31}-1).
\end{equation*}
The logged checkpoint timestep matches the corresponding terminal-evaluation timestep in every run. Reset matching does not reproduce the original quantum shot stream. The hybrid reference is its recorded evaluation, not a fresh quantum evaluation on the local stack.

As a calibration, we re-evaluate the 15 FCM checkpoints with their original deterministic bottleneck. Six local means differ from the recorded means by less than one return point. The median absolute difference is 14.59 points, the mean signed difference is $+27.78$, and the largest absolute difference is 392.44. The same four seeds remain above 4000. Table~\ref{tab:deploymentcalibration} reports the paired calibration results for all 15 seeds; the primary comparisons use the original FCM evaluations.

For this calibration, let $D_{i,e}=R^{\mathrm{local}}_{i,e}-R^{\mathrm{recorded}}_{i,e}$ and $\overline D_i=30^{-1}\sum_eD_{i,e}$. The standard error is $s_{D,i}/\sqrt{30}$, where $s_{D,i}$ uses the sample standard deviation with denominator 29. All mean differences are within two such \emph{paired-difference} standard errors; the maximum ratio is 1.41. This describes the size of the discrepancies relative to episode-level variation, not equivalence of the software stacks or bitwise replication. Seed 11, for example, has a recorded standard deviation of 6.96 but a local standard deviation of 690.16, so the discrepancy is not confined to policies that were already variable in the original evaluation. We therefore compare substitutes at the training-seed level and do not interpret small local--recorded differences as isolated feature effects.
\begin{table}[!htb]
\centering\small\setlength{\tabcolsep}{4pt}
\begin{tabular}{lrrrrr}
\toprule
Seed & Recorded & Local & $\overline D$ & $\mathrm{SE}(\overline D)$ & $|\overline D|/\mathrm{SE}$ \\
\midrule
0 & 654.1 & 654.3 & +0.17 & 0.22 & 0.77 \\
1 & 1,891.2 & 1,959.4 & +68.26 & 98.73 & 0.69 \\
2 & 2,664.8 & 2,692.4 & +27.56 & 54.52 & 0.51 \\
3 & 797.1 & 798.1 & +0.99 & 3.82 & 0.26 \\
4 & 400.8 & 401.4 & +0.54 & 1.48 & 0.37 \\
5 & 2,399.1 & 2,305.2 & -93.97 & 277.61 & 0.34 \\
6 & 4,379.9 & 4,617.7 & +237.74 & 168.39 & 1.41 \\
7 & 4,806.9 & 4,806.9 & -0.03 & 1.13 & 0.02 \\
8 & 1,253.5 & 1,268.1 & +14.59 & 27.01 & 0.54 \\
9 & 777.1 & 776.6 & -0.60 & 3.20 & 0.19 \\
10 & 3,394.4 & 3,786.8 & +392.44 & 292.31 & 1.34 \\
11 & 4,787.3 & 4,661.4 & -125.87 & 125.91 & 1.00 \\
12 & 2,563.4 & 2,445.3 & -118.12 & 298.12 & 0.40 \\
13 & 2,137.9 & 2,151.5 & +13.62 & 247.05 & 0.06 \\
14 & 4,720.5 & 4,719.9 & -0.63 & 0.54 & 1.16 \\
\bottomrule
\end{tabular}

\caption{FCM calibration on matched reset seeds, with 30 episodes per checkpoint. Recorded and local columns are episode-mean returns. The last two columns use the standard error of the paired episode differences, not the standard error of the local scores alone. The recorded values are the FCM reference evaluations used in the primary comparison.}
\label{tab:deploymentcalibration}
\end{table}

\subsection{Feature substitutions at fixed network weights}
\label{app:deploymentsubstitutes}
Writing $(u,c)=\operatorname{Pre}_\theta(s)$, every variant acts as $\operatorname{Post}_\psi(z,c)$, followed by clipping to the environment's action bounds. The six choices of $z$ are
\begin{equation*}
\begin{array}{ll}
\text{Surrogate mean:} & z=f_\phi(u),\\
\text{Rounded surrogate:} & z_k=\mathbf 1\{f_{\phi,k}(u)\ge 1/2\},\\
\text{Sampled surrogate:} & z_k\sim\operatorname{Bernoulli}(f_{\phi,k}(u))\ \text{independently},\\
\text{Constant half:} & z=\tfrac12\mathbf 1,\\
\text{Zeros:} & z=0,\\
\text{Fair bits:} & z_k\sim\operatorname{Bernoulli}(1/2)\ \text{independently}.
\end{array}
\end{equation*}
The sampled variants draw fresh features at each action. Their CPU random generators are initialized once per seed and variant, with seed $i$; these are not the historical quantum draws. Independent Bernoulli coordinates do not reproduce the circuit's joint readout law. Constants and fair bits also leave the learned, state-dependent bypass intact: they are not interventions on the entire state representation.

Table~\ref{tab:deploymentsummary} aggregates the 15 checkpoint scores; Table~\ref{tab:deploymentseeds} gives each checkpoint. The surrogate mean, rounded surrogate and sampled surrogate all retain the same ten high-return seeds as the recorded hybrid. Constant-half and fair-bit features retain nine, losing seed 1; zeros retain six, losing seeds 1, 4, 6 and 10. No substitute promotes any of the five originally lower-return seeds above 4000.
\begin{table}[!htb]
\centering\small\setlength{\tabcolsep}{4pt}
\begin{tabular}{lrrrrr}
\toprule
Features & Seeds & Mean & Median & $\ge4000$ & Median $\Delta_i$ \\
\midrule
Recorded hybrid & 15 & 3,619 & 4,588 & 10/15 & -- \\
Surrogate mean & 15 & 3,684 & 4,737 & 10/15 & +66 \\
Rounded surrogate & 15 & 3,600 & 4,613 & 10/15 & +34 \\
Sampled surrogate & 15 & 3,606 & 4,726 & 10/15 & +0 \\
Constant half & 15 & 3,486 & 4,535 & 9/15 & -151 \\
Fair bits & 15 & 3,375 & 4,513 & 9/15 & -4 \\
Zeros & 15 & 2,763 & 2,978 & 6/15 & -173 \\
\bottomrule
\end{tabular}

\caption{Deployment substitutions without further training. Each seed score averages 30 episodes; mean and median columns aggregate the 15 seed scores. $\Delta_i$ subtracts that seed's recorded hybrid score. The median difference is a median of differences, not a difference of medians. The recorded-hybrid row reports the 450 terminal episodes from the primary comparison; only the feature substitutes are evaluated in this deployment study.}
\label{tab:deploymentsummary}
\end{table}
\begin{table}[!htb]
\centering\small\setlength{\tabcolsep}{3.2pt}
\begin{tabular}{lrrrrrrr}
\toprule
Seed & Hybrid & Mean & Round & Sample & Half & Fair bits & Zeros \\
\midrule
0 & 4,453 & 4,737 & 4,629 & 4,726 & 4,840 & 4,513 & 4,458 \\
1 & 4,326 & 4,717 & 4,613 & 4,589 & 2,952 & 1,950 & 1,525 \\
2 & 901 & 868 & 1,068 & 1,022 & 786 & 1,000 & 924 \\
3 & 270 & 520 & 282 & 277 & 687 & 272 & 292 \\
4 & 5,376 & 5,377 & 5,373 & 5,376 & 5,223 & 5,237 & 2,978 \\
5 & 596 & 664 & 593 & 558 & 674 & 568 & 580 \\
6 & 5,355 & 5,421 & 5,390 & 5,238 & 5,444 & 5,405 & 3,415 \\
7 & 5,238 & 5,408 & 5,401 & 5,391 & 5,395 & 5,356 & 5,110 \\
8 & 4,751 & 4,887 & 4,898 & 4,891 & 4,877 & 4,750 & 4,864 \\
9 & 4,750 & 4,738 & 4,489 & 4,745 & 4,535 & 4,846 & 4,590 \\
10 & 4,835 & 4,831 & 4,713 & 4,833 & 4,591 & 4,829 & 1,997 \\
11 & 769 & 651 & 892 & 739 & 617 & 634 & 596 \\
12 & 5,231 & 5,265 & 5,282 & 5,285 & 5,033 & 5,227 & 4,434 \\
13 & 2,846 & 2,486 & 2,159 & 2,216 & 2,205 & 1,767 & 1,304 \\
14 & 4,588 & 4,684 & 4,223 & 4,200 & 4,426 & 4,272 & 4,385 \\
\bottomrule
\end{tabular}

\caption{All primary-checkpoint deployment means, rounded to integers. ``Mean'', ``Round'' and ``Sample'' refer to the three surrogate features defined above. Each checkpoint is evaluated with all six feature substitutes. Reset seeds match the recorded hybrid evaluation, but the local and historical execution stacks are not demonstrated to be identical.}
\label{tab:deploymentseeds}
\end{table}

\begin{samepage}
Among the ten originally high-return seeds, the median surrogate-minus-hybrid change is $+80.85$ points. This is a descriptive summary, not a non-inferiority test. Seeds 4 and 10 differ by $+1.09$ and $-4.09$ points, with both their local and recorded standard deviations below ten. For seed 10, the small negative difference exceeds two paired-difference standard errors; closeness in return does not mean indistinguishability at that precision. The hybrid mean exceeds every substitute for seeds 10 \emph{and} 13. Its margins over the best substitute are 2.22 and 360.42 points, respectively. These contrasts distinguish minor numerical changes from the larger loss for seed 13.\par
\end{samepage}

\paragraph{Classical deployment and training procedures.}
The stored surrogate mean supplies a deployable classical actor,
\begin{equation*}
 \widetilde\pi(s)=\operatorname{Post}_\psi\!\bigl(f_\phi(u_\theta(s)),c_\theta(s)\bigr),
\end{equation*}
without an additional fitting or distillation stage. Its layer shapes, activation functions and bypass match the configured FCM actor. The weights and training objectives differ: FCM learns the bottleneck through the policy loss, whereas the PQC surrogate learns circuit responses by BCE and the circuit supplies behavior during interaction. This follow-up therefore gives a same-architecture deployment comparison between two training procedures, with ten versus four high-return seeds. The primary tables retain their sampled-hybrid evaluations; the deterministic actor is evaluated separately here.

The substitutions support retaining strong policies without executing the circuit at deployment. Success of the constant and random substitutes for many checkpoints also limits any general claim that state-dependent sampled features are necessary for their high returns. Conversely, the losses for some seeds show that not every checkpoint is feature-insensitive. The experiments neither identify a causal benefit of quantum behavior during training nor show that it could not be replaced by another behavior channel; they change deployment features after training, not the training data-generation process.

\subsection{Channel characterisation on surrogate-policy rollouts}
\label{app:visitedcontrols}
For each primary PQC checkpoint, 64 control vectors are sampled without replacement from the state visits collected during the 30 surrogate-mean evaluation episodes. The sampling generator is seeded by the training-seed label. These vectors come from the \emph{classical update actor's rollouts}, not the hybrid policy's trajectories, the original training replay distribution, or the separate live-diagnostic runs. Controls within a seed may be dependent; the training seed remains the unit of outcome comparisons.

We characterize these controls using the ten-qubit, ten-layer circuit with bit-flip and phase-flip probabilities of $10^{-3}$ after each U3 gate, as in Appendix~\ref{app:entropy}. Density-matrix evolution computes the Born law up to floating-point arithmetic. At each control, we estimate the mode-output distribution from 20,000 bags of $S=100$ or $S=1000$ shots, using smallest-code tie-breaking. The sampler is reset to seed zero for each checkpoint. This study uses a specified tie convention; the training backend's convention was not recorded. Aggregated entropy is estimated using the plug-in value with the Miller--Madow correction. The Born calculation is therefore exact within the simulator, while aggregated entropy and the variance coefficient remain Monte Carlo estimates.

For seed $i$, let $u_{i,1},\ldots,u_{i,64}$ denote the sampled controls. Table~\ref{tab:behavior} reports the seed-level summaries
\begin{equation*}
 \widehat H_{i,1000}=\operatorname*{median}_{\ell=1,\ldots,64}
     \widehat H\!\left(P_{u_{i,\ell},1000}\right),\qquad
 \widehat{\sqrt v}_{i,1000}=\operatorname*{median}_{\ell=1,\ldots,64}
     \sqrt{\sum_{k=1}^{10}\widehat\mu_k(u_{i,\ell})\bigl(1-\widehat\mu_k(u_{i,\ell})\bigr)}.
\end{equation*}
These are medians of conditional quantities, not the entropy or variance of the pooled output distribution. Table~\ref{tab:visitedchannels} additionally gives both shot budgets and
$\widehat p^\star_S$, the estimated probability that the aggregate output equals the smallest-code Born argmax, summarized by its median over controls.

The per-seed median saturation fraction lies in $[0.945,0.995]$. This fraction counts all 220 controls below 0.05 or above 0.95, including routing controls, rather than measuring only rotation angles. Median Born entropies range from 7.66 to 9.88 bits. These ranges concern \emph{per-seed medians}; individual controls span a broader range. Saturation therefore coexists with substantially different readout laws across checkpoints, and is not by itself a diagnosis of an almost uniform output.
\begin{table}[!htb]
\centering\small\setlength{\tabcolsep}{3pt}
\begin{tabular}{lrrrrrrr}
\toprule
Seed & Sat. & $H(\pi_u)$ & $\widehat H_{100}$ & $\widehat H_{1000}$ & $\widehat{\sqrt v}_{100}$ & $\widehat{\sqrt v}_{1000}$ & $\widehat p^\star_{1000}$ \\
\midrule
0 & 0.977 & 8.10 & 6.26 & 4.54 & 1.22 & 1.03 & 0.1204 \\
1 & 0.964 & 8.02 & 5.98 & 4.29 & 1.21 & 1.01 & 0.0993 \\
2 & 0.991 & 9.61 & 8.58 & 8.86 & 1.46 & 1.49 & 0.0024 \\
3 & 0.995 & 9.88 & 9.03 & 8.92 & 1.48 & 1.49 & 0.0032 \\
4 & 0.973 & 8.14 & 6.34 & 5.39 & 1.23 & 1.14 & 0.0411 \\
5 & 0.991 & 9.61 & 8.52 & 8.59 & 1.45 & 1.47 & 0.0034 \\
6 & 0.977 & 8.57 & 6.98 & 5.43 & 1.29 & 1.14 & 0.0640 \\
7 & 0.964 & 7.96 & 5.37 & 3.78 & 1.13 & 0.95 & 0.1517 \\
8 & 0.964 & 7.66 & 4.67 & 2.75 & 1.03 & 0.79 & 0.2440 \\
9 & 0.982 & 8.79 & 7.23 & 6.52 & 1.32 & 1.26 & 0.0171 \\
10 & 0.945 & 7.66 & 4.45 & 2.47 & 1.01 & 0.74 & 0.3848 \\
11 & 0.982 & 9.11 & 7.97 & 7.41 & 1.41 & 1.35 & 0.0096 \\
12 & 0.964 & 7.81 & 5.37 & 3.37 & 1.12 & 0.90 & 0.2065 \\
13 & 0.950 & 7.96 & 5.47 & 4.11 & 1.14 & 0.99 & 0.1011 \\
14 & 0.986 & 9.40 & 8.43 & 8.45 & 1.45 & 1.45 & 0.0030 \\
\bottomrule
\end{tabular}

\caption{Visited-control characterisation. Each entry is a median over 64 controls for one primary checkpoint; entropies are in bits. ``Sat.'' is the fraction of all 220 controls outside $[0.05,0.95]$. $\widehat H_S$ and $\widehat{\sqrt v}_S$ concern the mode-aggregated law; $H(\pi_u)$ concerns individual shots. Controls are collected under the surrogate-mean policy, not the hybrid policy.}
\label{tab:visitedchannels}
\end{table}
\FloatBarrier

\paragraph{Association with recorded training outcomes.}
At $S=1000$, the medians across the ten high-return seeds and five other seeds are respectively 4.414 and 8.589 bits for $\widehat H_{i,1000}$, 1.018 and 1.472 for $\widehat{\sqrt v}_{i,1000}$, and 0.1099 and 0.0034 for $\widehat p^\star_{i,1000}$. The entropy--terminal-return Spearman coefficient is $-0.604$ ($n=15$, two-sided asymptotic $p=0.0172$); the exact two-sided Mann--Whitney comparison between the ten and five seed summaries gives $U=7$ and $p=0.0280$. The entropy--late-BCE coefficient is $+0.961$ ($p=1.30\times10^{-8}$). All are post-hoc, unadjusted associations; the 960 controls are not treated as 960 independent training outcomes. Entropy and the variance coefficient have the same seed ordering here and are not independent confirmations.

At $S=100$, the corresponding high-return/other-seed entropy medians are 6.120/8.522 bits and the variance-coefficient medians are 1.211/1.446. The two shot budgets characterize the same controls rather than policies retrained at those budgets. High-return seed 14 has entropy 8.45 bits at 1000 shots, while lower-return seed 13 has 4.11 bits. These exceptions, the non-hybrid control collection, and the absence of a training intervention preclude a claim that low entropy causes good learning. Joint entropy also differs from the marginal-entropy term in Proposition~\ref{prop:bce}; the measurements do not decompose the logged BCE into irreducible uncertainty and fitting error.

\paragraph{Data and reproducibility.}
We compute deployment scores from individual episode returns, matching checkpoint timesteps and environment-reset seeds to the primary terminal evaluations. We compute the channel summaries from per-control statistics, check the entropy-correction and variance identities, and associate each checkpoint with its late-training BCE using the training-seed label. The FCM calibration is reported separately and is not used to rescale returns.

The records used for this statistical analysis contain episode returns, per-control summaries and channel configurations, but do not include checkpoint weights, saved control vectors, full Born probability vectors or a local software-version manifest. They also lack source hashes for these deployment and visited-control executions. The package to be released upon acceptance will include the protocol scripts, experimental records and statistical-analysis code. These materials allow the reported summaries to be recomputed; the saved records alone do not support independent re-execution of the policy evaluations or density-matrix simulations.
\end{document}